\documentclass[%
 reprint,
 superscriptaddress,
 nofootinbib,
 amsmath,amssymb,
 aps,
 prd,
]{revtex4-2}

\usepackage{graphicx}
\usepackage{dcolumn}
\usepackage{bm}

\usepackage{booktabs}
\usepackage[english]{babel}
\usepackage{amsmath,amssymb,amsbsy,amstext, amsthm, simplewick}
\usepackage{hyperref}
\usepackage{graphicx}
\usepackage{amsfonts}
\usepackage{amssymb}
\usepackage{upgreek}
\usepackage{cancel}
\usepackage{color}
\usepackage{slashed}
\usepackage[dvipsnames]{xcolor}
\usepackage{tikz-cd}
\usepackage{feynmp-auto}
\usepackage{color}
\usepackage{soul}
\usepackage{dsfont}
\usepackage{verbatim}
\usepackage[utf8]{inputenc} 
\usepackage{soul}

\newcommand{\half}{\frac{1}{2}}

\usepackage{empheq}
\newlength\dlf  

\def\@fnsymbol#1{\ensuremath{\ifcase#1\or $\Re$\or $\Im$\or  \else\@ctrerr\fi}}
\renewcommand{\thefootnote}{\fnsymbol{footnote}}

\begin{document}

\title{The Hydrogen Mixing Portal, Its Origins, and Its Cosmological Effects}

\author{Lucas Johns}
\email{NASA Einstein Fellow (ljohns@berkeley.edu)}
\affiliation{Departments of Astronomy and Physics, University of California,
Berkeley, CA 94720, U.S.A.}
\affiliation{Department of Physics, University of California, San Diego, CA 92093, U.S.A.}
\author{Seth Koren}
\email{EFI Oehme Fellow (sethk@uchicago.edu)}
\affiliation{Enrico Fermi Institute, University of Chicago,
Chicago, IL 60637, U.S.A.}
\affiliation{Department of Physics, University of California,
Santa Barbara, CA 93106, U.S.A.}

\begin{abstract}
Hydrogen oscillation into a dark-sector state $H'$ has recently been proposed as a novel mechanism through which hydrogen can be cooled during the dark ages---without direct couplings between the Standard Model and dark matter. 
In this work we demonstrate that the requisite mixing can appear naturally from a microphysical theory, and argue that the startling deviations from standard cosmology are nonetheless consistent with observations.
A symmetric mirror model enforces the necessary degeneracy between $H$ and $H'$, and an additional `twisted' $B+L'$ symmetry dictates that $H$--$H'$ mixing is the leading connection between the sectors. We write down a UV completion where $\sim$ TeV-scale leptoquarks generate the partonic dimension-12 mixing operator, thus linking to the energy frontier. 
With half of all $H$ atoms oscillating into $H'$, the composition of the universe is scandalously different during part of its history. We qualitatively discuss structure formation: both the modifications to it in the Standard Model sector and the possibility of it in the mirror sector, which has recently been proposed as a resolution to the puzzle of early supermassive black holes. While the egregious loss of SM baryons mostly self-erases during reionization, to our knowledge this is the first model that suggests there \textit{should} be `missing baryons' in the late universe, and highly motivates a continued, robust observational program of high-precision searches for cosmic baryons.
\end{abstract}

\maketitle
\renewcommand{\thefootnote}{\arabic{footnote}}
\section{Introduction}

The dark ages of the universe---the hundreds of millions of years after recombination but before cosmic dawn, during which halos, galaxies, and finally the first stars take shape---remain one of the most observationally mysterious epochs of our universe's chronology. The era of 21 cm cosmology is upon us, however, and observatories both current and coming promise unprecedented access to this long expanse of cosmic history. 

Along with those promises comes the possibility for new fundamental physics discoveries. Scenarios predicting 21 cm signatures are diverse, encompassing dark matter (DM) decay, annihilation, and warmth \cite{Chen:2003gz,Furlanetto:2006wp,2007MNRAS.377..245V,2009PhRvD..80c5007B,Cumberbatch:2008rh,Finkbeiner:2008gw,Slatyer:2009yq,Natarajan:2009bm,Natarajan:2010dc,Yue:2012na,2013MNRAS.429.1705V,Dvorkin:2013cea,Evoli:2014pva,Tashiro:2014tsa,Oldengott:2016yjc,Rudakovskiy:2016ngi,Lopez-Honorez:2017csg}; cosmic strings \cite{Brandenberger:2010hn,2011JCAP...08..014H,2012PhRvD..85l3535T,2012JCAP...05..014P}; primordial black holes \cite{Ricotti:2007au,Mack:2008nv,Tashiro:2012qe,Belotsky:2014twa}; and dark energy \cite{Wyithe:2007rq,2009JCAP...04..002X,Kohri:2016bqx,Costa:2018aoy,Yang:2019nhz,Li:2019loh}. No doubt even more applications will be recognized as the field continues to advance.

New fundamental physics may have \textit{already} been discovered by the Experiment to Detect the Global Epoch of Reionization Signature (EDGES), which reported an absorption feature at redshift $z \sim 17$ with at least twice the depth of the standard $\Lambda$CDM prediction \cite{Bowman:2018yin}, a $3.8\sigma$ deviation \cite{Barkana:2018qrx,Barkana:2018lgd}. The publication of this result was followed by the quick realization that interactions of DM with electrons and/or protons could be responsible for cooling the hydrogen gas beyond the extent expected from adiabatic expansion alone \cite{Munoz:2018pzp,Barkana:2018lgd,Jia:2018csj,Liu:2019knx}. However, such a possibility faces severe constraints from a variety of sources. Interactions of DM with charged particles is constrained both by early-time measurements of the cosmic microwave background and by late-time direct and indirect detection experiments. Within the well-explored scenarios, the only way for such interactions to be responsible for the EDGES signal is for the interacting DM to be only a small subcomponent of the full DM density. Additional strict requirements on the mass and interaction strength, as well as some further contortions, are required to ensure that these scenarios are self-consistent (see e.g. \cite{Berlin:2018sjs,Barkana:2018qrx,Slatyer:2018aqg,Fraser:2018acy,Kovetz:2018zan,Liu:2019knx} for discussions).

A more conservative move is to attribute the surprising absorption amplitude to unknown astrophysics, but here, too, contortions are evidently required. These proposals posit the existence of high-$z$ sources that drive the radio background above the level from cosmic microwave background (CMB) photons. While such interpretations of the EDGES data draw support from results reported by ARCADE-2 and the Long Wavelength Array, the challenge is again self-consistency. Early star-forming galaxies can enhance the radio background, but to match the EDGES data they must be $\sim 1000$ times more efficient at emitting photons near 1.4 GHz than their low-$z$ counterparts, assuming that relations between radio emissivity and star-formation rate (SFR) in present-day galaxies extrapolate to higher redshifts \cite{10.1093/mnras/sty3260}. Furthermore, electrons accelerated by the supernovae of Population III stars---which could conceivably give rise to this emission via synchrotron radiation---are accompanied by cosmic rays that heat the intergalactic medium (IGM) and tend to offset the effect of making the backlight brighter \cite{Jana:2018gqk}. Another possible source of synchrotron-radiating electrons, radio-loud black hole accretion, must occur in highly obscured environments to avoid excessive heating of the IGM. Even then these sources are beset by inverse Compton scattering of electrons on CMB photons, though whether they must be orders of magnitude radio-louder than their counterparts today is a matter of ongoing debate \cite{Sharma:2018agu,Ewall-Wice:2018bzf,Ewall-Wice:2019may}. At this time, the question of the viability of a strong radio excess can only be said to be unsettled. 

The most conservative view of all is that the EDGES anomaly is an artifact of the data analysis. There has been spirited discussion regarding this possibility \cite{2018Natur.564E..32H,2018Natur.564E..35B,2019ApJ...874..153B,2019ApJ...880...26S,2020ApJ...897..132T,2019MNRAS.489.4007S,2020MNRAS.492...22S}. It appears that a decisive verdict will have to await (dis)confirmation from future experiments. Thankfully, the 21 cm revolution is just beginning, with numerous ambitious projects underway or under development. Future measurements of the sky-averaged signal by DARE \cite{2012AdSpR..49..433B}, LEDA \cite{2018MNRAS.478.4193P}, PRI$^Z$M  \cite{2019JAI.....850004P}, SARAS \cite{2018ExA....45..269S}, and REACH \cite{deLera:2019} will be able to confirm the presence of the anomalous absorption feature, while other upcoming experiments such as HERA \cite{2019BAAS...51g.241P}, OVRO-LWA \cite{2019AJ....158...84E}, and SKA1-LOW \cite{Mellema:2012ht} will measure the power spectrum of 21 cm fluctuations.  Measurements of the line intensity of other atoms, such as helium \cite{Visbal:2015sca,2009arXiv0905.1698B,2009PhRvD..80f3010M,10.1093/mnras/staa1951}, molecular hydrogen \cite{2013ApJ...768..130G} and deuterium \cite{Kosenko_2018,2006PhRvL..97i1301S} provide complementary views of cosmic dawn, as outlined in recent community reports \cite{Kovetz:2017agg,Kovetz:2019uss,Chang:2019xgc}. We refer to the recent work Ref.~\cite{2020PASP..132f2001L} for a review of observational methods and challenges with a focus on the key ingredient of data analysis. 

In the meantime the anomaly persists, and it is worth considering how else it might be produced. While beyond the Standard Model (BSM) scenarios have been proposed that generate a radio excess \cite{Fraser:2018acy,Moroi:2018vci,Pospelov:2018kdh,Choi:2019jwx,Brandenberger:2018dfj,Brandenberger:2019lfm}, in this work we embrace the baryon cooling interpretation. We have recently proposed hydrogen oscillations into a dark state as a novel mechanism through which baryon cooling may take place \cite{companion}. This allows such cooling without requiring any direct interactions between SM particles and DM. It is only \textit{neutral} atomic hydrogen which picks up an effective interaction with DM, and this interaction is only active over cosmological timescales and only in the relatively quiescent intergalactic medium---automatically restricting the effects to the dark ages between recombination and reionization. In \cite{companion} we established that this mechanism can indeed cool hydrogen to the level suggested by EDGES, but did not discuss the microphysical origins of such mixing nor its effects past the EDGES signal. We herein take up both these topics: grounding the mechanism in sensible particle physics and preliminarily exploring the effects of dark age mixing on the formation of structures and stars. 

This mechanism can in fact be naturally accommodated within the well-motivated framework of a mirror model, where the necessary features can be symmetry-protected. In a mirror model there is a $\mathbb{Z}_2$ symmetry relating the SM gauge groups and field content to another dark, `mirror' sector. Such models have a long and storied history, having been first suggested in passing by Lee and Yang in 1956 \cite{Lee:1956qn}. We refer the reader to the historical review by Okun \cite{Okun:2006eb} for further references, but mention that mirror matter was first connected to dark matter by Blinnikov and Khlopov in 1982 \cite{Blinnikov:1982eh}, was suggested to appear naturally from the heterotic string by the Princeton string quartet in 1984 \cite{Gross:1984dd}, and was first written down in full with unbroken symmetry by Foot, Lew, and Volkas in 1991 \cite{Foot:1991bp}. In recent years closely-related structures have seen intense study after being connected to the hierarchy problem by Chacko, Goh, and Harnik in 2005 \cite{Chacko:2005pe}. 

In a mirror model with unbroken $\mathbb{Z}_2$ symmetry, there is a mirror hydrogen state which is exactly degenerate with SM hydrogen. As a result, if these states have just a small amount of mass mixing, the SM hydrogen which forms during recombination will gradually oscillate into mirror hydrogen, bearing out the phenomenology discussed in \cite{companion}. Mirror hydrogen's interactions with dark matter are then effectively inherited by SM hydrogen over cosmologically-long timescales. In particular, over the dark ages hydrogen can lose energy to the dark sector. Below we will assume for simplicity that the mirror sector is unpopulated before hydrogen begins oscillating into it, though we note that our qualitative results are robust to the inclusion of a subdominant mirror sector component. With unbroken $\mathbb{Z}_2$ symmetry in the Lagrangian, this asymmetry is most easily set up by appealing to cosmic variance: it traces back to quantum fluctuations in the early universe in our Hubble patch, while the theory itself remains symmetric. 

In the scenario we consider, communication between the SM and dark sectors passes predominantly through the hydrogen portal.\footnote{We use this term to refer specifically to the hydrogen \textit{mixing} portal, which is distinct from the hydrogen \textit{decay} portal recently introduced in Ref.~\cite{PhysRevLett.125.231803}.} Being neutral in the SM, a hydrogen atom violates no gauge symmetries when it oscillates into mirror hydrogen. But the same is true of the neutron. We thus need to justify why the hydrogen portal might predominate---or, indeed, even be relevant---despite its suppressions from both proton--electron wavefunction overlap and the high mass dimension of the hydrogen--mirror hydrogen mixing operator. We argue below that such a theory is indeed sensible if a `twisted' $B+L'$ symmetry is imposed. We exhibit a plausible UV completion by introducing the appropriate leptoquarks such that hydrogen oscillations are permitted but neutron oscillations and proton decay are not. As it turns out, the very small mixing parameter that arises from leptoquarks with masses around the TeV-scale energy frontier puts us right in the neighborhood of cosmological timescales. 

The question, then, is whether \textit{in-medium} hydrogen oscillations are consistent with what we know about the cosmic timeline. In \cite{companion} we showed that there is indeed ample parameter space where in-medium mixing is efficient enough to account for the cooling suggested by EDGES. However, making a significant fraction of baryons temporarily disappear from the universe is a new idea, as far as we know, and a radical one on the face of it. The baryon density in the early universe is precisely known from cosmological measurements. In the local universe, tallying the various populations of baryons is challenging and uncertain. Still, in light of recent measurements of the warm--hot intergalactic medium, where a significant fraction of baryons now reside, reducing the local baryon density by a full factor of 2 is a no-go. If half of all baryons are to disappear during the dark ages, they must find their way back by the current era.

As it happens, reionization naturally shepherds most of the baryons back to our sector of the universe. If the mixing portal is in equilibrium during the epoch of reionization, it holds the abundances of SM and mirror hydrogen equal, so that the reionization of SM hydrogen leads to an effective depletion of the abundance of mirror hydrogen. As a result, the drastically different composition of the universe during the dark ages naturally hides itself. There will be some extant baryons trapped as relic mirror hydrogen in the late universe, but we do not expect them to conflict with the tallies mentioned above. On the contrary, they may even be connected to the part of the longstanding missing baryons problem that still remains. 

The temporary conversion of hydrogen into mirror hydrogen is likely to have other observable effects besides lowering the gas temperature at cosmic dawn. We take up this topic in a preliminary and qualitative way, surveying the relevance of $H_2$ cooling, mirror stars, direct-collapse black holes, streaming velocities, and probes of reionization. These issues are complex, intertwined, and deserving of further investigation.

We begin Sec.~\ref{sec:cosmology} with an overview of the pertinent parts of 21 cm cosmology, then analyze the physics of the hydrogen portal in the dark ages, providing additional detail omitted from \cite{companion}. In Sec.~\ref{sec:microphysics} we address how the desired mixing phenomenology might arise from an underlying particle physics model. We then turn in Sec.~\ref{sec:other} to a discussion of other astrophysical and cosmological implications, before concluding in Sec.~\ref{sec:conc}.

\section{Cosmological Evolution} \label{sec:cosmology}

Between the epochs of recombination and reionization, most of the baryons in the universe are in the form of neutral atomic hydrogen. The spin temperature $T_S$ is defined in relation to the density ratio of the $F=1$ and $F=0$ ground-state hyperfine levels:
\begin{equation}
\frac{n_1}{n_0} = 3 \exp \left( - \frac{T_\textrm{hf}}{T_S} \right),
\end{equation}
where $T_\textrm{hf} \cong 68$ mK is the hyperfine splitting expressed as a temperature and the factor of 3 is the degeneracy ratio of the two levels. Because $T_S$ is set by a competition among different interactions, it does not always track the brightness temperature $T_R$ of the radiation illuminating the gas. This makes hydrogen visible against its background at some redshifts \cite{Pritchard_2012}.

Experiments like EDGES are sensitive to the redshifted (and frequency-dependent) differential brightness temperature
\begin{equation}
T_{21} = \frac{T_\textrm{obs} - T_R}{1+z} = \frac{T_S - T_R}{1+z} \left( 1 - e^{-\tau} \right). \label{T21def}
\end{equation}
Here $T_\textrm{obs} = T_S \left( 1 - e^{-\tau} \right) + T_R e^{-\tau}$ is the observed brightness temperature and $\tau$ is the optical depth of the hydrogen gas. In words, $T_{21}$ measures the change in brightness temperature of the microwave background due to absorption and re-emission by hydrogen gas en route to Earth. While an absorption feature in $T_{21}$ was anticipated by EDGES, the depth of the feature was not. We are proposing that mixing between SM hydrogen $H$ and mirror hydrogen $H'$ can explain the anomaly. In this section we first describe qualitatively how mixing alters the cosmic timeline, then move on to a quantitative analysis of the dynamics of mixing and cooling.

\subsection{Cosmic Timeline and Evolution of the Hydrogen Spin Temperature} 

The spin temperature tracks the equilibrium set by the competition of induced hyperfine transitions, atomic collisions, and resonant scattering of Ly$\alpha$ photons \cite{4065250}:
\begin{equation}
T_S^{-1} = \frac{T_R^{-1} + x_c T_g^{-1} + x_\alpha T_\alpha^{-1}}{1 + x_c + x_\alpha}.
\end{equation}
$x_c$ and $x_\alpha$ are coefficients that quantify the relative importance of collisions and spin flip via Ly$\alpha$ transitions. They respectively couple $T_S$ to the kinetic temperature $T_g$ of the gas and to the color temperature $T_\alpha$ of the Ly$\alpha$ radiation field. 

\begin{figure*}
    \centering
    \includegraphics[trim = 2cm 2cm 3cm 2cm,
    width=.8\textwidth]{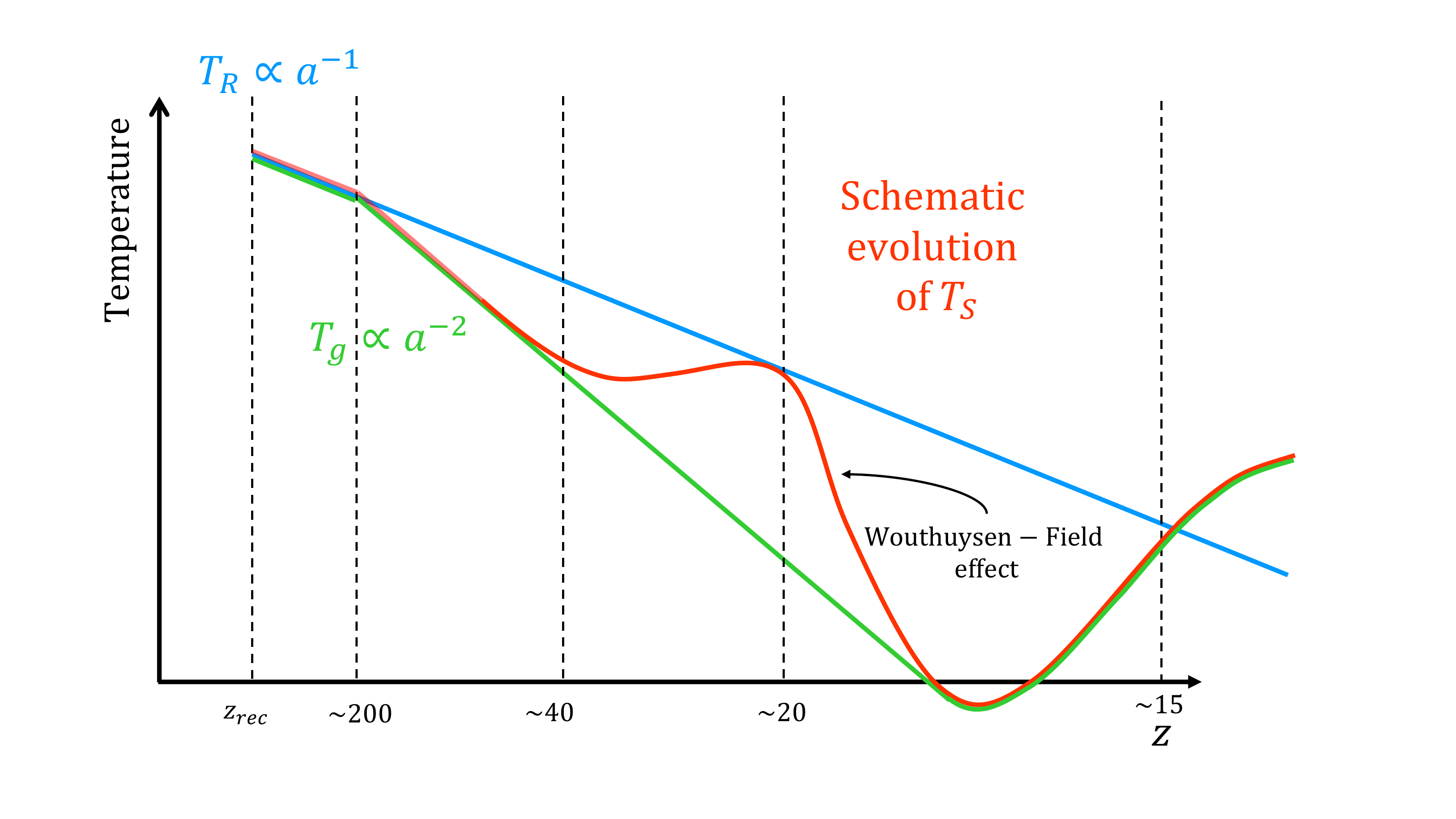}
    \caption{The schematic evolution of the spin temperature and its relation to the radiation and gas temperatures in $\Lambda$CDM. During the first part of the dark ages, the SM gas remains thermally coupled to the CMB. After this coupling becomes inefficient, the spin temperature collisionally couples to the gas as it evolves adiabatically, until this coupling becomes inefficient and the spin temperature recouples to the CMB. After stars turn on, the Wouthuysen-Field effect again couples the spin temperature to the gas temperature, then eventually the stars heat the gas above the CMB temperature. Further detail may be found in e.g. the reviews \cite{Barkana:2001avi,loeb2008first,Pritchard_2012}. In comparison, our model results in super-adiabatic cooling of the gas starting after $z \sim 200$ and ending before $z \sim 20$, resulting in the spin temperature recoupling to a colder $T_g$.}
    \label{fig:temperatures}
\end{figure*}

In the standard cosmology (see Fig.~\ref{fig:temperatures}), the sky-averaged spin temperature, which we'll continue to denote simply with $T_S$, passes through several qualitatively distinct stages as the universe expands. In the aftermath of recombination, CMB photons constitute the 21 cm radiation field and keep the hydrogen gas at $T_g = T_R = T_\textrm{CMB}$ through the combination of Compton scattering of CMB photons on residual free electrons and collisions of those electrons with hydrogen atoms. $T_S$ therefore follows $T_\textrm{R}$ until the gas drops out of thermal equilibrium with the CMB at $z \sim 150 - 200$. Past that point, adiabatic cooling causes $T_g$---and with it $T_S$---to drop below $T_R$. At $z \sim 30-40$, however, collisional coupling becomes inefficient, and subsequent evolution of $T_S$ is driven by interactions with the radiation field. 

The standard expectation is that at cosmic dawn hydrogen is visible in absorption, it being colder than the CMB backlight ($T_g \cong 7$~K vs. $T_\textrm{CMB} \cong 49$~K at $z = 17$, for example). The amplitude of the trough is maximized when $T_S$ is closely coupled to $T_g$ by the Wouthuysen--Field effect, which is to say by the resonant scattering of Ly$\alpha$ photons \cite{4065250,wouthuysen1952excitation}. In that extreme case, the depth at $z = 17$ is anticipated to be $T_{21} \cong - 210$~mK. EDGES, in contrast, observed a trough spanning $20 \gtrsim z \gtrsim 15$ and reaching a maximum amplitude of $T_{21} = -500^{+200}_{-500}$~mK  (99\% C.L.) at $z \cong 17$. This finding has been variously interpreted as revealing super-adiabatic cooling of the gas, non-CMB background sources, or systematic/experimental error. Our proposal falls in the first group.

In the presence of $H$--$H'$ mixing, oscillations begin once the ionization fraction drops at recombination.\footnote{Prior to and during recombination, the operator giving rise to mixing permits passage through the portal via $e+p \longrightarrow e'+p'$, but only at an extremely low level. Given the mixing parameters we focus on, conversion of $H$ to $H'$ during recombination is also negligible. See Sec.~\ref{sec:recomb} for more discussion.} Quantum superpositions of $H$ and $H'$ are continuously decohered by collisional processes. The equilibrium to which this tends is an equipartition of number density $n_H = n_{H'}$. If mixing equilibrates during the dark ages, the SM hydrogen density is thus cut in half.

Sometime after recombination, $H$--$H'$ mixing comes into equilibrium and equalizes the mirror and SM hydrogen densities. After the SM gas thermally decouples from the CMB, the mirror gas loses thermal energy through its interactions with DM. This cooling is carried over to the SM sector by the rapid interconversion of $H$ and $H'$, causing $T_g$ to dip below the green curve shown in Fig.~\ref{fig:temperatures}. By cosmic dawn, the SM gas has been cooled to the extent required to explain the EDGES anomaly. Finally, the universe reionizes. Provided that mixing continues to be in equilibrium during this last epoch, $n_{H'}$ will track the falling density of neutral SM atoms and the baryons will be returned to our sector.

We emphasize, in distinction with other DM-assisted cooling models, that at the end of the dark ages the SM hydrogen gas is both colder \textit{and} more dilute in the $H$--$H'$ mixing scenario than it is in the standard (non-BSM) one. Since the optical depth $\tau$ is small, $T_{21}$ is approximately proportional to
\begin{equation}
\tau = \frac{3 A_{10} n_{H}}{16 \nu_{21}^2 T_S H(z)},
\end{equation}
where $A_{10} = 2.85 \times 10^{-15}~\textrm{s}^{-1}$ is the spontaneous emission coefficient for the 21 cm hyperfine transition, $\nu_{21} = 1.42$~GHz is its rest-frame frequency, and $H(z)$ is the Hubble rate. (From here on we will be dropping the explicit argument of this last quantity, but it will be clear from the context whether the Hubble rate or the hydrogen symbol is being referred to.) Plugging this into Eq.~\eqref{T21def} and assuming that $T_S = T_g$, consistency with the EDGES best-fit amplitude implies that SM--DM interactions cool the gas down to a temperature~\cite{Barkana:2018qrx}
\begin{equation}
T_g^{\textrm{SM}-\textrm{DM}} (z = 17) \cong 3.3~\textrm{K}.
\end{equation}
Mixing equilibration, on the other hand, requires the gas to cool by an additional factor of $\sim 2$ to offset the reduction in gas density:
\begin{equation}
T_g^{H-H'} (z = 17) \cong 1.7~\textrm{K}.
\end{equation}
This number constrains the properties of DM, as we discuss below, but it makes no direct demands on $H$--$H'$ oscillations. All that is required of the latter is that they achieve mixing equilibrium at the appropriate times.

\subsection{Dynamics of the Hydrogen Portal}

The nonequilibrium dynamics of $H$--$H'$ mixing is described by the quantum kinetic equation
\begin{equation}
i \frac{d\rho}{dt} = \left[ \mathcal{H}, \rho \right] + i\mathcal{C}. \label{qke}
\end{equation}
$\rho$ is a $2 \times 2$ density matrix whose diagonal entries are proportional to the classical densities of ground-state $H$ and $H'$ and whose off-diagonal entries encode the quantum coherence between the two states. (We assume that the singlet and triplet hyperfine states do not substantially differ in their mixing with their respective mirror counterparts, though we will mention in Sec.~\ref{sec:microphysics} the possibility that they vary.) $\mathcal{H}$ is the Hamiltonian and $\mathcal{C}$ is the collision term.

The Hamiltonian has the form
\begin{equation}
\mathcal{H} = \begin{pmatrix} E_H^0 + \Delta V & \delta \\ \delta & E_{H'}^0 - \Delta V \end{pmatrix}, \label{oscham}
\end{equation}
where $E^0_{H(H')}$ is the energy of (mirror) hydrogen in vacuum; $\Delta V = (V_H - V_{H'}) / 2$ is the in-medium potential, with contribution $V_{H(H')}$ from $H$ ($H'$) forward scattering; and $\delta$ is the mixing parameter.\footnote{Strictly speaking, $H$ and $H'$ are in-vacuum energy eigenstates only when $\delta$ vanishes. $E^0_{H}$ and  $E^0_{H'}$ are the energies of the non-mixing theory.} Following the convention in the neutrino literature, we parameterize the mixing in terms of an in-medium oscillation frequency
\begin{equation}
\omega_m = \sqrt{\left( \Delta E^0 + \Delta V \right)^2 + \delta^2},
\end{equation}
where $\Delta E^0 = \left( E_H^0 - E_{H'}^0 \right) / 2$, and an in-medium mixing angle given by
\begin{equation}
\sin^2 2 \theta_m = \frac{\delta^2}{\omega_m^2}.
\end{equation}
Since the dark ages take place on the order of a hundred million years after the Big Bang, and we want oscillations to be effective on this timescale, an initial guess might be that $\delta \sim \mathcal{O}(10^{-40})$~GeV. An immediate takeaway is that the mixing is sensitive to exceedingly small effects.\footnote{For comparison, this estimate suggests that $\delta$ is about 20 orders of magnitude smaller than the comparable parameter in the mass mixing of neutrinos at MeV energies.} In particular, we must ensure that the SM and mirror hydrogen masses are highly degenerate, $|m_H - m_{H'}| \lesssim \delta$, to avoid the mixing angle being extremely small even in vacuum. Hereafter we adopt $E_H^0 = E_{H'}^0$, which holds if the $\mathbb{Z}_2$ symmetry relating SM to mirror fields is exact, since in that case the masses and atomic structures in the two sectors are identical.

The potentials $V_H$ and $V_{H'}$ quantify the refraction experienced by $H$ and $H'$ as they traverse the medium. The physics is essentially the same as for a classical wave. For a hydrogen atom with de Broglie wave number $k \approx m_H v$, the index of refraction is
\begin{equation}
n = 1 + \sum_i \frac{2 \pi n_i}{k^2} f_i (0),
\end{equation}
with the sum calculated over all scattering processes, each process $i$ having density of scatterers $n_i$ and elastic forward-scattering amplitude $f_i(0)$. The velocity in medium is related to that in vacuum by $v = v_0 / n$, or
\begin{equation}
v \approx v_0 \left( 1 - \sum_i \frac{2 \pi n_i}{k^2} f_i (0) \right),
\end{equation}
from which it follows that the potential due to the medium is
\begin{equation}
V_H = E_H - E_H^0 \approx - \sum_i \frac{2 \pi n_i}{m_H} f_i(0) \label{vhdef}
\end{equation}
to lowest order. 

The final term in the kinetic equation, the collision term $i \mathcal{C}$, receives contributions from collisional processes in both sectors, including $H'$--DM interactions. It has two crucial effects: redistribution of momentum and decoherence of the $H$--$H'$ system into interaction states.\footnote{Because $H$ and $H'$ have completely distinct sets of (non-gravitational) interactions, all scattering events that change the atom's momentum also change its coherence. This sort of statement is not universally true, though. Neutral-current scattering of neutrinos, for example, can change a neutrino's momentum without decohering it in flavor space.} Baryonic cooling is achieved by the cooperation of oscillations and collisions. Schematically, a hydrogen atom with momentum $p$ is brought down to momentum $p'$ by the sequence
\begin{equation}
H(p) \xrightarrow{\textrm{oscillation}} H'(p) \xrightarrow{\textrm{$H'$--DM}} H'(p') \xrightarrow{\textrm{oscillation}} H(p').
\end{equation}
For the entire sequence to be efficient on a cosmic timescale, in-medium, decoherence-affected oscillations and heat transfer from $H'$ to DM must \textit{both} be fast relative to the Hubble rate.

Under the assumption that the mixing channel is in equilibrium, the number densities and temperatures of the hydrogen species satisfy $n_H = n_{H'}$ and $T_g \equiv T_H = T_{H'}$, rendering the kinetic treatment superfluous. The evolution of the gas temperature is simply described (here as a function of scale factor $a$) by
\begin{equation}
H a \frac{d T_g}{da} = - 2 H T_g + \Gamma_C \left( T_\textrm{CMB} - T_g \right) + \frac{2}{3} \dot{Q}_g, \label{gasevol}
\end{equation}
written in terms of the Compton scattering rate $\Gamma_C$ and the heating rate $\dot{Q}_g$ due to $H'$--DM scattering. This formula is identical to the one appearing in cooling scenarios with SM--DM scattering. As in those models, $\dot{Q}_g$ includes dissipative heating associated with the bulk velocity of the gas---in our case the mirror-sector gas---relative to DM.

Since our aim is to demonstrate feasibility, we will simplify the problem even further and just establish that the desired hierarchy of timescales (specifically the timescales of mixing, heat transfer, and Hubble expansion) can be realized. We start with mixing. The equilibration rate of the mixing channel is \cite{KAINULAINEN1990191,PhysRevLett.68.3137,PhysRevLett.72.17,PhysRevD.62.093025,PhysRevD.100.083536}
\begin{equation}
\Gamma_\textrm{osc} \sim \frac{\Gamma_c}{4} \frac{\sin^2 2 \theta_m}{1 + \left( \Gamma_c / 2 \omega_m \right)^2}, \label{gosc}
\end{equation}
where the total rate of collisions $\Gamma_c = \Gamma_H + \Gamma_{H'}$ is the sum of the scattering rates of $H$ and $H'$. As shown in Ref.~\cite{PhysRevD.100.083536}, Eq.~\eqref{gosc} emerges from the quantum kinetic equation upon assuming that nonequilibrium deviations of the density matrix decay exponentially.\footnote{There are subtleties pertaining to the accuracy of Eq.~\eqref{gosc} near maximal mixing $\theta_m \sim \pi/4$, but the equation is adequate for our purposes.} The physics here can be understood by breaking down the right-hand side into three pieces. The appearance of $\sin^2 2\theta_m$ signals that equilibration takes longer if the mixing angle is small. The leading $\Gamma_c$ factor captures the intuition that a faster decoherence rate entails a faster $H'$ production rate. The term in the denominator demonstrates that this trend is only valid up to a point: If a quantum state is measured too quickly ($\Gamma_c \gg \omega_m$), it will never have time to develop coherence and will be stuck close to its initial state. This regime of mixing is known as the quantum Zeno limit. 

The task now is to estimate the various contributions to $\Gamma_c$ and $\Delta V$. Working very approximately, the $H-H$ scattering rate over $20 \lesssim z \lesssim 200$ is
\begin{align}
\Gamma_{H-H} (z) &\sim n_H \left\langle \sigma v \right\rangle_{H-H} \notag \\
&\sim \left( \eta n_\textrm{CMB} \right) \left( 4 \pi a_0^2 \right) \sqrt{\frac{3 T_g}{m_H}} \notag \\
&\sim \left( 1 \times 10^{-43}~\textrm{GeV} \right) \left( 1+z \right)^4,
\end{align}
where $\eta \sim 6 \times 10^{-10}$ is the cosmic baryon asymmetry, $a_0 = 3 \times 10^{-4}~\textrm{eV}^{-1}$ is the Bohr radius, and $n_\textrm{CMB}$ is the number density of CMB photons. 

Rayleigh scattering with CMB photons occurs at a rate
\begin{align}
\Gamma_{H-\textrm{CMB}} &\sim n_\textrm{CMB} \sigma_T \left( \frac{\left\langle \omega \right\rangle_\textrm{CMB}}{\omega_{\textrm{Ly}\alpha}} \right)^4 \notag \\
&\sim \left( 4 \times 10^{-52} ~\textrm{GeV} \right) \left( 1+z \right)^7,
\end{align}
where $\sigma_T \sim 2 \times 10^{-15}~\textrm{eV}^{-2}$ is the Thomson cross section, $\omega_{\textrm{Ly}\alpha}$ is the Ly$\alpha$ transition frequency, and $\left\langle \omega \right\rangle_\textrm{CMB}$ is the average frequency of a CMB photon. The severe $(1+z)^7$ scaling is a consequence of the CMB background diluting and redshifting. These interactions are always subdominant at $z \lesssim 200$, as is to be expected. For the moment we set aside interactions with photons from astrophysical sources, focusing first on the mixing over the course of the dark ages.

Appealing to Eq.~\eqref{vhdef} and estimating the amplitude of elastic $H-H$ forward scattering as $f_{H-H}(0) \sim -a_0$, the potential associated with this process is
\begin{equation}
V_{H-H} (z) \sim \frac{2 \pi n_H a_0}{m_H} \sim \left( 3 \times 10^{-42}~\textrm{GeV} \right) \left( 1+z \right)^3.
\end{equation}
By the same general formula, but using\footnote{This estimate can be justified from either a classical or a quantum effective theory. In the former case, we regard a bound electron as an oscillating dipole with natural frequency $\omega_{\textrm{Ly}\alpha}$ and dipole moment $p = e^2 E_0 / \left[ m_e \left( \omega^2 - \omega_{\textrm{Ly}\alpha}^2 \right) \right]$, where $\omega$ and $E_0$ are the frequency and electric field amplitude associated with the incident photons. The Poynting flux radiated along a given direction is proportional to $\omega^4 p^2$ and the incident energy flux is proportional to $E_0^2$. Hence at low photon energies $d\sigma / d \Omega \sim \alpha^4 a_0^2 \left( \omega / \omega_{\textrm{Ly}\alpha} \right)^4$, motivating Eq.~\eqref{ampcmb}. Alternatively, one can write down an effective Lagrangian for photon--hydrogen scattering and estimate the coefficients using the Bohr radius as the only relevant scale \cite{Kaplan:2005es}, resulting in $f(0) \sim a_0^3 (k \cdot v)^2$, where $k_\mu$ is the photon momentum and $v_\mu$ is the hydrogen velocity.}
\begin{equation}
f_{H-\textrm{CMB}} (0) \sim \alpha^2 a_0 \left( \frac{ \left\langle \omega \right\rangle_\textrm{CMB}}{\omega_{\textrm{Ly}\alpha}} \right)^2 \label{ampcmb}
\end{equation}
for the $H-\gamma$ amplitude, we find
\begin{equation}
V_{H-\textrm{CMB}} \sim - \left( 2 \times 10^{-45} ~\textrm{GeV} \right) \left( 1+z \right)^5
\end{equation}
for the potential generated by the CMB.

Another possible contribution to $V$ comes from magnetic fields. Fields in the IGM today are on the order of a nanogauss, which translates to
\begin{equation}
| V_B | \sim \mu_B B \sim 6 \times 10^{-27} ~\textrm{GeV},
\end{equation}
a magnitude large enough to take seriously. Considerable uncertainties surround the physical origin of these fields, however. We refer to Ref.~\cite{Subramanian:2019jyd} for a review of their status. It is conceivable that present-day magnetic fields are the products of standard cosmological and astrophysical processes, though even here the development over redshift is rather uncertain. The most concrete predictions find magnetic fields generated during recombination of size $B \sim 10^{-30} - 10^{-21} \ \text{G}$, and during reionization of size $B \sim 10^{-23} - 10^{-19} \text{ G}$ \cite{10.1111/j.1365-2966.2005.09442.x,PhysRevD.71.043502,Kulsrud_1997,Gnedin_2000,1994MNRAS.271L..15S,PhysRevLett.95.121301,PhysRevD.75.103501,10.1093/mnras/stx2007,10.1093/mnras/stv1578}. For comparison, $|V_B| \gtrsim |V_{H-H}|$ at $z \sim 1100$ for $B \gtrsim 10^{-15}~\textrm{G}$, or at $z \sim 20$ for $B \gtrsim 10^{-21}~\textrm{G}$. We hence neglect magnetic fields in our analysis of $\Gamma_\textrm{osc}$, assuming that they are insignificant for mixing throughout most of the IGM until after the universe has reionized. We do note, though, that the high sensitivity of $H$--$H'$ mixing to magnetic fields makes this an intriguing area for further study. 

As for photons from astrophysical sources, we do not attempt here to model their evolution. Instead, we show that it's reasonable to expect oscillations to be in equilibrium during reionization, using $z = 7$ as a representative redshift.  Assuming that there are $\mathcal{O}(10)$ ionizing photons per baryon at reionization, it follows that
\begin{equation}
V_{H-\gamma} (z = 7) \sim - 9 \times 10^{-43} ~\textrm{GeV}.
\end{equation}
For the scattering rate, we note that the photoionization cross section is well approximated by
\begin{equation}
\sigma_I \sim 7 \times 10^{-18} ~\textrm{cm}^2 \left( \frac{\omega_\gamma}{13.6 ~\textrm{eV}} \right)^{-3},
\end{equation}
which means that $\sigma_I \sim 10^7 \sigma_T$ at photon frequencies $\omega_{\gamma}$ near the ionization threshold. This makes photoionization the dominant contribution to $\Gamma_H$ during the epoch of reionization. We find
\begin{equation}
\Gamma_{H-\gamma} (z = 7) \sim 2 \times 10^{-34} ~\textrm{GeV}.
\end{equation}

\begin{figure*}
    \centering
    \includegraphics[width=0.8\textwidth]{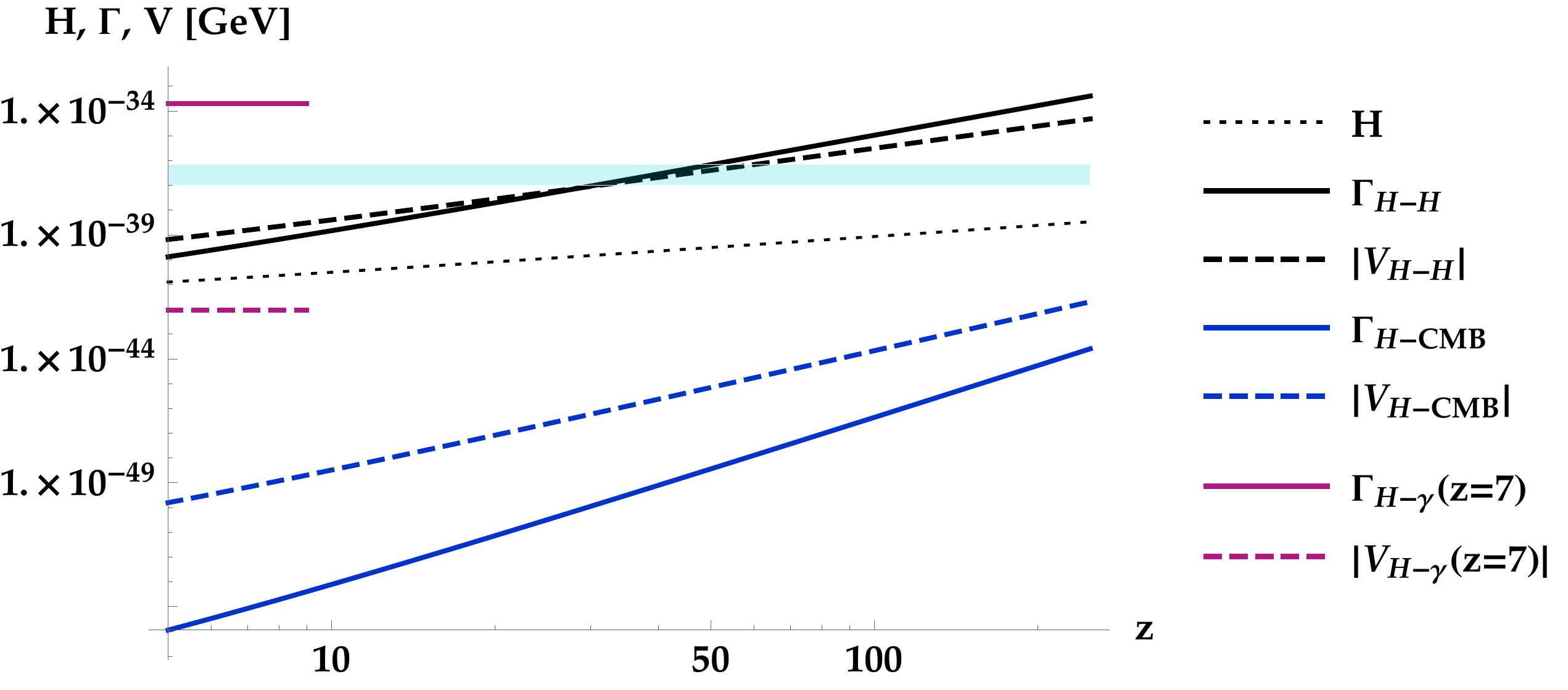}
    \caption{Comparison of rates affecting the mixing dynamics. Solid curves denote contributions to $\Gamma_c$, dashed curves to $\Delta V$. The dotted curve is the Hubble rate. Purple curves represent scattering of hydrogen on photons during the epoch of reionization, $z \sim 5 - 9$. Since we are not modeling the evolution of the radiation field from astrophysical sources, in estimating these terms we pick $z = 7$ and assume that there are $\sim 10$ ionizing photons per baryon. As a further point of comparison, the light blue band marks the range in $\delta$ that can account for the EDGES signal, assuming velocity-independent $\bar{\sigma}_{H'-X}$ (see Sec.~\ref{sec:minmax}).}
    \label{fig:rates}
\end{figure*}

Fig.~\ref{fig:rates} summarizes the in-medium effects on mixing. From recombination down to reionization, the dominant process by far is $H$--$H$ scattering. During the epoch of reionization, $H$--$\gamma$ scattering takes over as the dominant contribution to $\Gamma_c$. Forward scattering on photons, regardless of their source, remains subdominant. 

$H'$--$H'$ scattering contributes to $\Gamma_c$ and $\Delta V$ as well. It is described by the same formulas as $H$--$H$ scattering but with the replacement $H \rightarrow H'$. The conversion of $H$ into $H'$ decreases $\Gamma_{H-H}$ and $|V_{H-H}|$ and increases $\Gamma_{H'-H'}$ and $|V_{H'-H'}|$. These changes have no net effect on $\Gamma_c$, but they do on $\Delta V$, driving it to zero as $n_{H}$ and $n_{H'}$ equilibrate. Feedback of this kind is unimportant for our timescale analysis, however.

Adopting a phenomenological approach, we parametrize the $H'$--DM momentum-transfer cross section as
\begin{equation}
\bar{\sigma}_{H'-X} = \int d\cos\theta \left( 1 - \cos\theta \right) \frac{d \sigma_{H'-X}}{d \cos\theta} = \sigma_0 | \vec{v}_m |^n,
\end{equation}
where $\vec{v}_m$ is the relative velocity of the scattering particles. The heat-transfer rate $\dot{Q}_g$ appearing in Eq.~\eqref{gasevol} depends on $\bar{\sigma}_{H'-X}$ and has been studied elsewhere in relation to models with SM--DM interactions. Using this parametrization, we approximate the final terms in the mixing as
\begin{equation}
\Gamma_{H'-X} \sim \frac{\rho_X}{m_X} \bar{\sigma}_{H'-X} v_m
\end{equation}
and
\begin{equation}
V_{H'-X} \sim \frac{\rho_X}{m_H m_X} \sqrt{\pi \bar{\sigma}_{H'-X}},
\end{equation}
where $\rho_X$ is the DM energy density.

\subsection{Minimal Mixing, Maximal Cooling \label{sec:minmax}}

For simplicity, our strategy is to focus on $H'$--$X$ interactions that do not substantially change the mixing dynamics at any point (\textit{minimal mixing}) but successfully bring $H'$ and $X$ into thermal equilibrium (\textit{maximal cooling}). The same strategy was adopted in Ref.~\cite{companion}. Here we summarize the main points and provide further elaboration.

This region of parameter space is simple to treat for two reasons. First, the redshifts at which mixing becomes and ceases to be efficient are determined solely by SM parameters and $\delta$. Constraints are straightforward to apply when the mixing timeline itself does not depend on $\bar{\sigma}_{H'-X}$. Second, the requisite DM mass $m_X$ can be approximated by a simple function of the desired gas temperature if $H'$ cools to the furthest extent possible.

At high redshifts, $\Gamma_\textrm{osc}$ is enhanced by the large scattering rate but is suppressed to an even greater extent by the small in-medium mixing angle and the quantum Zeno effect. It reaches a peak at $| \Delta V | \sim \delta$. Then, after the transition to $\theta_m \sim \theta$, the mixing rate becomes independent of $\delta$ and falls off as $\Gamma_\textrm{osc} \approx \Gamma_c / 4$. One must take $\delta \gtrsim 4 \times 10^{-39}$~GeV to ensure that mixing comes into equilibrium prior to $z \sim 20$.

A more stringent lower limit comes from reionization. Needing $\Gamma_\textrm{osc}$ to exceed $H$ at $z = 7$---a coarse approximation of what it takes for $n_{H'}$ to track the falling neutral fraction of SM hydrogen---places a bound $\delta \gtrsim 1 \times 10^{-37}$~GeV. During the epoch of reionization, a significant hierarchy appears between $\Gamma_c$ and $|\Delta V|$, as seen in Fig.~\ref{fig:rates}, because ionization photons disproportionately contribute to decoherence versus forward scattering.

Restricting the mixing dynamics to be independent of $H'$--$X$ interactions translates to an upper bound on $\bar{\sigma}_{H'-X}$. We observe that
\begin{equation}
\left| \frac{V_{H'-X}}{V_{H-H}} \right| > \frac{\Gamma_{H'-X}}{\Gamma_{H-H}}
\end{equation}
for $n \leq 0$ at all relevant redshifts and cross sections, hence the more restrictive bound comes from $| V_{H'-X} | \lesssim | V_{H-H} |$ rather than from $\Gamma_{H'-X} \lesssim \Gamma_{H-H}$. The resulting constraint is
\begin{equation}
\frac{\bar{\sigma}_{H'-X} (z = 20)}{\sigma_{H-H}} \lesssim 0.2,
\end{equation}
written it as a fraction of the hard-sphere atomic cross section $\sigma_{H-H} = 4 \pi a_0^2$.

Thermalization of $H'$ and DM while mixing is in equilibrium leads to $T_g = T_{H'} = T_X$, the last of these being the DM temperature. The idealized best-case cooling scenario---in which both the DM thermal energy and the heating due to bulk relative velocities are neglected---has \cite{Barkana:2018lgd}
\begin{equation}
T_g = T_g^0 \frac{n_H + n_{H'}}{n_H + n_{H'} + n_X} \sim T_g^0 \frac{1}{1 + \frac{6~\textrm{GeV}}{m_X}}, \label{maxcooling}
\end{equation}
assuming that $H'$ and $X$ do not engage in number-changing interactions. ($T_g^0$ is the gas temperature due to adiabatic cooling alone.) This expression implies that the desired amount of cooling is achieved for $m_X \approx 2~\textrm{GeV}$. In reality the cooling will not be as efficient as Eq.~\eqref{maxcooling} assumes, but the DM mass it outputs is nonetheless a reasonable fiducial value. We take it as such for the remainder of this subsection.

Continuing to neglect the DM thermal energy, the heating rate of the gas is approximately \cite{Dvorkin:2013cea, Munoz:2015bca, Munoz:2017qpy}
\begin{equation}
\dot{Q}_g \sim - \frac{\rho_X \sigma_0 m_H}{\left( m_H + m_X \right)^2} \left( \frac{T_g}{m_H} \right)^{\frac{n+1}{2}} \frac{2^{\frac{5+n}{2}} \Gamma \left( 3 + \frac{n}{2} \right)}{\sqrt{\pi}} T_g. \label{qgdot}
\end{equation}
Our criterion for achieving maximal cooling of $T_g$ is that $| \dot{Q}_g | H^{-1}$ is at least comparable to the thermal energy of the gas. We now consider $n = 0$ and $n = -4$ separately. The case of $n=-4$ is a useful benchmark as it has a simple interpretation as milli(mirror)charged DM. Furthermore, this extreme growth as temperatures cool is \textit{necessary} in the conventional hydrogen-cooling models \cite{Berlin:2018sjs,Barkana:2018qrx,Slatyer:2018aqg,Fraser:2018acy,Kovetz:2018zan,Liu:2019knx} to avoid issues from scattering efficiently around recombination. As papers have also often mentioned $n=-2$ (e.g. electric dipole DM) and $n=-1$ (e.g. a Yukawa interaction), we choose $n=0$ to show by comparison how much more freedom exists for the form of $H'$--DM scattering in our scenario. 

Beginning with $n = 0$, we note that cooling becomes more efficient, relative to the Hubble timescale, at higher redshifts. The condition to impose is thus that cooling becomes inefficient only after mixing has come into equilibrium. To get a sense for the constraints arising from this condition, we note that imposing efficient cooling at $z \gtrsim 20$ entails
\begin{equation}
\frac{\bar{\sigma}_{H'-X}}{\sigma_{H-H}} \gtrsim 1 \times 10^{-3}. \label{sigmalower0}
\end{equation}
The lower bound is more lenient by a factor of $\sim 300$ at $\delta \sim 7 \times 10^{-37}~\textrm{GeV}$, the largest mixing parameter at which $n = 0$ is viable. (For larger values, DM thermally couples to the CMB at $z \gtrsim 200$.)

For $n = -4$, cooling is relatively more efficient at lower redshifts. The requisite condition now is that $H'$ and DM thermally equilibrate sometime before $z \sim 20$. Thus
\begin{equation}
\frac{\bar{\sigma}_{H'-X} (z = 20)}{\sigma_{H-H}} \gtrsim 1 \times 10^{-3}. \label{sigmalower4}
\end{equation}
Unlike in the previous case, here the upper limit on $\delta$ stemming from the mixing timeline only applies to
\begin{equation}
\frac{\bar{\sigma}_{H'-X} (z = 20)}{\sigma_{H-H}} \gtrsim 4 \times 10^{-2}
\end{equation}
because cross sections below this threshold are unable to transfer heat efficiently between $H'$ and $X$ at $z \gtrsim 200$. As a result, it is acceptable to have mixing equilibrate during this early phase.

Fig.~1 of Ref.~\cite{companion} depicts the parameter space in which we expect hydrogen to be cooled to the temperature needed to explain the EDGES absorption feature, \textit{assuming that} $m_X \approx 2~\textrm{GeV}$\textit{, that cooling is maximally efficient in the sense specified above Eq.~\eqref{maxcooling}, and that $H'$--DM interactions are always subdominant in the mixing dynamics.} It is notable that $n = -4$ does not have a much more expansive parameter space than $n = 0$, as one might expect in a model where the cooling is caused by SM--DM scattering. Although the $v_m^{-4}$ scaling does make cooling more efficient at later times, $V_{H'-X}$ is also enhanced, threatening to push us out of the limit of minimal mixing. In the end, having $m_X$ be comparable to $m_H$ means that $\bar{\sigma}_{H'-X}$ cannot deviate too much from $\sigma_{H-H}$.

Relaxing the maximal-cooling assumption potentially enlarges the viable parameter space quite significantly because $m_X$ is no longer anchored to the GeV scale. Two possibilities present themselves. The first is that $H'$ and $X$ still thermally equilibrate but the initial thermal energy of $X$ is not negligible. Then
\begin{equation}
T_g \sim T_g^0 \frac{1}{1 + \frac{6~\textrm{GeV}}{m_X}} + T_X^0 \frac{1}{1 + \frac{m_X}{6~\textrm{GeV}}},
\label{efficientcooling}
\end{equation}
where $(3/2) T_X^0$ is the average kinetic energy per DM particle in the baseline scenario without heat exchange between DM and the gas. $T_X^0$ may or may not be an actual temperature. DM masses much below 2 GeV can be consistent with the gas temperature inferred from EDGES if $T_X^0$ is itself approximately equal to that temperature. The mass cannot be made arbitrarily small, however, because it is limited by the effects of DM free streaming on structure formation.

The second possibility is that $H'$ and $X$ do not thermally equilibrate. Again $m_X$ must be below $\sim 2$~GeV, but there is more leeway here for DM to be very light, if it is initially nonthermal and continues to be so even as the gas transfers heat. When dark matter is not produced thermally, the phase space distribution is no longer universal. Different production mechanisms will give rise to different phase space distributions, and as far as we are aware there have been few attempts to understand such distributions analytically, though we point to the recent work \cite{Ballesteros:2020adh} which has studied nonthermal production in detail and finds fitting functions for numerically-calculated nonthermal phase space distribution. While light dark matter has the advantage of allowing far larger number densities and so makes it easier for the rate of interactions to rise above the Hubble rate, each interaction will generically sap less kinetic energy from the hydrogen. A detailed comparison of the prospects of nonthermal dark matter compared to thermal dark matter will likely depend sensitively on the phase space density assumed, and so we leave this for future work.

Given the approximate nature of our analysis, delineating the parameter space completely and precisely demands a more careful treatment. Nonetheless, it appears that the hydrogen-portal scenario, in the limit where mixing and cooling are at their simplest, is a viable way to account for the EDGES anomaly.

\section{Microphysical Origins} \label{sec:microphysics}

\subsection{Mirror Matter EFT}

While the cosmological analysis of Section \ref{sec:cosmology} depends only on the state hydrogen mixes with, the near-exact degeneracy required demands symmetry-enforced protection. To the tiny degree of splitting we require, all SM particles contribute to the mass of hydrogen, so we impose a $\mathbb{Z}_2$ symmetry which exchanges every SM particle with a mirror copy charged under a mirror version of the SM gauge group. As a result, the SM and mirror sector spectra contain exactly degenerate hydrogen bound states. 

Mixing of hydrogen with \textit{anti}hydrogen has been considered previously in the context of grand unified theories (GUTs) where baryon number $B$ and lepton number $L$ are no longer accidental global symmetries. As a result, baryon and lepton number may be broken either explicitly or, often, spontaneously, as with the majoron \cite{Chikashige:1980ui}. Since violation by a single unit is strongly constrained by the requirement of proton metastability, this motivates study of the violation of baryon and/or lepton number by two units, as in hydrogen-antihydrogen mixing or neutron-antineutron mixing \cite{1970JETPL..12..228K,Feinberg:1978sd,Misra:1982mg,Arnellos:1982nt,Mohapatra:1982aj,Caswell:1982qs}. These may be constrained by both laboratory tests and cosmological observations \cite{Mohapatra:2009wp,Phillips:2014fgb,Grossman:2018rdg}.

In the context of mirror models, the mixing of neutrons with mirror neutrons has seen detailed study and bounds exist from dedicated searches \cite{Berezhiani:2005hv,Berezhiani:2006je,Berezhiani:2008bc,Berezhiani:2017azg,Berezhiani:2017jkn,Berezhiani:2018eds,Ban:2007tp,Altarev:2009tg,Serebrov:2007gw,Mohapatra:2017lqw}, but to our knowledge the mixing of hydrogen with mirror hydrogen has not previously been considered. On general grounds one might expect that $n-n'$ mixing is always far more important---not only does it come from a lower-dimension operator, but $H$--$H'$ mixing from a contact operator gets an enormous relative penalty from the wavefunction overlap of the electron with the proton. 

However, if our mirror model incorporates a `twisted' $B+L'$ and/or $B'+L$ symmetry,\footnote{The choice of sign in $B \pm L'$ and/or $B'\pm L$ merely dictates whether SM hydrogen mixes with mirror hydrogen or mirror antihydrogen. In either case, gauging this symmetry in the UV would require additional heavy fermions to cancel anomalies.} $n-n'$ oscillation is disallowed and hydrogen-mirror hydrogen mixing can be the leading connection between the two sectors.\footnote{Of course the marginal kinetic mixing portal and Higgs portal interactions are also allowed, but are not compulsory and are not generated appreciably by the toy UV completion of Section \ref{sec:UVcomp}. A large Higgs portal interaction in the $\mathbb{Z}_2$-symmetric theory may violate Higgs coupling measurements \cite{Burdman:2014zta,Craig:2015pha} unless it is implemented somewhat exotically \cite{Csaki:2019qgb}. The other challenge with introducing these additional couplings is the prospect that they will equilibrate the sectors in the early universe \cite{Craig:2016lyx}, but this can be avoided if the couplings are small enough \cite{Koren:2019iuv} and/or the reheating temperature is low enough \cite{Chacko:2016hvu,Craig:2016lyx}. It would be an interesting direction to augment this work with a version of the twin Higgs mechanism \cite{Chacko:2005pe} and so connect to the physics of the hierarchy problem. We refer to \cite{Koren:2020biu} for a recent pedagogical review thereof.} We thus consider an effective theory which is a mirror model with these symmetries imposed.\footnote{We note parenthetically that one could introduce a small $n - n'$ mixing which would be technically natural as the only source of $B+L'$-breaking. It may be interesting to construct scenarios where both mixings occur and $n-n'$ mixing is responsible for other observed puzzles (e.g. \cite{Berezhiani:2018eds}) but we do not consider this possibility further.}

In the effective theory above $\Lambda_{\text{QCD}}$, hydrogen-mirror hydrogen mixing begins its life as an 8-fermion operator with all the constituent elementary partons. This is a dimension-12 operator of the schematic form
\begin{equation}
\mathcal{O}_\text{partonic} \sim \frac{1}{{ \Lambda}^8} \ \bar e' e \ \bar u' u \ \bar u' u \ \bar d' d + \text{ h.c.}, 
\end{equation}
where we give solely the constituent fields and note that there are many different spinor contractions possible. To find the low-energy mass mixing such partonic operators induce, we must first map these to hadronic operators such as
\begin{equation}
\mathcal{O}_\text{hadronic} \sim \frac{(4 \pi)^2 \Lambda_\text{QCD}^6}{{ \Lambda}^8} \ \bar e' \Gamma e \ \bar p' \Gamma p + \text{ h.c.}, 
\end{equation}
where $\Gamma \in \lbrace \mathds{1}, \gamma_5, \gamma_\mu, \gamma_\mu \gamma_5, \sigma_{\mu\nu}, \sigma_{\mu\nu} \gamma_5 \rbrace$. We have matched the interpolating operator $u u d \sim 4\pi \Lambda_\text{QCD}^3 p$ solely using dimensional analysis, with $\Lambda_\text{QCD}$ the only relevant scale and $4\pi$ a strong coupling factor. As far as we are aware it is not known how to calculate the map from these partonic operators to the hadronic operators in full detail. Finally, these operators lead to mixing of a strength 
\begin{align}
    \delta_{n,F} &= \left \langle H_{n,F} \left| {O}_\text{hadronic} \right| H'_{n,F}\right \rangle \nonumber \\
    \delta_{n,F} &\sim \frac{(4 \pi)^2 \Lambda_\text{QCD}^6}{{ \Lambda}^8} \frac{1}{n^3 a_0^3},\label{eqn:dimanalmixing}\\
    \delta_F \equiv \delta_{1,F} &\sim 10^{-37} \text{ GeV } \left(\frac{\Lambda_{\text{QCD}}}{250 \text{ MeV}}\right)^6 \left(\frac{260 \text{ GeV}}{\Lambda}\right)^8
\end{align}
where $n$ is the principal quantum number, $F$ is the hyperfine quantum number, and the factors of the Bohr radius $a_0$ are a contact operator penalty for finding the electron inside the proton. Since it is a contact operator no mixing occurs for the $\ell > 0$ states, in which the electron wavefunction has no support at the origin. 

By relating the hydrogen wavefunction to scattering states, it can be easily shown that a variety of hadronic operators allow for non-trivial mixing structure in hyperfine-space, and a judicious such choice may allow for mixing to solely (or mainly) affect the $F=1$ state. We defer consideration of such scenarios to later work. 

\subsection{Toy UV Completion} \label{sec:UVcomp}

We can write down a toy model of a UV completion which gives this mixing by adding leptoquarks with the appropriate quantum numbers. Leptoquarks are bosons carrying both baryon and lepton number and are in general among the most well-motivated extensions to the SM, appearing ubiquitously in grand unified theories as a consequence of unifying quarks and leptons (e.g. \cite{Frampton:1991ay}). Lighter, TeV-scale leptoquarks are predicted in models ranging from supersymmetric extensions of the SM which violate R-parity (e.g. \cite{Hall:1983id,Zwirner:1984is,Dawson:1985vr}) to models of extended technicolor to models where our familiar fermions are composite (e.g. \cite{Schrempp:1984nj}). In recent years light leptoquarks have been suggested as sources of various flavor anomalies (e.g. \cite{Gripaios:2014tna}). Our leptoquarks differ from the standard ones solely by being charged under the twisted global symmetries.  Technically, this global symmetry assignment means half of ours might be called `diquarks' as their only allowed coupling to SM fermions is to two quarks, but we'll continue to use `leptoquarks' as a general label because their only conserved quantum number mixes baryon and lepton numbers.

Multiple choices of gauge charges and spin are possible, but for simplicity we use pairs of SM and mirror scalars with the same gauge charges but different charges under the global symmetries, as listed in Table \ref{tab:leptoquarks}. We'll consider further adding $N_\omega$ of each type. We eschew embedding them in a particular UV model and introduce their couplings solely as a proof of principle. We emphasize that including these states does not lead to mixing of other states at appreciable levels---neither of fundamental states with the photon kinetic mixing portal or the Higgs portal, nor of other composite states such as neutrons or pions. 

\begin{table}[]
\centering
\begin{tabular}{|l|l|l|l|l|l|}
\hline
Field  & $SU(3)$ & $SU(2)$ & $U(1)$ & $B+L'$ & $B'+L$ \\ \hline
$\omega_q$  &  3 & 1 & $-\frac{1}{3}$ & $-\frac{2}{3}$  & 0 \\ \hline
$\omega_\ell$ & 3 & 1 & $-\frac{1}{3}$ & $\frac{1}{3}$ & 1  \\ \hline
$\omega'_q$ & $3'$ & 1 & $-\frac{1}{3}'$ & 0 & $-\frac{2}{3}$ \\ \hline
$\omega'_\ell$ & $3'$ & 1 & $-\frac{1}{3}'$  & 1 & $\frac{1}{3}$ \\ \hline
\end{tabular}
\caption{A toy UV completion which generates $H$--$H'$ mixing via the addition of two SM leptoquarks and two mirror leptoquarks with different $B+L'$ charges. The upper two rows have only SM gauge charges and the lower two rows have only mirror gauge charges. The global charges allow $\omega_q$ to interact solely with pairs of quarks, while $\omega_\ell$ interacts with a quark and a lepton. $\omega_q$ may alternatively be called a `diquark' from the perspective of its SM couplings.}
\label{tab:leptoquarks}
\end{table}

\begin{figure}
    \centering
    \includegraphics[trim = 0.3cm 1cm 0.3cm 1.2cm, width=0.45\textwidth]{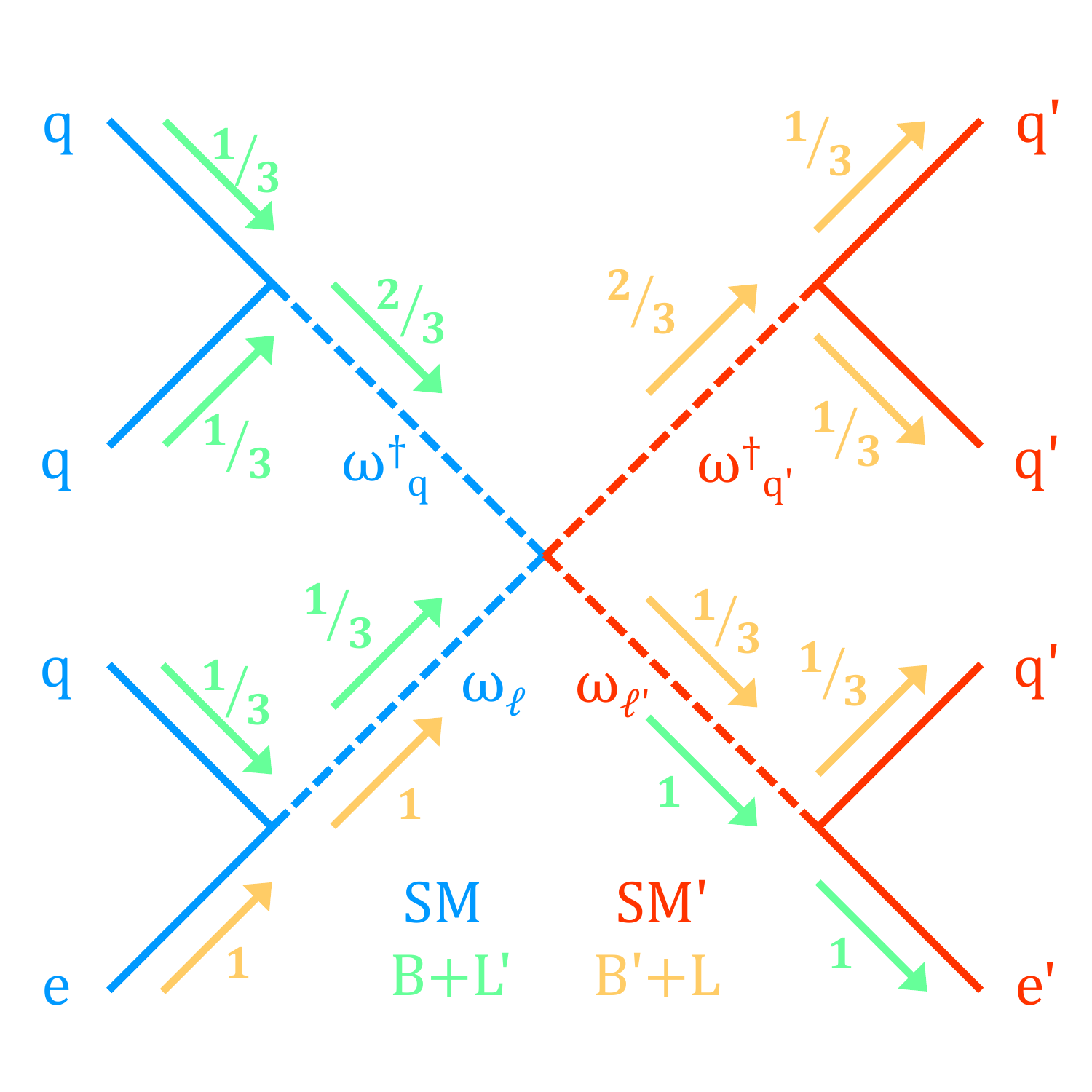}
    \caption{The Feynman diagram generating $\mathcal{O}_{\text{partonic}}$ in the toy UV completion of Section \ref{sec:microphysics}. Particles carrying SM gauge charges are in blue, while those with mirror gauge charges are in red. Arrows track global charge flow, with green arrows giving $B+L'$ and orange arrows giving $B'+L$.}
    \label{fig:uvmixing}
\end{figure}

These scalars allow a mixed quartic coupling 
\begin{equation}
\mathcal{L} \supset \lambda_{ijkl} \omega_q^{i \dagger} \omega_\ell^{j} \omega'{}_q^k \omega'{}_\ell^{l \dagger} + \text{ h.c.}
\end{equation}
where $i,j,k,l=1..N_\omega$, and for simplicity we'll take $\lambda_{ijkl}\equiv 1$.\footnote{We mention parenthetically that including generic scalar portal interactions $\sim |\omega|^2 |\omega'|^2$ will allow two-loop diagrams mixing other neutral bound states such as pions or positronium. As a result of suppression from the UV masses, the resulting oscillation timescales are safely far longer than their lifetimes.} Their global charges are also chosen such that each SM leptoquark couples either to a $B=2/3,L=0$ or $B=1/3,L=1$ fermion current (and the same on the mirror side), which is necessary to prevent rapid proton decay. Their interactions have the schematic form
\begin{align} \label{eqn:leptoquarklag}
    \mathcal{L} &\supset 
    \lambda_q \omega_q^{\dagger A} \epsilon_{ABC} \left(\bar d_R^B u^{cC}_R + \bar q^B_L i \sigma_2 q^{cC}_L\right) \nonumber\\
    &+ \lambda_\ell \omega_\ell^{\dagger} \left(\bar e^c_R u_R + \bar q^c_L i \sigma_2 \ell_L\right) \nonumber\\  &+ \lambda_q \omega'{}_q^{\dagger A'} \epsilon_{A'B'C'}  \left( \bar d'{}_R^{B'} u'{}^{cC'}_R + \bar q^{B'}_L i \sigma_2 q^{cC'}_L\right) \nonumber\\
    &+ \lambda_\ell \omega'{}_\ell^{\dagger} \left(\bar e'{}^c_R u'_R + \bar q^c_L i \sigma_2 \ell_L\right) \nonumber\\
    &+ \text{ h.c.}
\end{align}
where capital Latin letters are used as color indices, we have left off the species indices for compactness, and we have mostly followed the notation of \cite{deBlas:2017xtg} (though the field with these gauge quantum numbers is denoted `$\omega_1$' in their work). These interactions allow a tree-level diagram generating the operator $\mathcal{O}_\text{partonic}$ at low energies with $\Lambda \simeq M_{\omega}$, as seen in Figure \ref{fig:uvmixing}. The multiplicity $N_\omega$ of the leptoquarks allows for $N_\omega^4$ such diagrams connecting the sectors, which has the effect of lowering the effective scale suppressing the higher-dimensional operator for a given leptoquark mass $\Lambda \simeq M_{\omega}/\sqrt{N_\omega}$.

In Section \ref{sec:cosmology} we found that the approximate size of the mixing necessary to account for the EDGES anomaly was $\delta \gtrsim 10^{-37} \text{ GeV}$. Inverting our approximate relationship between the partonic operator and the mixing, and setting all the couplings to unity, we have
\begin{align}
    M_{\omega}^8 &\lesssim \frac{(4 \pi)^2 \Lambda^6_{\text{QCD}} N_\omega^4}{a_0^3 \ \delta} \\
    M_{\omega} &\lesssim 260 \text{ GeV } \left(\frac{\Lambda_{\text{QCD}}}{250 \text{ MeV}}\right)^{3/4} \sqrt{N_\omega}
\end{align}

As our toy UV completion contains new colored states, the LHC imposes constraints on their masses which are in tension with this upper bound for $N_\omega \simeq 1$. There are lower bounds from the LHC on the masses of leptoquarks $\omega_\ell$ decaying to a quark and lepton pair \cite{Aad:2020iuy,Sirunyan:2018btu,Aaboud:2019jcc}. These leptoquarks admit couplings to both the left-handed and right-handed quark and lepton pairs, and the lower bound on their mass depends on their branching ratio to charged leptons. The limit ranges from $M_{\omega_\ell} \gtrsim 1800 \text{ GeV}$ if all decays are to charged leptons, to  $M_{\omega_\ell} \gtrsim 1400 \text{ GeV}$ if half the decays are to neutrinos \cite{Aad:2020iuy}. These bounds are independent of the overall size of the couplings, as the bounds come from pair-production through gluons at the LHC (though see the recent \cite{Greljo:2020nud} for evidence that the electron parton distribution function of the proton can provide sizeable resonant production).

Diquarks are, however, generally far more constrained due to the possibility of single-production at the LHC. Limits are already in the range of multiple TeV for $\mathcal{O}(1)$ couplings \cite{Aad:2019hjw,Sirunyan:2018xlo}, though a precise bound would require a dedicated reinterpretation of those studies. Understanding the bound as a function of both $M_\omega$ and $\lambda_q$ may reveal slightly more parameter space, as the production cross-section and $\delta$ depend upon these in different ways.

Furthermore, with many degenerate or near-degenerate scalars their total differential cross-section to dijet pairs may greatly deviate from the Breit-Wigner lineshape. This may result in lowered efficiency for finding these signals when searches rely explicitly on fitting the data to resonance signal shapes, or when estimating backgrounds by analyzing sidebands which may have been contaminated with signal events. Reinterpreting those studies to found bounds in the $(M_\omega,\lambda_q,N_\omega)$ space may be an interesting, albeit challenging, exercise.

As our interest is solely in providing a proof of concept we don't analyze these possibilities further, and simply note that the constraint in the simplest case is $M_\omega \equiv \sqrt{M_{\omega_\ell} M_{\omega_q}} \gtrsim 3 \text{ TeV}$. As such, in this simple, toy UV completion it is necessary to have a large number of each species, perhaps $N_\omega \sim 10^2$. At the cost of $4(N_\omega-1)$ additional particles, the mixing increases by a factor of $N_\omega^4$. As for an ultraviolet reason for this increased multiplicity, it may naturally result from the leptoquarks carrying an additional quantum number of a (broken) non-Abelian group. In such a case, at weak coupling their Yukawa interactions with SM fermions would have to be loop-suppressed, whereby the reduced coupling would have them easily evade collider bounds. Embedding the scenario into extended technicolor or some sort of SM fermion compositeness might give rise to large such Yukawa couplings, but would likely produce a wide variety of such effects, which may be dangerous.

\section{Other Aspects of the Cosmology and Astrophysics \label{sec:other}}

In the foregoing sections we have established that the hydrogen portal has viable parameter space and can be situated within a plausible UV completion. We now turn to several cosmological and astrophysical issues raised by the scenario, proceeding in roughly chronological order.

\subsection{Mirror Sector Cosmology}

The cosmology of mirror sectors has been explored extensively in many directions, and we refer readers to the reviews \cite{Okun:2006eb,Ciarcelluti:2010zz,Foot:2014mia}. Much of this literature breaks the $\mathbb{Z}_2$ symmetry either explicitly or spontaneously--either to set up the requisite dearth of mirror matter, or to allow some species of mirror matter to be the dark matter.  However, we require that the $\mathbb{Z}_2$ symmetry remains \textit{very, very} good, while still populating only the SM sector and a dark matter sector which communicates only with the unpopulated mirror sector. 

The simplest option to populate the SM is to keep the $\mathbb{Z}_2$ exact at the level of the theory, and to rely on cosmic variance giving effectively asymmetric initial conditions in our Hubble patch. In a model with a pair of independent inflatons, there are regions of the universe where, by chance, the SM inflaton takes on large field values while the mirror inflaton is near the origin of field space. Such a Hubble patch may then contain primarily SM matter after reheating. This was first discussed long ago in the context of `old inflation', but can be adapted to other inflationary scenarios \cite{Kolb:1985bf,Hodges:1993yb,Berezinsky:1999az,Roux:2020wkp}.

It is not necessary that mirror matter be \textit{entirely} absent in the early universe, but precision measurements of the CMB strongly constrain the energy density in relativistic particles during recombination \cite{Aghanim:2018eyx}. For the mirror sector temperature this requires $T_{\text{mirror}} \lesssim \half T_{\text{SM}}$ \cite{Roux:2020wkp}, which leads to a large suppression of mirror energy density $\rho\propto T^4$. While we have taken the initial mirror energy density to vanish for simplicity, it would be interesting to explore cosmological histories which do have some early nonzero mirror energy density, either from reheating or produced via feeble kinetic mixing (see e.g.  \cite{Khlopov:1989fj,Berezhiani:2000gw,Berezhiani:2000gh,Berezhiani:2003wj,Ciarcelluti:2003wm,Ignatiev:2003js,Ciarcelluti:2004ik,Ciarcelluti:2004ip,Ciarcelluti:2008qk,Ciarcelluti:2010zz,Foot:2011ve,Foot:2014mia,Craig:2015pha,Craig:2016lyx,Koren:2019iuv} for some related work).

The same sort of reasoning also provides the simplest sort of dark matter model to construct, where again cosmic variance is relied upon. A simple toy example would be to extend both the SM and mirror sectors with an axion(-like particle). The production of dark matter may take place through the well-known `misalignment mechanism' with a large initial displacement or velocity for the mirror axion field \cite{Dine:1982ah,Abbott:1982af,Preskill:1982cy,Co:2019jts}. But it's entirely possible that in our Hubble patch the SM axion field had an initial position near its minimum, so that SM axions constitute a negligible portion of dark matter. The physics of cooling hydrogen from axion dark matter has recently been studied in \cite{Houston:2018vrf,Houston:2018vbk,Sikivie:2018tml,Lawson:2018qkc,Lambiase:2018lhs,Moroi:2018vci,Choi:2019jwx}. Of course an even wider range of possibilities would come with two additional dark \textit{sectors}---each of which interacts either with the SM or the mirror sector---arranged such that solely the mirror dark sector is populated during reheating. This gives the most freedom in choosing the interactions between mirror hydrogen and dark matter, but comes at the cost of minimality.

It is in principle possible that some acceptably-small $\mathbb{Z}_2$-breaking could be responsible for these effective asymmetries. However, such scenarios will be quite constrained. As an example, if we posit that DM has $\mathbb{Z}_2$-violating couplings such that it interacts solely with $H'$, then $H'$--$X$ scattering is forbidden from being elastic. This is simply because the DM legs in a such a scattering diagram can be connected to form an irreducible one-loop contribution to the $H'$ mass, which can be estimated to be far too large no matter the DM mass. 

However, inelastic scattering between mirror hydrogen and dark matter does not necessarily induce such a mass splitting. There is then a conceivable scenario with $\mathbb{Z}_2$ violation where mirror hydrogen scatters off of dark matter into another state, whether that's an excitation of a composite dark matter particle, another fundamental dark sector state, or a (de-)excitation of mirror hydrogen. Models where dark matter scatters inelastically with SM particles have been considered many times, and have a rich phenomenology (e.g. \cite{TuckerSmith:2001hy,Finkbeiner:2007kk,Chang:2008gd,Khlopov:2008ty,Batell:2009vb,Graham:2010ca,Alves:2010dd,McCullough:2013jma,Barello:2014uda,Blennow:2016gde}). 

Since the mechanism under consideration allows very general sorts of $H'$--DM interactions, as we saw explicitly in Section \ref{sec:cosmology}, we eschew studying any explicit realization and merely offer the above as qualitative guidance for future model building.

\subsection{Recombination \label{sec:recomb}}

Hydrogen oscillations only begin at the start of recombination, as a considerable abundance of neutral hydrogen begins to form.\footnote{It is simple to estimate the rate for free $e+p \rightarrow e'+p'$ reactions before recombination and find that it could only be cosmologically relevant at enormously high temperatures, which the universe may never have attained and which in any case are far above the validity of the theory of Section \ref{sec:microphysics}. In the late universe, the cross-section for $e+p \rightarrow e'+p'$ can be estimated as $\sigma \sim 10^{-85} \text{ cm}^2$ for a cutoff $\Lambda\sim 1 \text{ TeV}$, making this possible disappearance never phenomenologically relevant.} Recombination has been precisely measured using CMB data and exactingly studied using the nonequilibrium theory of atomic transitions and radiative transfer. Is $H$--$H'$ mixing significant at the level of observation?

The concern is that the change in $n_{H}$ (by which, as usual, we mean the \textit{neutral} atomic number density) due to oscillations might alter the course of recombination and the evolution of the visibility function. As is well known, recombination does not proceed directly through the formation of hydrogen atoms in the ground state. The neutral fraction is inhibited from following the Saha equilibrium value because the universe promptly becomes opaque to Lyman continuum photons: each recombination directly to the $1s$ level is counterbalanced by photoionization of another atom. Having $H$ oscillate (perhaps just partially) on the photoionization timescale would attenuate this bottleneck to some degree.\footnote{Incidentally, similar considerations do \textit{not} apply to the slowdown caused by the optical thickness of Ly$n$ lines. As noted in Sec.~\ref{sec:microphysics}, the $l>0$ excited levels do not mix. Hydrogen atoms at these readily ionized levels are unshielded by oscillations.} On the other hand, because $H'$ oscillates back into $H$, the bottleneck would not be broken through entirely. The timing of recombination would still hinge on the redshifting of lines away from resonance and the production of off-resonance photons from two-photon decay \cite{peebles1968recombination,zeldovich1969recombination}.

The crucial point, though, is that the timescale for mixing is simply too long. The CMB has a finite width $\Delta z$, which affects the magnitude of polarization on large scales \cite{Hu:1994jd,Zaldarriaga:1995gi}. A Gaussian fit to the visibility function during recombination gives a last scattering surface centered at $z \simeq 1100$ with a width of $\Delta z \sim 90$ \cite{Hadzhiyska:2018mwh,Aghanim:2018eyx}, corresponding to a cosmic time of $\text{few} \times 10^4 \text{ yrs}$. For $\delta$ in the range we have been focusing on, the oscillation timescale in the dense medium of the early universe exceeds this duration because of mixing suppression by $\sin^2 2\theta_m$ and the quantum Zeno effect.

\subsection{Structure, Stars, and Reionization}

Galaxies and stars form during the dark ages and ultimately bring the era to a close. The full problem is a unique one in which particle oscillations and gravitational dynamics are intertwined. In principle, the evolution should be studied with the quantum kinetic equation for the density matrix in the space of SM and mirror hydrogen states. We leave detailed calculation to future work, here only making some qualitative comments in terms of classical mixtures. Such a treatment is at least partially justified when discussing the evolution of collapsing gas clouds, as the densities climb to sufficiently high values that $H$--$H'$ mixing effectively shuts off. Once this point is reached, the SM and mirror components of a gas cloud can be regarded as evolving independently.

\begin{figure}
    \centering
    \includegraphics[width=0.44\textwidth]{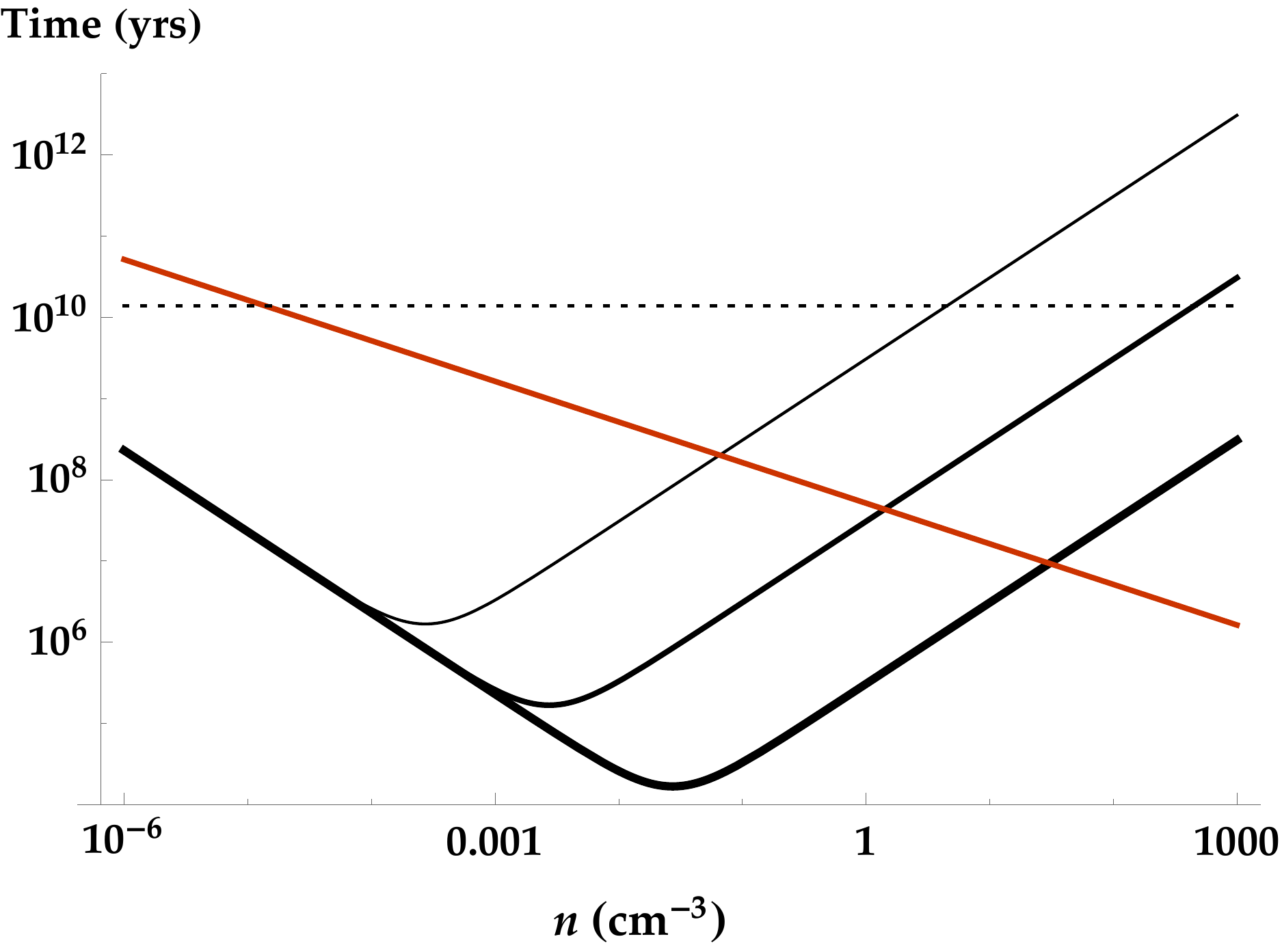}
    \caption{Comparing the oscillation time scale $\Gamma_\textrm{osc}^{-1}$ in a gas cloud of density $n$ and temperature $T \sim 10^4$~K (black curves)  with the gravitational free-fall time $t_\textrm{grav}$ (red) and the present age of the universe (dotted). The values of $\delta$ are $5 \times 10^{-38}$~GeV (thin), $5 \times 10^{-37}$~GeV (medium), and $5 \times 10^{-36}$~GeV (thick).}
    \label{fig:freefall}
\end{figure}

Fig.~\ref{fig:freefall} illustrates the shutting-off of mixing at high densities, showing how the oscillation timescale $\Gamma_\textrm{osc}^{-1}$ compares to the gravitational free-fall time $t_\textrm{grav}$ and the present age of the universe, for three values of $\delta$ and for a gas temperature of $\sim 10^4$~K. The free-fall time is
\begin{equation}
t_\textrm{grav} = \left( \frac{3 \pi}{32 G \rho} \right)^{\frac{1}{2}} \sim \frac{5 \times 10^7~\textrm{yrs}}{n^{1/2}},
\end{equation}
where $\rho$ is the mass density and $n$ is the number density in units of cm$^{-3}$. For the mixing parameters we are considering, oscillations evidently become dynamically unimportant at densities that are fairly low from the standpoint of collapsing gas clouds. For point of reference, the crossing $\Gamma_\textrm{osc}^{-1} \sim t_\textrm{grav}$ corresponds to a Jeans mass on the order of $10^8~M_\odot$ for $\delta = 5 \times 10^{-37}$~GeV. As in the analysis of Sec.~\ref{sec:cosmology}, we neglect the growth of magnetic fields, which---like gas densities---are capable of suppressing mixing.

At the level of linear cosmology, fluids of hydrogen or mirror hydrogen will behave approximately the same as one another, having the same sound speed and Jeans scales. However, significant radiative cooling is necessary for the gas to collapse and fragment into star-forming regions \cite{hoyle1953fragmentation,1977MNRAS.179..541R}. The contraction of a pressure-supported cloud must proceed quickly enough to avoid the quasistatic compression regime, where the gas remains in equilibrium and the pressure continually readjusts to support the cloud against collapse. This requires an ordering of timescales
\begin{equation}\label{eqn:cooling}
    H^{-1} \gtrsim t_\text{grav} \gtrsim t_\text{cool}
\end{equation}
where $H^{-1}$ is the Hubble time at collapse and $t_\text{cool}$ is the cooling timescale.

It is at this stage that the fluids behave differently, as the SM matter has a leg up on cooling as a result of the small residual free electron fraction left over from recombination. Free electrons catalyze the formation of molecular hydrogen via
\begin{align*}
    H + e^- &\rightarrow H^- + \gamma \\
    H^- + H &\rightarrow H_2 + e^-.
\end{align*}
While atomic hydrogen only cools efficiently down to $T \sim 10^4 \ \text{K}$, whereupon Ly$\alpha$ excitations shut off, $H_2$ and other molecules facilitate cooling to lower temperatures by means of their rotational and vibrational levels \cite{1961Obs....81..240M,1968ApJ...154..891P,2005ApJ...624..794S,Barkana:2001avi}. The earliest stars form in minihaloes with mass $M \sim 10^6 \ M_\odot$, where gas is first able to cool and collapse. These haloes have virial temperatures $T_{\text{vir}} \sim \text{ few } \times 10^3 K$, indicating the crucial role played in their development by $H_2$  \cite{1997ApJ...474....1T,2001MNRAS.328..969B,2012ApJ...760....4O}. To make contact with Fig.~\ref{fig:freefall}, we note that the critical density at which the levels of $H_2$ reach local thermodynamic equilibrium is $n \sim 10^4$~cm$^{-3}$. This is well above the density at which $H$--$H'$ mixing becomes inefficient, consistent with our assertion that, by this phase of collapse, gas clouds of $H$ and $H'$ can be regarded as evolving separately.

As heavier halos form, the situation changes because $H_2$ is weakly bound and gets collisionally dissociated at temperatures $T \gtrsim 5\times 10^4 \ \text{K}$. This erases the SM advantage and puts the sectors on equal footing in massive halos that start with virial temperatures much above this threshold. As already mentioned, such halos can only cool down to $T \sim 10^4 \ \text{K}$ through atomic hydrogen. However, during structure formation, shocks can form in these halos with large enough speed to ionize atomic hydrogen. Afterwards, recombinations are out-of-equilibrium as the gas behind the shock cools, and the ionization fraction can be well above the equilibrium value as it cools to below $T \sim 10^4 \ \text{K}$ \cite{Shapiro:1987kan,Oh2002:Hai,Johnson2006:Bro}. At this point molecular hydrogen can again form by electron catalysis. However, accretion-induced shocks affect SM and mirror hydrogen in largely the same way, and so prepare both fluids to cool and fragment.

Predicting when and which objects form in the mirror sector is a task beyond the scope of this study, but we point to a few ideas that have been addressed elsewhere (though not in connection to the hydrogen mixing portal). The authors of Refs.~\cite{DAmico:2017lqj,Latif:2018kqv} studied structure formation in the context of a subdominant mirror sector which is colder than the SM. Quite interestingly, they show that this leads to a mirror ionization fraction which is far lower than in the SM, and so mirror structure formation is modified because the mirror sector halos cannot cool efficiently. Not only does their work provide quantitative evidence for the modified mirror structure formation history we argued for above, but importantly it shows that these qualitative arguments are robust to the inclusion of a subdominant initial mirror sector density. In particular, it will remain the case that SM stars will form first and SM reionization will occur first. They also demonstrated that in low-mass mirror halos, rather than forming stars, this creates the possibility for direct collapse to form the seeds of the supermassive black holes observed at $z \sim 6 - 7$. Rather more generally, the formation of these early black hole seeds has been proposed to occur with a general subdominant component of dark matter having quite strong self-interactions \cite{Pollack:2014rja,Choquette:2018lvq}, as this model will produce.

In higher-mass mirror halos, shocks from gravitational in-fall may be strong enough to ionize mirror hydrogen, which may then form mirror $H_2$ and facilitate efficient cooling \cite{DAmico:2017lqj,Latif:2018kqv}. In such halos, evolution will proceed qualitatively similarly to SM halos, so it is natural to expect the formation of mirror stars. Mirror stars have been considered in Refs.~\cite{Mohapatra:1996yy,Foot:1999hm,Mohapatra:1999ih,Foot:2000vy,Mohapatra:2000qx,Mohapatra:2001sx,Foot:2000tp,Berezhiani:2003xm,Ignatiev:2003js,Berezhiani:2005vv}, among others. They have also seen recent study predicting striking observational signatures \cite{Curtin:2019lhm,Curtin:2019ngc,Curtin:2020tkm}. In the late universe the relic fraction of baryons trapped in the mirror sector becomes a dark component which has self-interactions and dissipative dynamics. Such a component gives rise to the possibility of larger dark bound structures (e.g. \cite{Fan:2013yva,Fan:2013tia,McCullough:2013jma,Buckley:2017ttd}) and other non-WIMP-like behavior. In fact Refs.~\cite{Mohapatra:2000qx,Mohapatra:2001sx} proposed (broken-$\mathbb{Z}_2$) mirror hydrogen atoms as a self-interacting DM candidate to address the `cusp v. core' problem already two decades ago. We refer to \cite{Tulin:2017ara} for a recent general review of self-interacting DM and its connections to small scale structure issues, though there has been less study of the effects of a subdominant self-interacting component than is perhaps warranted.

Returning to the SM sector, it is likely significant that the total fraction of SM gas to DM is half what it would be in the standard cosmology. How this affects the timing of star formation---whether, for example, the formation of the first stars is delayed because a longer period of time is needed to accrete cold gas---is difficult to say a priori and will depend on when mixing comes into equilibrium and when, as a cloud begins to collapse, mixing shuts off.

An important piece of physics affecting structure formation in \textit{both} sectors is the relative motion between baryons and DM, which traces back to baryon acoustic oscillations at the time of recombination \cite{PhysRevD.82.083520}. As photons decouple from the baryonic fluid, the sound speed drops precipitously to $\sim 6$~km/s, well below the RMS relative velocity of baryons and DM, $\sim 30$~km/s. The advection of baryonic density perturbations across DM potential wells results in a number of important effects, including suppression of the growth of small-scale structure \cite{10.1111/j.1365-2966.2011.19541.x,Greif_2011,Naoz_2012,Naoz_2012b,10.1093/mnras/stz013,schauer2020influence}. Streaming velocities are particularly important for models with SM--DM scattering in the post-recombination universe because the bulk relative motion is collisionally dissipated, acting as a heat source for both baryons and DM \cite{Dvorkin:2013cea,Munoz:2015bca}. In our scenario, streaming velocities are damped by $H'$--$X$ scattering, but with mixing in equilibrium, the generated heat is partially transferred to the SM gas. Beyond affecting the sky-averaged thermal evolution, the dissipation of streaming velocities and the conveyance of heat from $H'$ to $H$ alters fluctuations over the sky.

To recap this subsection so far: A no-shortcuts cosmological study of the hydrogen-portal scenario would account for the ``freezing-out'' of mixing in high-density regions, the different chemical dynamics and collapsed-object formation taking place in the two sectors, the alteration to SM galaxy and star formation due to the dilution of the gas density, and the relative motion between baryons and DM.

Reionization is downstream of these effects, with possible consequences for its timing and tomography. The reionization history of the universe remains quite uncertain, both theoretically and observationally. As one measure of this uncertainty, \cite{Greig:2016wjs} studies the constraints placed on a simple, popular three-parameter effective model of the epoch of reionization by a variety of observations. Very roughly, this leads to a $2\sigma$ uncertainty window on the point at which the universe was half reionized of $z \sim 6 \sim 10$, which is $\sim 500$ million years wide in cosmic time. See also e.g. \cite{Bouwens:2015vha,Mitra:2015yqa,Gorce:2017glg,Park:2018ljd} for the effects of a variety of data sets on constraining the timing of reionization. Without carefully assessing the various ways in which hydrogen mixing feeds into the relevant cosmology and astrophysics, it is difficult to say how and at what level reionization will be affected.

The inverse relationship---how reionization affects mixing---also deserves closer study. In Sec.~\ref{sec:cosmology} we argued that the high rate of photoionization successfully maintains mixing equilibrium in the IGM as the neutral hydrogen fraction drops. The actual efficiency with which mirror hydrogen is reconverted is undoubtedly imperfect, though, and depends both on the details of reionization itself and on what fraction of mirror hydrogen is protected by the suppression of mixing in high-density regions.

The incomplete reconversion of mirror hydrogen to SM hydrogen may be related to the long-standing puzzle of `missing baryons' in the late universe. For decades, observations had only been able to account for 60-70\% of the number density of baryons that was measured in the early universe from BBN and the CMB \cite{Fukugita:1997bi,Fukugita:2004ee,2012ApJ...759...23S}. Over the past decade, attention has turned increasingly to an undetected component of the warm-hot intergalactic medium (WHIM) as a potential source. And indeed, in the last two years and due to the combined efforts of many groups, observations have finally been able to confirm such a component \cite{2018Natur.558..406N,deGraaff2019,Johnson_2019,Kov_cs_2019,Macquart:2020lln}. However, while this new component has been confirmed to exist, the amount detected thus far does not incontrovertibly resolve the discrepancy. A recent analysis including these observations concluded that $18 \pm 16\%$ of the baryons are still missing \cite{deGraaff2019}. 

The implications of these new data toward locating all of the missing baryons have sometimes been interpreted without reference to the finite precision of the searches. This is sensible given that---to our knowledge---there has been no `alternative hypothesis' previously put forth. Although we do not here attempt to quantify what fraction of mirror hydrogen remains in that sector down to low redshift, it is clear that, within the hydrogen-portal scenario, some baryons \textit{should} be missing. This scenario thus provides clear motivation for a continued, robust program of searches for baryons in the late universe. Increasing the precision with which the know the late-time inventory of cosmic baryons may be a useful way to falsify this proposal.

\section{Conclusion \label{sec:conc}}

In this work we have proposed that hydrogen is cooled during the dark ages on characteristic timescales $t \sim$ million years due a dimension-12 operator generated at energies $E \sim \hbar / (10^{-28} \text{ seconds})$ which effectively causes hydrogen to \textit{disappear} during the dark ages and \textit{reappear} during reionization. The particle physics underlying the scenario is conventional, simple, and requires no fine-tuning. The enormous deviations from standard cosmology appear not to be constrained, but are falsifiable in multiple ways with upcoming and proposed experiments. 

Our work takes advantage of a mechanism for connecting the SM to a dark sector which has not been previously studied, namely the oscillation of hydrogen into a mirror state over cosmological timescales. Not only does this mechanism provide a portal for hydrogen to be cooled down by dark matter, but it very naturally operates solely during the dark ages. Despite changing the makeup of the universe substantially during that time, it then naturally reverts these changes during the course of reionization. After studying the cosmological evolution of the hydrogen temperature during the dark ages, we have conducted an initial exploration of the qualitative effects of oscillations on the epoch of reionization more generally. 

We have explicitly constructed an EFT realization of the necessary mixing in which the necessary features are symmetry-protected in a mirror model. We have then given an example of the lowest-lying states in a UV completion, using states that appear naturally in many well-studied models of physics beyond the standard model. As the particular interaction of mirror hydrogen with dark matter is not crucial for understanding the mechanism, we have taken a phenomenological approach and used a simple parametrization of the interactions.

It seems unlikely that this mixing could ever be directly probed in late-universe terrestrial or astrophysical settings. In any situation where hydrogen is in a bound state such as a molecule, or confined in a space by walls of SM matter, or in a medium dense in SM matter, the potentials felt by hydrogen and mirror hydrogen will differ \textit{far} more than the mixing, $|\Delta V| \gg \delta$, which will heavily suppress oscillations, $\theta_m \ll 1$. This makes cosmological settings the only regime in which the mixing will be appreciable. Likely the main avenue of exploring this model will be studying further detailed behavior of hydrogen in the early universe. Understanding the precision cosmology of this model motivates further study of a variety of features of which we have only pursued initial explorations. 

Such theory efforts will be rewarded in the near-future, as an abundance of 21 cm data will be available from a variety of experiments. Future measurements of the sky-averaged signal by DARE \cite{2012AdSpR..49..433B}, LEDA \cite{2018MNRAS.478.4193P}, PRI$^Z$M  \cite{2019JAI.....850004P}, SARAS \cite{2018ExA....45..269S}, and REACH \cite{deLera:2019} will be able to confirm the presence of the anomalous absorption feature and measure its shape in further detail. Other upcoming experiments such as HERA \cite{2019BAAS...51g.241P}, OVRO-LWA \cite{2019AJ....158...84E}, and SKA1-LOW \cite{Mellema:2012ht} will measure the power spectrum of 21 cm fluctuations, which will provide a humongous wealth of information about the dark ages.  

Of course, the real smoking gun signal of this mechanism is that it affects solely neutral hydrogen atoms, so that efforts to probe the epoch of reionization via the line intensity of other atoms, such as helium \cite{Visbal:2015sca,2009arXiv0905.1698B,2009PhRvD..80f3010M,10.1093/mnras/staa1951}, molecular hydrogen \cite{2013ApJ...768..130G} and deuterium \cite{Kosenko_2018,2006PhRvL..97i1301S} will provide a complementary perspective. Measurements of these transitions are challenging, as the abundances of these elements are far lower than that of atomic hydrogen. However, if the anomalously cool spin temperature of hydrogen is confirmed by the current generation of reionization experiments, this would immediately become an important place to look. 

As mentioned in Section \ref{sec:cosmology}, the (dis)agreement of between early and late-time measurements of the number density of baryons is another interesting probe of this model. To our knowledge, this is the first proposal that there \textit{should} be some fraction of missing baryons in the late universe. Of course it's difficult to imagine a positive `detection' of a particular fraction of missing baryons, but a higher-precision inventory of the baryons at $z \simeq 0$ seems a fruitful avenue for falsifying this mechanism.

In comparison to prior efforts to find models reproducing the EDGES data---which seem to require many contortions---this is a model which \textit{wants} to produce such a signal. While empirical evidence will of course require further detailed study and await future 21cm experiments, it is remarkable how many non-trivial checks this model satisfies with a quite minimal set of input ingredients. There is indeed a modicum of tension between the natural mass scale for the UV completion, $\Lambda \lesssim \text{ TeV}/\text{ few }$ and the LHC constraints $\Lambda \gtrsim \text{ few } \times \text{ TeV}$. Given the number of things which go right, as well as the fact that the effect of interest appears $\sim 40$ orders of magnitude below the scale of the UV completion and seemingly by chance the natural value is solely in $\sim 1$ order of magnitude of tension, we think this mechanism is more than worth further exploration.

In this initial exploration we have left open many interesting lines of inquiry. Our discussions have been conservative in restricting to the most simple scenarios where we could semi-analytically compute. It would be interesting to understand what occurs outside of the case of complete thermal equilibration between $H'$ and DM which has no interplay with the $H-H'$ mixing, including the case of inelastic $H'$-DM interactions. We have furthermore restricted to the case of no initial mirror sector density, but allowing a nonzero initial density may well allow for richer structure formation effects in the mirror sector.

There are furthermore aspects of the cosmological history which we studied solely qualitatively or by comparing timescales, which are clear targets for further detailed study and computation. These range from understanding in detail how superpositions of $H-H'$ behave while clumping, to simulating the structure formation history in both sectors with modified number densities of various species, to computing the rates of ionizing radiation and the ensuing flow of $H'$ back into SM baryons, which would allow for direct connection to the missing baryons in the late universe. Understanding in detail the dynamics in the mirror sector may reveal the production of early black holes \cite{DAmico:2017lqj}, and predicting the spectrum of mirror stars produced could allow us to connect to spectacular astronomical signatures \cite{Curtin:2019lhm,Curtin:2019ngc}.

On the side of particle physics, it is clearly of interest to further understand UV complete scenarios in which this mechanism is embedded. This would be useful not just to generate the mixing operator without relying on a multiplicity of scalars, but also in suggesting DM candidates, or in finding a natural way for mixing to depend on the hyperfine quantum number. As we've mentioned, there are a variety of potential connections to other puzzles that could be made more explicit, from leptoquarks being related to grand unification or flavor anomalies, to supersymmetry helping keep $\mathbb{Z}_2$ violation small. A particularly intriguing direction is the recent advent of a twin Higgs model which does not require $\mathbb{Z}_2$-breaking \cite{Csaki:2019qgb}, which may allow one to connect to the physics of the hierarchy problem (see e.g. \cite{Koren:2020biu} for a recent introduction and review thereof).

Clearly, the desire to explore this scenario motivates a wide variety of directions for more-detailed study by cosmologists and particle physicists alike. More generally, to the extent that early universe particle cosmologists tend to regard the evolution after BBN as being fixed, it is surprising that this humongous change in the behavior of the universe during the dark ages seems unconstrained. It would be interesting to push on this possibility, both on the theoretical side in exploring the space of allowed modifications and on the observational side in thinking about new probes of this era. \\

\section*{Acknowledgements}\addcontentsline{toc}{section}{Acknowledgements}

The authors thank Samuel Alipour-fard, Guido D'Amico, Robert McGehee, Paolo Panci, and Yiming Zhong for comments on a draft of this manuscript. LJ thanks Anna Schauer for insights into first-star formation and for suggesting a connection to direct-collapse black holes. SK thanks Vera Gluscevic for presenting an enlightening seminar on 21cm cosmology at the KITP in December 2019.

The work of LJ was supported by NSF Grant No. PHY-1914242 and by NASA through the NASA Hubble Fellowship grant \# HST-HF2-51461.001-A awarded by the Space Telescope Science Institute, which is operated by the Association of Universities for Research in Astronomy, Incorporated, under NASA contract NAS5-26555. The work of SK was supported in part by the Department of Energy under the grant DE-SC0250757, and by a Mafalda and Reinhard Oehme Postdoctoral Fellowship from the Enrico Fermi Institute at the University of Chicago.

\bibliography{hydrogen}

\begin{thebibliography}{241}%
\makeatletter
\providecommand \@ifxundefined [1]{%
 \@ifx{#1\undefined}
}%
\providecommand \@ifnum [1]{%
 \ifnum #1\expandafter \@firstoftwo
 \else \expandafter \@secondoftwo
 \fi
}%
\providecommand \@ifx [1]{%
 \ifx #1\expandafter \@firstoftwo
 \else \expandafter \@secondoftwo
 \fi
}%
\providecommand \natexlab [1]{#1}%
\providecommand \enquote  [1]{``#1''}%
\providecommand \bibnamefont  [1]{#1}%
\providecommand \bibfnamefont [1]{#1}%
\providecommand \citenamefont [1]{#1}%
\providecommand \href@noop [0]{\@secondoftwo}%
\providecommand \href [0]{\begingroup \@sanitize@url \@href}%
\providecommand \@href[1]{\@@startlink{#1}\@@href}%
\providecommand \@@href[1]{\endgroup#1\@@endlink}%
\providecommand \@sanitize@url [0]{\catcode `\\12\catcode `\$12\catcode
  `\&12\catcode `\#12\catcode `\^12\catcode `\_12\catcode `\%12\relax}%
\providecommand \@@startlink[1]{}%
\providecommand \@@endlink[0]{}%
\providecommand \url  [0]{\begingroup\@sanitize@url \@url }%
\providecommand \@url [1]{\endgroup\@href {#1}{\urlprefix }}%
\providecommand \urlprefix  [0]{URL }%
\providecommand \Eprint [0]{\href }%
\providecommand \doibase [0]{https://doi.org/}%
\providecommand \selectlanguage [0]{\@gobble}%
\providecommand \bibinfo  [0]{\@secondoftwo}%
\providecommand \bibfield  [0]{\@secondoftwo}%
\providecommand \translation [1]{[#1]}%
\providecommand \BibitemOpen [0]{}%
\providecommand \bibitemStop [0]{}%
\providecommand \bibitemNoStop [0]{.\EOS\space}%
\providecommand \EOS [0]{\spacefactor3000\relax}%
\providecommand \BibitemShut  [1]{\csname bibitem#1\endcsname}%
\let\auto@bib@innerbib\@empty
\bibitem [{\citenamefont {Chen}\ and\ \citenamefont
  {Kamionkowski}(2004)}]{Chen:2003gz}%
  \BibitemOpen
  \bibfield  {author} {\bibinfo {author} {\bibfnamefont {X.-L.}\ \bibnamefont
  {Chen}}\ and\ \bibinfo {author} {\bibfnamefont {M.}~\bibnamefont
  {Kamionkowski}},\ }\bibfield  {title} {\bibinfo {title} {{Particle decays
  during the cosmic dark ages}},\ }\href
  {https://doi.org/10.1103/PhysRevD.70.043502} {\bibfield  {journal} {\bibinfo
  {journal} {Phys. Rev. D}\ }\textbf {\bibinfo {volume} {70}},\ \bibinfo
  {pages} {043502} (\bibinfo {year} {2004})},\ \Eprint
  {https://arxiv.org/abs/astro-ph/0310473} {arXiv:astro-ph/0310473}
  \BibitemShut {NoStop}%
\bibitem [{\citenamefont {Furlanetto}\ \emph {et~al.}(2006)\citenamefont
  {Furlanetto}, \citenamefont {Oh},\ and\ \citenamefont
  {Pierpaoli}}]{Furlanetto:2006wp}%
  \BibitemOpen
  \bibfield  {author} {\bibinfo {author} {\bibfnamefont {S.~R.}\ \bibnamefont
  {Furlanetto}}, \bibinfo {author} {\bibfnamefont {S.}~\bibnamefont {Oh}},\
  and\ \bibinfo {author} {\bibfnamefont {E.}~\bibnamefont {Pierpaoli}},\
  }\bibfield  {title} {\bibinfo {title} {{The Effects of Dark Matter Decay and
  Annihilation on the High-Redshift 21 cm Background}},\ }\href
  {https://doi.org/10.1103/PhysRevD.74.103502} {\bibfield  {journal} {\bibinfo
  {journal} {Phys. Rev. D}\ }\textbf {\bibinfo {volume} {74}},\ \bibinfo
  {pages} {103502} (\bibinfo {year} {2006})},\ \Eprint
  {https://arxiv.org/abs/astro-ph/0608385} {arXiv:astro-ph/0608385}
  \BibitemShut {NoStop}%
\bibitem [{\citenamefont {{Vald{\'e}s}}\ \emph {et~al.}(2007)\citenamefont
  {{Vald{\'e}s}}, \citenamefont {{Ferrara}}, \citenamefont {{Mapelli}},\ and\
  \citenamefont {{Ripamonti}}}]{2007MNRAS.377..245V}%
  \BibitemOpen
  \bibfield  {author} {\bibinfo {author} {\bibfnamefont {M.}~\bibnamefont
  {{Vald{\'e}s}}}, \bibinfo {author} {\bibfnamefont {A.}~\bibnamefont
  {{Ferrara}}}, \bibinfo {author} {\bibfnamefont {M.}~\bibnamefont
  {{Mapelli}}},\ and\ \bibinfo {author} {\bibfnamefont {E.}~\bibnamefont
  {{Ripamonti}}},\ }\bibfield  {title} {\bibinfo {title} {{Constraining dark
  matter through 21-cm observations}},\ }\href
  {https://doi.org/10.1111/j.1365-2966.2007.11594.x} {\bibfield  {journal}
  {\bibinfo  {journal} {Mon. Not. Roy. Astron. Soc.}\ }\textbf {\bibinfo
  {volume} {377}},\ \bibinfo {pages} {245} (\bibinfo {year} {2007})},\ \Eprint
  {https://arxiv.org/abs/astro-ph/0701301} {arXiv:astro-ph/0701301 [astro-ph]}
  \BibitemShut {NoStop}%
\bibitem [{\citenamefont {{Belikov}}\ and\ \citenamefont
  {{Hooper}}(2009)}]{2009PhRvD..80c5007B}%
  \BibitemOpen
  \bibfield  {author} {\bibinfo {author} {\bibfnamefont {A.~V.}\ \bibnamefont
  {{Belikov}}}\ and\ \bibinfo {author} {\bibfnamefont {D.}~\bibnamefont
  {{Hooper}}},\ }\bibfield  {title} {\bibinfo {title} {{How dark matter
  reionized the Universe}},\ }\href
  {https://doi.org/10.1103/PhysRevD.80.035007} {\bibfield  {journal} {\bibinfo
  {journal} {Phys. Rev. D}\ }\textbf {\bibinfo {volume} {80}},\ \bibinfo {eid}
  {035007} (\bibinfo {year} {2009})},\ \Eprint
  {https://arxiv.org/abs/0904.1210} {arXiv:0904.1210 [hep-ph]} \BibitemShut
  {NoStop}%
\bibitem [{\citenamefont {Cumberbatch}\ \emph {et~al.}(2010)\citenamefont
  {Cumberbatch}, \citenamefont {Lattanzi}, \citenamefont {Silk}, \citenamefont
  {Lattanzi},\ and\ \citenamefont {Silk}}]{Cumberbatch:2008rh}%
  \BibitemOpen
  \bibfield  {author} {\bibinfo {author} {\bibfnamefont {D.~T.}\ \bibnamefont
  {Cumberbatch}}, \bibinfo {author} {\bibfnamefont {M.}~\bibnamefont
  {Lattanzi}}, \bibinfo {author} {\bibfnamefont {J.}~\bibnamefont {Silk}},
  \bibinfo {author} {\bibfnamefont {M.}~\bibnamefont {Lattanzi}},\ and\
  \bibinfo {author} {\bibfnamefont {J.}~\bibnamefont {Silk}},\ }\bibfield
  {title} {\bibinfo {title} {{Signatures of clumpy dark matter in the global 21
  cm Background Signal}},\ }\href {https://doi.org/10.1103/PhysRevD.82.103508}
  {\bibfield  {journal} {\bibinfo  {journal} {Phys. Rev. D}\ }\textbf {\bibinfo
  {volume} {82}},\ \bibinfo {pages} {103508} (\bibinfo {year} {2010})},\
  \Eprint {https://arxiv.org/abs/0808.0881} {arXiv:0808.0881 [astro-ph]}
  \BibitemShut {NoStop}%
\bibitem [{\citenamefont {Finkbeiner}\ \emph {et~al.}(2008)\citenamefont
  {Finkbeiner}, \citenamefont {Padmanabhan},\ and\ \citenamefont
  {Weiner}}]{Finkbeiner:2008gw}%
  \BibitemOpen
  \bibfield  {author} {\bibinfo {author} {\bibfnamefont {D.~P.}\ \bibnamefont
  {Finkbeiner}}, \bibinfo {author} {\bibfnamefont {N.}~\bibnamefont
  {Padmanabhan}},\ and\ \bibinfo {author} {\bibfnamefont {N.}~\bibnamefont
  {Weiner}},\ }\bibfield  {title} {\bibinfo {title} {{CMB and 21-cm Signals for
  Dark Matter with a Long-Lived Excited State}},\ }\href
  {https://doi.org/10.1103/PhysRevD.78.063530} {\bibfield  {journal} {\bibinfo
  {journal} {Phys. Rev. D}\ }\textbf {\bibinfo {volume} {78}},\ \bibinfo
  {pages} {063530} (\bibinfo {year} {2008})},\ \Eprint
  {https://arxiv.org/abs/0805.3531} {arXiv:0805.3531 [astro-ph]} \BibitemShut
  {NoStop}%
\bibitem [{\citenamefont {Slatyer}\ \emph {et~al.}(2009)\citenamefont
  {Slatyer}, \citenamefont {Padmanabhan},\ and\ \citenamefont
  {Finkbeiner}}]{Slatyer:2009yq}%
  \BibitemOpen
  \bibfield  {author} {\bibinfo {author} {\bibfnamefont {T.~R.}\ \bibnamefont
  {Slatyer}}, \bibinfo {author} {\bibfnamefont {N.}~\bibnamefont
  {Padmanabhan}},\ and\ \bibinfo {author} {\bibfnamefont {D.~P.}\ \bibnamefont
  {Finkbeiner}},\ }\bibfield  {title} {\bibinfo {title} {{CMB Constraints on
  WIMP Annihilation: Energy Absorption During the Recombination Epoch}},\
  }\href {https://doi.org/10.1103/PhysRevD.80.043526} {\bibfield  {journal}
  {\bibinfo  {journal} {Phys. Rev. D}\ }\textbf {\bibinfo {volume} {80}},\
  \bibinfo {pages} {043526} (\bibinfo {year} {2009})},\ \Eprint
  {https://arxiv.org/abs/0906.1197} {arXiv:0906.1197 [astro-ph.CO]}
  \BibitemShut {NoStop}%
\bibitem [{\citenamefont {Natarajan}\ and\ \citenamefont
  {Schwarz}(2009)}]{Natarajan:2009bm}%
  \BibitemOpen
  \bibfield  {author} {\bibinfo {author} {\bibfnamefont {A.}~\bibnamefont
  {Natarajan}}\ and\ \bibinfo {author} {\bibfnamefont {D.~J.}\ \bibnamefont
  {Schwarz}},\ }\bibfield  {title} {\bibinfo {title} {{Dark matter annihilation
  and its effect on CMB and Hydrogen 21 cm observations}},\ }\href
  {https://doi.org/10.1103/PhysRevD.80.043529} {\bibfield  {journal} {\bibinfo
  {journal} {Phys. Rev. D}\ }\textbf {\bibinfo {volume} {80}},\ \bibinfo
  {pages} {043529} (\bibinfo {year} {2009})},\ \Eprint
  {https://arxiv.org/abs/0903.4485} {arXiv:0903.4485 [astro-ph.CO]}
  \BibitemShut {NoStop}%
\bibitem [{\citenamefont {Natarajan}\ and\ \citenamefont
  {Schwarz}(2010)}]{Natarajan:2010dc}%
  \BibitemOpen
  \bibfield  {author} {\bibinfo {author} {\bibfnamefont {A.}~\bibnamefont
  {Natarajan}}\ and\ \bibinfo {author} {\bibfnamefont {D.~J.}\ \bibnamefont
  {Schwarz}},\ }\bibfield  {title} {\bibinfo {title} {{Distinguishing standard
  reionization from dark matter models}},\ }\href
  {https://doi.org/10.1103/PhysRevD.81.123510} {\bibfield  {journal} {\bibinfo
  {journal} {Phys. Rev. D}\ }\textbf {\bibinfo {volume} {81}},\ \bibinfo
  {pages} {123510} (\bibinfo {year} {2010})},\ \Eprint
  {https://arxiv.org/abs/1002.4405} {arXiv:1002.4405 [astro-ph.CO]}
  \BibitemShut {NoStop}%
\bibitem [{\citenamefont {Yue}\ and\ \citenamefont {Chen}(2012)}]{Yue:2012na}%
  \BibitemOpen
  \bibfield  {author} {\bibinfo {author} {\bibfnamefont {B.}~\bibnamefont
  {Yue}}\ and\ \bibinfo {author} {\bibfnamefont {X.}~\bibnamefont {Chen}},\
  }\bibfield  {title} {\bibinfo {title} {{Reionization in the Warm Dark Matter
  Model}},\ }\href {https://doi.org/10.1088/0004-637X/747/2/127} {\bibfield
  {journal} {\bibinfo  {journal} {Astrophys. J.}\ }\textbf {\bibinfo {volume}
  {747}},\ \bibinfo {pages} {127} (\bibinfo {year} {2012})},\ \Eprint
  {https://arxiv.org/abs/1201.3686} {arXiv:1201.3686 [astro-ph.CO]}
  \BibitemShut {NoStop}%
\bibitem [{\citenamefont {{Vald{\'e}s}}\ \emph {et~al.}(2013)\citenamefont
  {{Vald{\'e}s}}, \citenamefont {{Evoli}}, \citenamefont {{Mesinger}},
  \citenamefont {{Ferrara}},\ and\ \citenamefont
  {{Yoshida}}}]{2013MNRAS.429.1705V}%
  \BibitemOpen
  \bibfield  {author} {\bibinfo {author} {\bibfnamefont {M.}~\bibnamefont
  {{Vald{\'e}s}}}, \bibinfo {author} {\bibfnamefont {C.}~\bibnamefont
  {{Evoli}}}, \bibinfo {author} {\bibfnamefont {A.}~\bibnamefont {{Mesinger}}},
  \bibinfo {author} {\bibfnamefont {A.}~\bibnamefont {{Ferrara}}},\ and\
  \bibinfo {author} {\bibfnamefont {N.}~\bibnamefont {{Yoshida}}},\ }\bibfield
  {title} {\bibinfo {title} {{The nature of dark matter from the global
  high-redshift H I 21 cm signal}},\ }\href
  {https://doi.org/10.1093/mnras/sts458} {\bibfield  {journal} {\bibinfo
  {journal} {Mon. Not. Roy. Astron. Soc.}\ }\textbf {\bibinfo {volume} {429}},\
  \bibinfo {pages} {1705} (\bibinfo {year} {2013})},\ \Eprint
  {https://arxiv.org/abs/1209.2120} {arXiv:1209.2120 [astro-ph.CO]}
  \BibitemShut {NoStop}%
\bibitem [{\citenamefont {Dvorkin}\ \emph {et~al.}(2014)\citenamefont
  {Dvorkin}, \citenamefont {Blum},\ and\ \citenamefont
  {Kamionkowski}}]{Dvorkin:2013cea}%
  \BibitemOpen
  \bibfield  {author} {\bibinfo {author} {\bibfnamefont {C.}~\bibnamefont
  {Dvorkin}}, \bibinfo {author} {\bibfnamefont {K.}~\bibnamefont {Blum}},\ and\
  \bibinfo {author} {\bibfnamefont {M.}~\bibnamefont {Kamionkowski}},\
  }\bibfield  {title} {\bibinfo {title} {{Constraining Dark Matter-Baryon
  Scattering with Linear Cosmology}},\ }\href
  {https://doi.org/10.1103/PhysRevD.89.023519} {\bibfield  {journal} {\bibinfo
  {journal} {Phys. Rev. D}\ }\textbf {\bibinfo {volume} {89}},\ \bibinfo
  {pages} {023519} (\bibinfo {year} {2014})},\ \Eprint
  {https://arxiv.org/abs/1311.2937} {arXiv:1311.2937 [astro-ph.CO]}
  \BibitemShut {NoStop}%
\bibitem [{\citenamefont {Evoli}\ \emph {et~al.}(2014)\citenamefont {Evoli},
  \citenamefont {Mesinger},\ and\ \citenamefont {Ferrara}}]{Evoli:2014pva}%
  \BibitemOpen
  \bibfield  {author} {\bibinfo {author} {\bibfnamefont {C.}~\bibnamefont
  {Evoli}}, \bibinfo {author} {\bibfnamefont {A.}~\bibnamefont {Mesinger}},\
  and\ \bibinfo {author} {\bibfnamefont {A.}~\bibnamefont {Ferrara}},\
  }\bibfield  {title} {\bibinfo {title} {{Unveiling the nature of dark matter
  with high redshift 21 cm line experiments}},\ }\href
  {https://doi.org/10.1088/1475-7516/2014/11/024} {\bibfield  {journal}
  {\bibinfo  {journal} {JCAP}\ }\textbf {\bibinfo {volume} {11}},\ \bibinfo
  {pages} {024}},\ \Eprint {https://arxiv.org/abs/1408.1109} {arXiv:1408.1109
  [astro-ph.HE]} \BibitemShut {NoStop}%
\bibitem [{\citenamefont {Tashiro}\ \emph {et~al.}(2014)\citenamefont
  {Tashiro}, \citenamefont {Kadota},\ and\ \citenamefont
  {Silk}}]{Tashiro:2014tsa}%
  \BibitemOpen
  \bibfield  {author} {\bibinfo {author} {\bibfnamefont {H.}~\bibnamefont
  {Tashiro}}, \bibinfo {author} {\bibfnamefont {K.}~\bibnamefont {Kadota}},\
  and\ \bibinfo {author} {\bibfnamefont {J.}~\bibnamefont {Silk}},\ }\bibfield
  {title} {\bibinfo {title} {{Effects of dark matter-baryon scattering on
  redshifted 21 cm signals}},\ }\href
  {https://doi.org/10.1103/PhysRevD.90.083522} {\bibfield  {journal} {\bibinfo
  {journal} {Phys. Rev. D}\ }\textbf {\bibinfo {volume} {90}},\ \bibinfo
  {pages} {083522} (\bibinfo {year} {2014})},\ \Eprint
  {https://arxiv.org/abs/1408.2571} {arXiv:1408.2571 [astro-ph.CO]}
  \BibitemShut {NoStop}%
\bibitem [{\citenamefont {Oldengott}\ \emph {et~al.}(2016)\citenamefont
  {Oldengott}, \citenamefont {Boriero},\ and\ \citenamefont
  {Schwarz}}]{Oldengott:2016yjc}%
  \BibitemOpen
  \bibfield  {author} {\bibinfo {author} {\bibfnamefont {I.~M.}\ \bibnamefont
  {Oldengott}}, \bibinfo {author} {\bibfnamefont {D.}~\bibnamefont {Boriero}},\
  and\ \bibinfo {author} {\bibfnamefont {D.~J.}\ \bibnamefont {Schwarz}},\
  }\bibfield  {title} {\bibinfo {title} {{Reionization and dark matter
  decay}},\ }\href {https://doi.org/10.1088/1475-7516/2016/08/054} {\bibfield
  {journal} {\bibinfo  {journal} {JCAP}\ }\textbf {\bibinfo {volume} {08}},\
  \bibinfo {pages} {054}},\ \Eprint {https://arxiv.org/abs/1605.03928}
  {arXiv:1605.03928 [astro-ph.CO]} \BibitemShut {NoStop}%
\bibitem [{\citenamefont {Rudakovskiy}\ and\ \citenamefont
  {Iakubovskyi}(2016)}]{Rudakovskiy:2016ngi}%
  \BibitemOpen
  \bibfield  {author} {\bibinfo {author} {\bibfnamefont {A.}~\bibnamefont
  {Rudakovskiy}}\ and\ \bibinfo {author} {\bibfnamefont {D.}~\bibnamefont
  {Iakubovskyi}},\ }\bibfield  {title} {\bibinfo {title} {{Influence of
  \textasciitilde{}7 keV sterile neutrino dark matter on the process of
  reionization}},\ }\href {https://doi.org/10.1088/1475-7516/2016/06/017}
  {\bibfield  {journal} {\bibinfo  {journal} {JCAP}\ }\textbf {\bibinfo
  {volume} {06}},\ \bibinfo {pages} {017}},\ \Eprint
  {https://arxiv.org/abs/1604.01341} {arXiv:1604.01341 [astro-ph.CO]}
  \BibitemShut {NoStop}%
\bibitem [{\citenamefont {Lopez-Honorez}\ \emph {et~al.}(2017)\citenamefont
  {Lopez-Honorez}, \citenamefont {Mena}, \citenamefont {Palomares-Ruiz},\ and\
  \citenamefont {Villanueva-Domingo}}]{Lopez-Honorez:2017csg}%
  \BibitemOpen
  \bibfield  {author} {\bibinfo {author} {\bibfnamefont {L.}~\bibnamefont
  {Lopez-Honorez}}, \bibinfo {author} {\bibfnamefont {O.}~\bibnamefont {Mena}},
  \bibinfo {author} {\bibfnamefont {S.}~\bibnamefont {Palomares-Ruiz}},\ and\
  \bibinfo {author} {\bibfnamefont {P.}~\bibnamefont {Villanueva-Domingo}},\
  }\bibfield  {title} {\bibinfo {title} {{Warm dark matter and the ionization
  history of the Universe}},\ }\href
  {https://doi.org/10.1103/PhysRevD.96.103539} {\bibfield  {journal} {\bibinfo
  {journal} {Phys. Rev. D}\ }\textbf {\bibinfo {volume} {96}},\ \bibinfo
  {pages} {103539} (\bibinfo {year} {2017})},\ \Eprint
  {https://arxiv.org/abs/1703.02302} {arXiv:1703.02302 [astro-ph.CO]}
  \BibitemShut {NoStop}%
\bibitem [{\citenamefont {Brandenberger}\ \emph {et~al.}(2010)\citenamefont
  {Brandenberger}, \citenamefont {Danos}, \citenamefont {Hernandez},\ and\
  \citenamefont {Holder}}]{Brandenberger:2010hn}%
  \BibitemOpen
  \bibfield  {author} {\bibinfo {author} {\bibfnamefont {R.~H.}\ \bibnamefont
  {Brandenberger}}, \bibinfo {author} {\bibfnamefont {R.~J.}\ \bibnamefont
  {Danos}}, \bibinfo {author} {\bibfnamefont {O.~F.}\ \bibnamefont
  {Hernandez}},\ and\ \bibinfo {author} {\bibfnamefont {G.~P.}\ \bibnamefont
  {Holder}},\ }\bibfield  {title} {\bibinfo {title} {{The 21 cm Signature of
  Cosmic String Wakes}},\ }\href
  {https://doi.org/10.1088/1475-7516/2010/12/028} {\bibfield  {journal}
  {\bibinfo  {journal} {JCAP}\ }\textbf {\bibinfo {volume} {12}},\ \bibinfo
  {pages} {028}},\ \Eprint {https://arxiv.org/abs/1006.2514} {arXiv:1006.2514
  [astro-ph.CO]} \BibitemShut {NoStop}%
\bibitem [{\citenamefont {{Hern{\'a}ndez}}\ \emph {et~al.}(2011)\citenamefont
  {{Hern{\'a}ndez}}, \citenamefont {{Wang}}, \citenamefont {{Brandenberger}},\
  and\ \citenamefont {{Fong}}}]{2011JCAP...08..014H}%
  \BibitemOpen
  \bibfield  {author} {\bibinfo {author} {\bibfnamefont {O.~F.}\ \bibnamefont
  {{Hern{\'a}ndez}}}, \bibinfo {author} {\bibfnamefont {Y.}~\bibnamefont
  {{Wang}}}, \bibinfo {author} {\bibfnamefont {R.}~\bibnamefont
  {{Brandenberger}}},\ and\ \bibinfo {author} {\bibfnamefont {J.}~\bibnamefont
  {{Fong}}},\ }\bibfield  {title} {\bibinfo {title} {{Angular 21 cm power
  spectrum of a scaling distribution of cosmic string wakes}},\ }\href
  {https://doi.org/10.1088/1475-7516/2011/08/014} {\bibfield  {journal}
  {\bibinfo  {journal} {J. Cosmol. Astropart. Phys.}\ }\textbf {\bibinfo
  {volume} {2011}}\bibfield  {number} {\bibinfo  {number} { (8)},\ \bibinfo
  {eid} {014}},\ }\Eprint {https://arxiv.org/abs/1104.3337} {arXiv:1104.3337
  [astro-ph.CO]} \BibitemShut {NoStop}%
\bibitem [{\citenamefont {{Tashiro}}\ \emph {et~al.}(2012)\citenamefont
  {{Tashiro}}, \citenamefont {{Sabancilar}},\ and\ \citenamefont
  {{Vachaspati}}}]{2012PhRvD..85l3535T}%
  \BibitemOpen
  \bibfield  {author} {\bibinfo {author} {\bibfnamefont {H.}~\bibnamefont
  {{Tashiro}}}, \bibinfo {author} {\bibfnamefont {E.}~\bibnamefont
  {{Sabancilar}}},\ and\ \bibinfo {author} {\bibfnamefont {T.}~\bibnamefont
  {{Vachaspati}}},\ }\bibfield  {title} {\bibinfo {title} {{Constraints on
  superconducting cosmic strings from early reionization}},\ }\href
  {https://doi.org/10.1103/PhysRevD.85.123535} {\bibfield  {journal} {\bibinfo
  {journal} {Phys. Rev. D}\ }\textbf {\bibinfo {volume} {85}},\ \bibinfo {eid}
  {123535} (\bibinfo {year} {2012})},\ \Eprint
  {https://arxiv.org/abs/1204.3643} {arXiv:1204.3643 [astro-ph.CO]}
  \BibitemShut {NoStop}%
\bibitem [{\citenamefont {{Pagano}}\ and\ \citenamefont
  {{Brandenberger}}(2012)}]{2012JCAP...05..014P}%
  \BibitemOpen
  \bibfield  {author} {\bibinfo {author} {\bibfnamefont {M.}~\bibnamefont
  {{Pagano}}}\ and\ \bibinfo {author} {\bibfnamefont {R.}~\bibnamefont
  {{Brandenberger}}},\ }\bibfield  {title} {\bibinfo {title} {{The 21 cm
  signature of a cosmic string loop}},\ }\href
  {https://doi.org/10.1088/1475-7516/2012/05/014} {\bibfield  {journal}
  {\bibinfo  {journal} {J. Cosmol. Astropart. Phys.}\ }\textbf {\bibinfo
  {volume} {2012}}\bibfield  {number} {\bibinfo  {number} { (5)},\ \bibinfo
  {eid} {014}},\ }\Eprint {https://arxiv.org/abs/1201.5695} {arXiv:1201.5695
  [astro-ph.CO]} \BibitemShut {NoStop}%
\bibitem [{\citenamefont {Ricotti}\ \emph {et~al.}(2008)\citenamefont
  {Ricotti}, \citenamefont {Ostriker},\ and\ \citenamefont
  {Mack}}]{Ricotti:2007au}%
  \BibitemOpen
  \bibfield  {author} {\bibinfo {author} {\bibfnamefont {M.}~\bibnamefont
  {Ricotti}}, \bibinfo {author} {\bibfnamefont {J.~P.}\ \bibnamefont
  {Ostriker}},\ and\ \bibinfo {author} {\bibfnamefont {K.~J.}\ \bibnamefont
  {Mack}},\ }\bibfield  {title} {\bibinfo {title} {{Effect of Primordial Black
  Holes on the Cosmic Microwave Background and Cosmological Parameter
  Estimates}},\ }\href {https://doi.org/10.1086/587831} {\bibfield  {journal}
  {\bibinfo  {journal} {Astrophys. J.}\ }\textbf {\bibinfo {volume} {680}},\
  \bibinfo {pages} {829} (\bibinfo {year} {2008})},\ \Eprint
  {https://arxiv.org/abs/0709.0524} {arXiv:0709.0524 [astro-ph]} \BibitemShut
  {NoStop}%
\bibitem [{\citenamefont {Mack}\ and\ \citenamefont
  {Wesley}(2008)}]{Mack:2008nv}%
  \BibitemOpen
  \bibfield  {author} {\bibinfo {author} {\bibfnamefont {K.~J.}\ \bibnamefont
  {Mack}}\ and\ \bibinfo {author} {\bibfnamefont {D.~H.}\ \bibnamefont
  {Wesley}},\ }\bibfield  {title} {\bibinfo {title} {{Primordial black holes in
  the Dark Ages: Observational prospects for future 21cm surveys}},\
  }\href@noop {} {\  (\bibinfo {year} {2008})},\ \Eprint
  {https://arxiv.org/abs/0805.1531} {arXiv:0805.1531 [astro-ph]} \BibitemShut
  {NoStop}%
\bibitem [{\citenamefont {Tashiro}\ and\ \citenamefont
  {Sugiyama}(2013)}]{Tashiro:2012qe}%
  \BibitemOpen
  \bibfield  {author} {\bibinfo {author} {\bibfnamefont {H.}~\bibnamefont
  {Tashiro}}\ and\ \bibinfo {author} {\bibfnamefont {N.}~\bibnamefont
  {Sugiyama}},\ }\bibfield  {title} {\bibinfo {title} {{The effect of
  primordial black holes on 21 cm fluctuations}},\ }\href
  {https://doi.org/10.1093/mnras/stt1493} {\bibfield  {journal} {\bibinfo
  {journal} {Mon. Not. Roy. Astron. Soc.}\ }\textbf {\bibinfo {volume} {435}},\
  \bibinfo {pages} {3001} (\bibinfo {year} {2013})},\ \Eprint
  {https://arxiv.org/abs/1207.6405} {arXiv:1207.6405 [astro-ph.CO]}
  \BibitemShut {NoStop}%
\bibitem [{\citenamefont {Belotsky}\ and\ \citenamefont
  {Kirillov}(2015)}]{Belotsky:2014twa}%
  \BibitemOpen
  \bibfield  {author} {\bibinfo {author} {\bibfnamefont {K.}~\bibnamefont
  {Belotsky}}\ and\ \bibinfo {author} {\bibfnamefont {A.}~\bibnamefont
  {Kirillov}},\ }\bibfield  {title} {\bibinfo {title} {{Primordial black holes
  with mass $10^{16}-10^{17}$ g and reionization of the Universe}},\ }\href
  {https://doi.org/10.1088/1475-7516/2015/01/041} {\bibfield  {journal}
  {\bibinfo  {journal} {JCAP}\ }\textbf {\bibinfo {volume} {01}},\ \bibinfo
  {pages} {041}},\ \Eprint {https://arxiv.org/abs/1409.8601} {arXiv:1409.8601
  [astro-ph.CO]} \BibitemShut {NoStop}%
\bibitem [{\citenamefont {Wyithe}\ \emph {et~al.}(2008)\citenamefont {Wyithe},
  \citenamefont {Loeb},\ and\ \citenamefont {Geil}}]{Wyithe:2007rq}%
  \BibitemOpen
  \bibfield  {author} {\bibinfo {author} {\bibfnamefont {S.}~\bibnamefont
  {Wyithe}}, \bibinfo {author} {\bibfnamefont {A.}~\bibnamefont {Loeb}},\ and\
  \bibinfo {author} {\bibfnamefont {P.}~\bibnamefont {Geil}},\ }\bibfield
  {title} {\bibinfo {title} {{Baryonic Acoustic Oscillations in 21cm Emission:
  A Probe of Dark Energy out to High Redshifts}},\ }\href
  {https://doi.org/10.1111/j.1365-2966.2007.12631.x} {\bibfield  {journal}
  {\bibinfo  {journal} {Mon. Not. Roy. Astron. Soc.}\ }\textbf {\bibinfo
  {volume} {383}},\ \bibinfo {pages} {1195} (\bibinfo {year} {2008})},\ \Eprint
  {https://arxiv.org/abs/0709.2955} {arXiv:0709.2955 [astro-ph]} \BibitemShut
  {NoStop}%
\bibitem [{\citenamefont {{Xia}}\ and\ \citenamefont
  {{Viel}}(2009)}]{2009JCAP...04..002X}%
  \BibitemOpen
  \bibfield  {author} {\bibinfo {author} {\bibfnamefont {J.-Q.}\ \bibnamefont
  {{Xia}}}\ and\ \bibinfo {author} {\bibfnamefont {M.}~\bibnamefont {{Viel}}},\
  }\bibfield  {title} {\bibinfo {title} {{Early dark energy at high redshifts:
  status and perspectives}},\ }\href
  {https://doi.org/10.1088/1475-7516/2009/04/002} {\bibfield  {journal}
  {\bibinfo  {journal} {J. Cosmol. Astropart. Phys.}\ }\textbf {\bibinfo
  {volume} {2009}}\bibfield  {number} {\bibinfo  {number} { (4)},\ \bibinfo
  {eid} {002}},\ }\Eprint {https://arxiv.org/abs/0901.0605} {arXiv:0901.0605
  [astro-ph.CO]} \BibitemShut {NoStop}%
\bibitem [{\citenamefont {Kohri}\ \emph {et~al.}(2017)\citenamefont {Kohri},
  \citenamefont {Oyama}, \citenamefont {Sekiguchi},\ and\ \citenamefont
  {Takahashi}}]{Kohri:2016bqx}%
  \BibitemOpen
  \bibfield  {author} {\bibinfo {author} {\bibfnamefont {K.}~\bibnamefont
  {Kohri}}, \bibinfo {author} {\bibfnamefont {Y.}~\bibnamefont {Oyama}},
  \bibinfo {author} {\bibfnamefont {T.}~\bibnamefont {Sekiguchi}},\ and\
  \bibinfo {author} {\bibfnamefont {T.}~\bibnamefont {Takahashi}},\ }\bibfield
  {title} {\bibinfo {title} {{Elucidating Dark Energy with Future 21 cm
  Observations at the Epoch of Reionization}},\ }\href
  {https://doi.org/10.1088/1475-7516/2017/02/024} {\bibfield  {journal}
  {\bibinfo  {journal} {JCAP}\ }\textbf {\bibinfo {volume} {02}},\ \bibinfo
  {pages} {024}},\ \Eprint {https://arxiv.org/abs/1608.01601} {arXiv:1608.01601
  [astro-ph.CO]} \BibitemShut {NoStop}%
\bibitem [{\citenamefont {Costa}\ \emph {et~al.}(2018)\citenamefont {Costa},
  \citenamefont {Landim}, \citenamefont {Wang},\ and\ \citenamefont
  {Abdalla}}]{Costa:2018aoy}%
  \BibitemOpen
  \bibfield  {author} {\bibinfo {author} {\bibfnamefont {A.~A.}\ \bibnamefont
  {Costa}}, \bibinfo {author} {\bibfnamefont {R.~C.}\ \bibnamefont {Landim}},
  \bibinfo {author} {\bibfnamefont {B.}~\bibnamefont {Wang}},\ and\ \bibinfo
  {author} {\bibfnamefont {E.}~\bibnamefont {Abdalla}},\ }\bibfield  {title}
  {\bibinfo {title} {{Interacting Dark Energy: Possible Explanation for 21-cm
  Absorption at Cosmic Dawn}},\ }\href
  {https://doi.org/10.1140/epjc/s10052-018-6237-7} {\bibfield  {journal}
  {\bibinfo  {journal} {Eur. Phys. J. C}\ }\textbf {\bibinfo {volume} {78}},\
  \bibinfo {pages} {746} (\bibinfo {year} {2018})},\ \Eprint
  {https://arxiv.org/abs/1803.06944} {arXiv:1803.06944 [astro-ph.CO]}
  \BibitemShut {NoStop}%
\bibitem [{\citenamefont {Yang}\ \emph {et~al.}(2019)\citenamefont {Yang},
  \citenamefont {Pan}, \citenamefont {Vagnozzi}, \citenamefont {Di~Valentino},
  \citenamefont {Mota},\ and\ \citenamefont {Capozziello}}]{Yang:2019nhz}%
  \BibitemOpen
  \bibfield  {author} {\bibinfo {author} {\bibfnamefont {W.}~\bibnamefont
  {Yang}}, \bibinfo {author} {\bibfnamefont {S.}~\bibnamefont {Pan}}, \bibinfo
  {author} {\bibfnamefont {S.}~\bibnamefont {Vagnozzi}}, \bibinfo {author}
  {\bibfnamefont {E.}~\bibnamefont {Di~Valentino}}, \bibinfo {author}
  {\bibfnamefont {D.~F.}\ \bibnamefont {Mota}},\ and\ \bibinfo {author}
  {\bibfnamefont {S.}~\bibnamefont {Capozziello}},\ }\bibfield  {title}
  {\bibinfo {title} {{Dawn of the dark: unified dark sectors and the EDGES
  Cosmic Dawn 21-cm signal}},\ }\href
  {https://doi.org/10.1088/1475-7516/2019/11/044} {\bibfield  {journal}
  {\bibinfo  {journal} {JCAP}\ }\textbf {\bibinfo {volume} {11}},\ \bibinfo
  {pages} {044}},\ \Eprint {https://arxiv.org/abs/1907.05344} {arXiv:1907.05344
  [astro-ph.CO]} \BibitemShut {NoStop}%
\bibitem [{\citenamefont {Li}\ \emph {et~al.}(2020)\citenamefont {Li},
  \citenamefont {Ren}, \citenamefont {Khurshudyan},\ and\ \citenamefont
  {Cai}}]{Li:2019loh}%
  \BibitemOpen
  \bibfield  {author} {\bibinfo {author} {\bibfnamefont {C.}~\bibnamefont
  {Li}}, \bibinfo {author} {\bibfnamefont {X.}~\bibnamefont {Ren}}, \bibinfo
  {author} {\bibfnamefont {M.}~\bibnamefont {Khurshudyan}},\ and\ \bibinfo
  {author} {\bibfnamefont {Y.-F.}\ \bibnamefont {Cai}},\ }\bibfield  {title}
  {\bibinfo {title} {{Implications of the possible 21-cm line excess at cosmic
  dawn on dynamics of interacting dark energy}},\ }\href
  {https://doi.org/10.1016/j.physletb.2019.135141} {\bibfield  {journal}
  {\bibinfo  {journal} {Phys. Lett. B}\ }\textbf {\bibinfo {volume} {801}},\
  \bibinfo {pages} {135141} (\bibinfo {year} {2020})},\ \Eprint
  {https://arxiv.org/abs/1904.02458} {arXiv:1904.02458 [astro-ph.CO]}
  \BibitemShut {NoStop}%
\bibitem [{\citenamefont {Bowman}\ \emph {et~al.}(2018)\citenamefont {Bowman},
  \citenamefont {Rogers}, \citenamefont {Monsalve}, \citenamefont {Mozdzen},\
  and\ \citenamefont {Mahesh}}]{Bowman:2018yin}%
  \BibitemOpen
  \bibfield  {author} {\bibinfo {author} {\bibfnamefont {J.~D.}\ \bibnamefont
  {Bowman}}, \bibinfo {author} {\bibfnamefont {A.~E.~E.}\ \bibnamefont
  {Rogers}}, \bibinfo {author} {\bibfnamefont {R.~A.}\ \bibnamefont
  {Monsalve}}, \bibinfo {author} {\bibfnamefont {T.~J.}\ \bibnamefont
  {Mozdzen}},\ and\ \bibinfo {author} {\bibfnamefont {N.}~\bibnamefont
  {Mahesh}},\ }\bibfield  {title} {\bibinfo {title} {{An absorption profile
  centred at 78 megahertz in the sky-averaged spectrum}},\ }\href
  {https://doi.org/10.1038/nature25792} {\bibfield  {journal} {\bibinfo
  {journal} {Nature}\ }\textbf {\bibinfo {volume} {555}},\ \bibinfo {pages}
  {67} (\bibinfo {year} {2018})},\ \Eprint {https://arxiv.org/abs/1810.05912}
  {arXiv:1810.05912 [astro-ph.CO]} \BibitemShut {NoStop}%
\bibitem [{\citenamefont {Barkana}\ \emph {et~al.}(2018)\citenamefont
  {Barkana}, \citenamefont {Outmezguine}, \citenamefont {Redigolo},\ and\
  \citenamefont {Volansky}}]{Barkana:2018qrx}%
  \BibitemOpen
  \bibfield  {author} {\bibinfo {author} {\bibfnamefont {R.}~\bibnamefont
  {Barkana}}, \bibinfo {author} {\bibfnamefont {N.~J.}\ \bibnamefont
  {Outmezguine}}, \bibinfo {author} {\bibfnamefont {D.}~\bibnamefont
  {Redigolo}},\ and\ \bibinfo {author} {\bibfnamefont {T.}~\bibnamefont
  {Volansky}},\ }\bibfield  {title} {\bibinfo {title} {{Strong constraints on
  light dark matter interpretation of the EDGES signal}},\ }\href
  {https://doi.org/10.1103/PhysRevD.98.103005} {\bibfield  {journal} {\bibinfo
  {journal} {Phys. Rev. D}\ }\textbf {\bibinfo {volume} {98}},\ \bibinfo
  {pages} {103005} (\bibinfo {year} {2018})},\ \Eprint
  {https://arxiv.org/abs/1803.03091} {arXiv:1803.03091 [hep-ph]} \BibitemShut
  {NoStop}%
\bibitem [{\citenamefont {Barkana}(2018)}]{Barkana:2018lgd}%
  \BibitemOpen
  \bibfield  {author} {\bibinfo {author} {\bibfnamefont {R.}~\bibnamefont
  {Barkana}},\ }\bibfield  {title} {\bibinfo {title} {{Possible interaction
  between baryons and dark-matter particles revealed by the first stars}},\
  }\href {https://doi.org/10.1038/nature25791} {\bibfield  {journal} {\bibinfo
  {journal} {Nature}\ }\textbf {\bibinfo {volume} {555}},\ \bibinfo {pages}
  {71} (\bibinfo {year} {2018})},\ \Eprint {https://arxiv.org/abs/1803.06698}
  {arXiv:1803.06698 [astro-ph.CO]} \BibitemShut {NoStop}%
\bibitem [{\citenamefont {Mu\~noz}\ and\ \citenamefont
  {Loeb}(2018)}]{Munoz:2018pzp}%
  \BibitemOpen
  \bibfield  {author} {\bibinfo {author} {\bibfnamefont {J.~B.}\ \bibnamefont
  {Mu\~noz}}\ and\ \bibinfo {author} {\bibfnamefont {A.}~\bibnamefont {Loeb}},\
  }\bibfield  {title} {\bibinfo {title} {{A small amount of mini-charged dark
  matter could cool the baryons in the early Universe}},\ }\href
  {https://doi.org/10.1038/s41586-018-0151-x} {\bibfield  {journal} {\bibinfo
  {journal} {Nature}\ }\textbf {\bibinfo {volume} {557}},\ \bibinfo {pages}
  {684} (\bibinfo {year} {2018})},\ \Eprint {https://arxiv.org/abs/1802.10094}
  {arXiv:1802.10094 [astro-ph.CO]} \BibitemShut {NoStop}%
\bibitem [{\citenamefont {Jia}(2019)}]{Jia:2018csj}%
  \BibitemOpen
  \bibfield  {author} {\bibinfo {author} {\bibfnamefont {L.-B.}\ \bibnamefont
  {Jia}},\ }\bibfield  {title} {\bibinfo {title} {{Dark photon portal dark
  matter with the 21-cm anomaly}},\ }\href
  {https://doi.org/10.1140/epjc/s10052-019-6542-9} {\bibfield  {journal}
  {\bibinfo  {journal} {Eur. Phys. J. C}\ }\textbf {\bibinfo {volume} {79}},\
  \bibinfo {pages} {80} (\bibinfo {year} {2019})},\ \Eprint
  {https://arxiv.org/abs/1804.07934} {arXiv:1804.07934 [hep-ph]} \BibitemShut
  {NoStop}%
\bibitem [{\citenamefont {Liu}\ \emph {et~al.}(2019)\citenamefont {Liu},
  \citenamefont {Outmezguine}, \citenamefont {Redigolo},\ and\ \citenamefont
  {Volansky}}]{Liu:2019knx}%
  \BibitemOpen
  \bibfield  {author} {\bibinfo {author} {\bibfnamefont {H.}~\bibnamefont
  {Liu}}, \bibinfo {author} {\bibfnamefont {N.~J.}\ \bibnamefont
  {Outmezguine}}, \bibinfo {author} {\bibfnamefont {D.}~\bibnamefont
  {Redigolo}},\ and\ \bibinfo {author} {\bibfnamefont {T.}~\bibnamefont
  {Volansky}},\ }\bibfield  {title} {\bibinfo {title} {{Reviving Millicharged
  Dark Matter for 21-cm Cosmology}},\ }\href
  {https://doi.org/10.1103/PhysRevD.100.123011} {\bibfield  {journal} {\bibinfo
   {journal} {Phys. Rev. D}\ }\textbf {\bibinfo {volume} {100}},\ \bibinfo
  {pages} {123011} (\bibinfo {year} {2019})},\ \Eprint
  {https://arxiv.org/abs/1908.06986} {arXiv:1908.06986 [hep-ph]} \BibitemShut
  {NoStop}%
\bibitem [{\citenamefont {Berlin}\ \emph {et~al.}(2018)\citenamefont {Berlin},
  \citenamefont {Hooper}, \citenamefont {Krnjaic},\ and\ \citenamefont
  {McDermott}}]{Berlin:2018sjs}%
  \BibitemOpen
  \bibfield  {author} {\bibinfo {author} {\bibfnamefont {A.}~\bibnamefont
  {Berlin}}, \bibinfo {author} {\bibfnamefont {D.}~\bibnamefont {Hooper}},
  \bibinfo {author} {\bibfnamefont {G.}~\bibnamefont {Krnjaic}},\ and\ \bibinfo
  {author} {\bibfnamefont {S.~D.}\ \bibnamefont {McDermott}},\ }\bibfield
  {title} {\bibinfo {title} {{Severely Constraining Dark Matter Interpretations
  of the 21-cm Anomaly}},\ }\href
  {https://doi.org/10.1103/PhysRevLett.121.011102} {\bibfield  {journal}
  {\bibinfo  {journal} {Phys. Rev. Lett.}\ }\textbf {\bibinfo {volume} {121}},\
  \bibinfo {pages} {011102} (\bibinfo {year} {2018})},\ \Eprint
  {https://arxiv.org/abs/1803.02804} {arXiv:1803.02804 [hep-ph]} \BibitemShut
  {NoStop}%
\bibitem [{\citenamefont {Slatyer}\ and\ \citenamefont
  {Wu}(2018)}]{Slatyer:2018aqg}%
  \BibitemOpen
  \bibfield  {author} {\bibinfo {author} {\bibfnamefont {T.~R.}\ \bibnamefont
  {Slatyer}}\ and\ \bibinfo {author} {\bibfnamefont {C.-L.}\ \bibnamefont
  {Wu}},\ }\bibfield  {title} {\bibinfo {title} {{Early-Universe constraints on
  dark matter-baryon scattering and their implications for a global 21 cm
  signal}},\ }\href {https://doi.org/10.1103/PhysRevD.98.023013} {\bibfield
  {journal} {\bibinfo  {journal} {Phys. Rev. D}\ }\textbf {\bibinfo {volume}
  {98}},\ \bibinfo {pages} {023013} (\bibinfo {year} {2018})},\ \Eprint
  {https://arxiv.org/abs/1803.09734} {arXiv:1803.09734 [astro-ph.CO]}
  \BibitemShut {NoStop}%
\bibitem [{\citenamefont {Fraser}\ \emph {et~al.}(2018)\citenamefont {Fraser}
  \emph {et~al.}}]{Fraser:2018acy}%
  \BibitemOpen
  \bibfield  {author} {\bibinfo {author} {\bibfnamefont {S.}~\bibnamefont
  {Fraser}} \emph {et~al.},\ }\bibfield  {title} {\bibinfo {title} {{The EDGES
  21 cm Anomaly and Properties of Dark Matter}},\ }\href
  {https://doi.org/10.1016/j.physletb.2018.08.035} {\bibfield  {journal}
  {\bibinfo  {journal} {Phys. Lett. B}\ }\textbf {\bibinfo {volume} {785}},\
  \bibinfo {pages} {159} (\bibinfo {year} {2018})},\ \Eprint
  {https://arxiv.org/abs/1803.03245} {arXiv:1803.03245 [hep-ph]} \BibitemShut
  {NoStop}%
\bibitem [{\citenamefont {Kovetz}\ \emph {et~al.}(2018)\citenamefont {Kovetz},
  \citenamefont {Poulin}, \citenamefont {Gluscevic}, \citenamefont {Boddy},
  \citenamefont {Barkana},\ and\ \citenamefont
  {Kamionkowski}}]{Kovetz:2018zan}%
  \BibitemOpen
  \bibfield  {author} {\bibinfo {author} {\bibfnamefont {E.~D.}\ \bibnamefont
  {Kovetz}}, \bibinfo {author} {\bibfnamefont {V.}~\bibnamefont {Poulin}},
  \bibinfo {author} {\bibfnamefont {V.}~\bibnamefont {Gluscevic}}, \bibinfo
  {author} {\bibfnamefont {K.~K.}\ \bibnamefont {Boddy}}, \bibinfo {author}
  {\bibfnamefont {R.}~\bibnamefont {Barkana}},\ and\ \bibinfo {author}
  {\bibfnamefont {M.}~\bibnamefont {Kamionkowski}},\ }\bibfield  {title}
  {\bibinfo {title} {{Tighter limits on dark matter explanations of the
  anomalous EDGES 21 cm signal}},\ }\href
  {https://doi.org/10.1103/PhysRevD.98.103529} {\bibfield  {journal} {\bibinfo
  {journal} {Phys. Rev. D}\ }\textbf {\bibinfo {volume} {98}},\ \bibinfo
  {pages} {103529} (\bibinfo {year} {2018})},\ \Eprint
  {https://arxiv.org/abs/1807.11482} {arXiv:1807.11482 [astro-ph.CO]}
  \BibitemShut {NoStop}%
\bibitem [{\citenamefont {Mirocha}\ and\ \citenamefont
  {Furlanetto}(2018)}]{10.1093/mnras/sty3260}%
  \BibitemOpen
  \bibfield  {author} {\bibinfo {author} {\bibfnamefont {J.}~\bibnamefont
  {Mirocha}}\ and\ \bibinfo {author} {\bibfnamefont {S.~R.}\ \bibnamefont
  {Furlanetto}},\ }\bibfield  {title} {\bibinfo {title} {{What does the first
  highly redshifted 21-cm detection tell us about early galaxies?}},\ }\href
  {https://doi.org/10.1093/mnras/sty3260} {\bibfield  {journal} {\bibinfo
  {journal} {Monthly Notices of the Royal Astronomical Society}\ }\textbf
  {\bibinfo {volume} {483}},\ \bibinfo {pages} {1980} (\bibinfo {year}
  {2018})},\ \Eprint
  {https://arxiv.org/abs/https://academic.oup.com/mnras/article-pdf/483/2/1980/27140778/sty3260.pdf}
  {https://academic.oup.com/mnras/article-pdf/483/2/1980/27140778/sty3260.pdf}
  \BibitemShut {NoStop}%
\bibitem [{\citenamefont {Jana}\ \emph {et~al.}(2019)\citenamefont {Jana},
  \citenamefont {Nath},\ and\ \citenamefont {Biermann}}]{Jana:2018gqk}%
  \BibitemOpen
  \bibfield  {author} {\bibinfo {author} {\bibfnamefont {R.}~\bibnamefont
  {Jana}}, \bibinfo {author} {\bibfnamefont {B.~B.}\ \bibnamefont {Nath}},\
  and\ \bibinfo {author} {\bibfnamefont {P.~L.}\ \bibnamefont {Biermann}},\
  }\bibfield  {title} {\bibinfo {title} {{Radio background and IGM heating due
  to Pop III supernova explosions}},\ }\href
  {https://doi.org/10.1093/mnras/sty3426} {\bibfield  {journal} {\bibinfo
  {journal} {Mon. Not. Roy. Astron. Soc.}\ }\textbf {\bibinfo {volume} {483}},\
  \bibinfo {pages} {5329} (\bibinfo {year} {2019})},\ \Eprint
  {https://arxiv.org/abs/1812.07404} {arXiv:1812.07404 [astro-ph.HE]}
  \BibitemShut {NoStop}%
\bibitem [{\citenamefont {Sharma}(2018)}]{Sharma:2018agu}%
  \BibitemOpen
  \bibfield  {author} {\bibinfo {author} {\bibfnamefont {P.}~\bibnamefont
  {Sharma}},\ }\bibfield  {title} {\bibinfo {title} {{Astrophysical radio
  background cannot explain the EDGES 21-cm signal: constraints from cooling of
  non-thermal electrons}},\ }\href {https://doi.org/10.1093/mnrasl/sly147}
  {\bibfield  {journal} {\bibinfo  {journal} {Mon. Not. Roy. Astron. Soc.}\
  }\textbf {\bibinfo {volume} {481}},\ \bibinfo {pages} {L6} (\bibinfo {year}
  {2018})},\ \Eprint {https://arxiv.org/abs/1804.05843} {arXiv:1804.05843
  [astro-ph.HE]} \BibitemShut {NoStop}%
\bibitem [{\citenamefont {Ewall-Wice}\ \emph {et~al.}(2018)\citenamefont
  {Ewall-Wice}, \citenamefont {Chang}, \citenamefont {Lazio}, \citenamefont
  {Dore}, \citenamefont {Seiffert},\ and\ \citenamefont
  {Monsalve}}]{Ewall-Wice:2018bzf}%
  \BibitemOpen
  \bibfield  {author} {\bibinfo {author} {\bibfnamefont {A.}~\bibnamefont
  {Ewall-Wice}}, \bibinfo {author} {\bibfnamefont {T.-C.}\ \bibnamefont
  {Chang}}, \bibinfo {author} {\bibfnamefont {J.}~\bibnamefont {Lazio}},
  \bibinfo {author} {\bibfnamefont {O.}~\bibnamefont {Dore}}, \bibinfo {author}
  {\bibfnamefont {M.}~\bibnamefont {Seiffert}},\ and\ \bibinfo {author}
  {\bibfnamefont {R.}~\bibnamefont {Monsalve}},\ }\bibfield  {title} {\bibinfo
  {title} {{Modeling the Radio Background from the First Black Holes at Cosmic
  Dawn: Implications for the 21 cm Absorption Amplitude}},\ }\href
  {https://doi.org/10.3847/1538-4357/aae51d} {\bibfield  {journal} {\bibinfo
  {journal} {Astrophys. J.}\ }\textbf {\bibinfo {volume} {868}},\ \bibinfo
  {pages} {63} (\bibinfo {year} {2018})},\ \Eprint
  {https://arxiv.org/abs/1803.01815} {arXiv:1803.01815 [astro-ph.CO]}
  \BibitemShut {NoStop}%
\bibitem [{\citenamefont {Ewall-Wice}\ \emph {et~al.}(2020)\citenamefont
  {Ewall-Wice}, \citenamefont {Chang},\ and\ \citenamefont
  {Lazio}}]{Ewall-Wice:2019may}%
  \BibitemOpen
  \bibfield  {author} {\bibinfo {author} {\bibfnamefont {A.}~\bibnamefont
  {Ewall-Wice}}, \bibinfo {author} {\bibfnamefont {T.-C.}\ \bibnamefont
  {Chang}},\ and\ \bibinfo {author} {\bibfnamefont {T.~J.~W.}\ \bibnamefont
  {Lazio}},\ }\bibfield  {title} {\bibinfo {title} {{The Radio Scream from
  Black Holes at Cosmic Dawn: A Semi-Analytic Model for the Impact of Radio
  Loud Black-Holes on the 21 cm Global Signal}},\ }\href
  {https://doi.org/10.1093/mnras/stz3501} {\bibfield  {journal} {\bibinfo
  {journal} {Mon. Not. Roy. Astron. Soc.}\ }\textbf {\bibinfo {volume} {492}},\
  \bibinfo {pages} {6086} (\bibinfo {year} {2020})},\ \Eprint
  {https://arxiv.org/abs/1903.06788} {arXiv:1903.06788 [astro-ph.GA]}
  \BibitemShut {NoStop}%
\bibitem [{\citenamefont {{Hills}}\ \emph {et~al.}(2018)\citenamefont
  {{Hills}}, \citenamefont {{Kulkarni}}, \citenamefont {{Meerburg}},\ and\
  \citenamefont {{Puchwein}}}]{2018Natur.564E..32H}%
  \BibitemOpen
  \bibfield  {author} {\bibinfo {author} {\bibfnamefont {R.}~\bibnamefont
  {{Hills}}}, \bibinfo {author} {\bibfnamefont {G.}~\bibnamefont {{Kulkarni}}},
  \bibinfo {author} {\bibfnamefont {P.~D.}\ \bibnamefont {{Meerburg}}},\ and\
  \bibinfo {author} {\bibfnamefont {E.}~\bibnamefont {{Puchwein}}},\ }\bibfield
   {title} {\bibinfo {title} {{Concerns about modelling of the EDGES data}},\
  }\href {https://doi.org/10.1038/s41586-018-0796-5} {\bibfield  {journal}
  {\bibinfo  {journal} {Nature}\ }\textbf {\bibinfo {volume} {564}},\ \bibinfo
  {pages} {E32} (\bibinfo {year} {2018})},\ \Eprint
  {https://arxiv.org/abs/1805.01421} {arXiv:1805.01421 [astro-ph.CO]}
  \BibitemShut {NoStop}%
\bibitem [{\citenamefont {{Bowman}}\ \emph {et~al.}(2018)\citenamefont
  {{Bowman}}, \citenamefont {{Rogers}}, \citenamefont {{Monsalve}},
  \citenamefont {{Mozdzen}},\ and\ \citenamefont
  {{Mahesh}}}]{2018Natur.564E..35B}%
  \BibitemOpen
  \bibfield  {author} {\bibinfo {author} {\bibfnamefont {J.~D.}\ \bibnamefont
  {{Bowman}}}, \bibinfo {author} {\bibfnamefont {A.~E.~E.}\ \bibnamefont
  {{Rogers}}}, \bibinfo {author} {\bibfnamefont {R.~A.}\ \bibnamefont
  {{Monsalve}}}, \bibinfo {author} {\bibfnamefont {T.~J.}\ \bibnamefont
  {{Mozdzen}}},\ and\ \bibinfo {author} {\bibfnamefont {N.}~\bibnamefont
  {{Mahesh}}},\ }\bibfield  {title} {\bibinfo {title} {{Reply to Hills et
  al.}},\ }\href {https://doi.org/10.1038/s41586-018-0797-4} {\bibfield
  {journal} {\bibinfo  {journal} {Nature}\ }\textbf {\bibinfo {volume} {564}},\
  \bibinfo {pages} {E35} (\bibinfo {year} {2018})}\BibitemShut {NoStop}%
\bibitem [{\citenamefont {{Bradley}}\ \emph {et~al.}(2019)\citenamefont
  {{Bradley}}, \citenamefont {{Tauscher}}, \citenamefont {{Rapetti}},\ and\
  \citenamefont {{Burns}}}]{2019ApJ...874..153B}%
  \BibitemOpen
  \bibfield  {author} {\bibinfo {author} {\bibfnamefont {R.~F.}\ \bibnamefont
  {{Bradley}}}, \bibinfo {author} {\bibfnamefont {K.}~\bibnamefont
  {{Tauscher}}}, \bibinfo {author} {\bibfnamefont {D.}~\bibnamefont
  {{Rapetti}}},\ and\ \bibinfo {author} {\bibfnamefont {J.~O.}\ \bibnamefont
  {{Burns}}},\ }\bibfield  {title} {\bibinfo {title} {{A Ground Plane Artifact
  that Induces an Absorption Profile in Averaged Spectra from Global 21 cm
  Measurements, with Possible Application to EDGES}},\ }\href
  {https://doi.org/10.3847/1538-4357/ab0d8b} {\bibfield  {journal} {\bibinfo
  {journal} {Astrophys. J.}\ }\textbf {\bibinfo {volume} {874}},\ \bibinfo
  {eid} {153} (\bibinfo {year} {2019})},\ \Eprint
  {https://arxiv.org/abs/1810.09015} {arXiv:1810.09015 [astro-ph.IM]}
  \BibitemShut {NoStop}%
\bibitem [{\citenamefont {{Singh}}\ and\ \citenamefont
  {{Subrahmanyan}}(2019)}]{2019ApJ...880...26S}%
  \BibitemOpen
  \bibfield  {author} {\bibinfo {author} {\bibfnamefont {S.}~\bibnamefont
  {{Singh}}}\ and\ \bibinfo {author} {\bibfnamefont {R.}~\bibnamefont
  {{Subrahmanyan}}},\ }\bibfield  {title} {\bibinfo {title} {{The Redshifted 21
  cm Signal in the EDGES Low-band Spectrum}},\ }\href
  {https://doi.org/10.3847/1538-4357/ab2879} {\bibfield  {journal} {\bibinfo
  {journal} {Astrophys. J.}\ }\textbf {\bibinfo {volume} {880}},\ \bibinfo
  {eid} {26} (\bibinfo {year} {2019})},\ \Eprint
  {https://arxiv.org/abs/1903.04540} {arXiv:1903.04540 [astro-ph.CO]}
  \BibitemShut {NoStop}%
\bibitem [{\citenamefont {{Tauscher}}\ \emph {et~al.}(2020)\citenamefont
  {{Tauscher}}, \citenamefont {{Rapetti}},\ and\ \citenamefont
  {{Burns}}}]{2020ApJ...897..132T}%
  \BibitemOpen
  \bibfield  {author} {\bibinfo {author} {\bibfnamefont {K.}~\bibnamefont
  {{Tauscher}}}, \bibinfo {author} {\bibfnamefont {D.}~\bibnamefont
  {{Rapetti}}},\ and\ \bibinfo {author} {\bibfnamefont {J.~O.}\ \bibnamefont
  {{Burns}}},\ }\bibfield  {title} {\bibinfo {title} {{Formulating and
  Critically Examining the Assumptions of Global 21 cm Signal Analyses: How to
  Avoid the False Troughs That Can Appear in Single-spectrum Fits}},\ }\href
  {https://doi.org/10.3847/1538-4357/ab9a3f} {\bibfield  {journal} {\bibinfo
  {journal} {Astrophys. J.}\ }\textbf {\bibinfo {volume} {897}},\ \bibinfo
  {eid} {132} (\bibinfo {year} {2020})},\ \Eprint
  {https://arxiv.org/abs/2005.00034} {arXiv:2005.00034 [astro-ph.CO]}
  \BibitemShut {NoStop}%
\bibitem [{\citenamefont {{Spinelli}}\ \emph {et~al.}(2019)\citenamefont
  {{Spinelli}}, \citenamefont {{Bernardi}},\ and\ \citenamefont
  {{Santos}}}]{2019MNRAS.489.4007S}%
  \BibitemOpen
  \bibfield  {author} {\bibinfo {author} {\bibfnamefont {M.}~\bibnamefont
  {{Spinelli}}}, \bibinfo {author} {\bibfnamefont {G.}~\bibnamefont
  {{Bernardi}}},\ and\ \bibinfo {author} {\bibfnamefont {M.~G.}\ \bibnamefont
  {{Santos}}},\ }\bibfield  {title} {\bibinfo {title} {{On the contamination of
  the global 21-cm signal from polarized foregrounds}},\ }\href
  {https://doi.org/10.1093/mnras/stz2425} {\bibfield  {journal} {\bibinfo
  {journal} {Mon. Not. Roy. Astron. Soc.}\ }\textbf {\bibinfo {volume} {489}},\
  \bibinfo {pages} {4007} (\bibinfo {year} {2019})},\ \Eprint
  {https://arxiv.org/abs/1908.05303} {arXiv:1908.05303 [astro-ph.CO]}
  \BibitemShut {NoStop}%
\bibitem [{\citenamefont {{Sims}}\ and\ \citenamefont
  {{Pober}}(2020)}]{2020MNRAS.492...22S}%
  \BibitemOpen
  \bibfield  {author} {\bibinfo {author} {\bibfnamefont {P.~H.}\ \bibnamefont
  {{Sims}}}\ and\ \bibinfo {author} {\bibfnamefont {J.~C.}\ \bibnamefont
  {{Pober}}},\ }\bibfield  {title} {\bibinfo {title} {{Testing for calibration
  systematics in the EDGES low-band data using Bayesian model selection}},\
  }\href {https://doi.org/10.1093/mnras/stz3388} {\bibfield  {journal}
  {\bibinfo  {journal} {Mon. Not. Roy. Astron. Soc.}\ }\textbf {\bibinfo
  {volume} {492}},\ \bibinfo {pages} {22} (\bibinfo {year} {2020})},\ \Eprint
  {https://arxiv.org/abs/1910.03165} {arXiv:1910.03165 [astro-ph.CO]}
  \BibitemShut {NoStop}%
\bibitem [{\citenamefont {{Burns}}\ \emph {et~al.}(2012)\citenamefont
  {{Burns}}, \citenamefont {{Lazio}}, \citenamefont {{Bale}}, \citenamefont
  {{Bowman}}, \citenamefont {{Bradley}}, \citenamefont {{Carilli}},
  \citenamefont {{Furlanetto}}, \citenamefont {{Harker}}, \citenamefont
  {{Loeb}},\ and\ \citenamefont {{Pritchard}}}]{2012AdSpR..49..433B}%
  \BibitemOpen
  \bibfield  {author} {\bibinfo {author} {\bibfnamefont {J.~O.}\ \bibnamefont
  {{Burns}}}, \bibinfo {author} {\bibfnamefont {J.}~\bibnamefont {{Lazio}}},
  \bibinfo {author} {\bibfnamefont {S.}~\bibnamefont {{Bale}}}, \bibinfo
  {author} {\bibfnamefont {J.}~\bibnamefont {{Bowman}}}, \bibinfo {author}
  {\bibfnamefont {R.}~\bibnamefont {{Bradley}}}, \bibinfo {author}
  {\bibfnamefont {C.}~\bibnamefont {{Carilli}}}, \bibinfo {author}
  {\bibfnamefont {S.}~\bibnamefont {{Furlanetto}}}, \bibinfo {author}
  {\bibfnamefont {G.}~\bibnamefont {{Harker}}}, \bibinfo {author}
  {\bibfnamefont {A.}~\bibnamefont {{Loeb}}},\ and\ \bibinfo {author}
  {\bibfnamefont {J.}~\bibnamefont {{Pritchard}}},\ }\bibfield  {title}
  {\bibinfo {title} {{Probing the first stars and black holes in the early
  Universe with the Dark Ages Radio Explorer (DARE)}},\ }\href
  {https://doi.org/10.1016/j.asr.2011.10.014} {\bibfield  {journal} {\bibinfo
  {journal} {Advances in Space Research}\ }\textbf {\bibinfo {volume} {49}},\
  \bibinfo {pages} {433} (\bibinfo {year} {2012})},\ \Eprint
  {https://arxiv.org/abs/1106.5194} {arXiv:1106.5194 [astro-ph.CO]}
  \BibitemShut {NoStop}%
\bibitem [{\citenamefont {Price}\ \emph {et~al.}(2018)\citenamefont {Price}
  \emph {et~al.}}]{2018MNRAS.478.4193P}%
  \BibitemOpen
  \bibfield  {author} {\bibinfo {author} {\bibfnamefont {D.~C.}\ \bibnamefont
  {Price}} \emph {et~al.},\ }\bibfield  {title} {\bibinfo {title} {{Design and
  characterization of the Large-aperture Experiment to Detect the Dark Age
  (LEDA) radiometer systems}},\ }\href {https://doi.org/10.1093/mnras/sty1244}
  {\bibfield  {journal} {\bibinfo  {journal} {Mon. Not. Roy. Astron. Soc.}\
  }\textbf {\bibinfo {volume} {478}},\ \bibinfo {pages} {4193} (\bibinfo {year}
  {2018})},\ \Eprint {https://arxiv.org/abs/1709.09313} {arXiv:1709.09313
  [astro-ph.IM]} \BibitemShut {NoStop}%
\bibitem [{\citenamefont {Philip}\ \emph {et~al.}(2019)\citenamefont {Philip}
  \emph {et~al.}}]{2019JAI.....850004P}%
  \BibitemOpen
  \bibfield  {author} {\bibinfo {author} {\bibfnamefont {L.}~\bibnamefont
  {Philip}} \emph {et~al.},\ }\bibfield  {title} {\bibinfo {title} {{Probing
  Radio Intensity at High-Z from Marion: 2017 Instrument}},\ }\href
  {https://doi.org/10.1142/S2251171719500041} {\bibfield  {journal} {\bibinfo
  {journal} {Journal of Astronomical Instrumentation}\ }\textbf {\bibinfo
  {volume} {8}},\ \bibinfo {eid} {1950004} (\bibinfo {year} {2019})},\ \Eprint
  {https://arxiv.org/abs/1806.09531} {arXiv:1806.09531 [astro-ph.IM]}
  \BibitemShut {NoStop}%
\bibitem [{\citenamefont {{Singh}}\ \emph {et~al.}(2018)\citenamefont
  {{Singh}}, \citenamefont {{Subrahmanyan}}, \citenamefont {{Shankar}},
  \citenamefont {{Rao}}, \citenamefont {{Girish}}, \citenamefont
  {{Raghunathan}}, \citenamefont {{Somashekar}},\ and\ \citenamefont
  {{Srivani}}}]{2018ExA....45..269S}%
  \BibitemOpen
  \bibfield  {author} {\bibinfo {author} {\bibfnamefont {S.}~\bibnamefont
  {{Singh}}}, \bibinfo {author} {\bibfnamefont {R.}~\bibnamefont
  {{Subrahmanyan}}}, \bibinfo {author} {\bibfnamefont {N.~U.}\ \bibnamefont
  {{Shankar}}}, \bibinfo {author} {\bibfnamefont {M.~S.}\ \bibnamefont
  {{Rao}}}, \bibinfo {author} {\bibfnamefont {B.~S.}\ \bibnamefont {{Girish}}},
  \bibinfo {author} {\bibfnamefont {A.}~\bibnamefont {{Raghunathan}}}, \bibinfo
  {author} {\bibfnamefont {R.}~\bibnamefont {{Somashekar}}},\ and\ \bibinfo
  {author} {\bibfnamefont {K.~S.}\ \bibnamefont {{Srivani}}},\ }\bibfield
  {title} {\bibinfo {title} {{SARAS 2: a spectral radiometer for probing cosmic
  dawn and the epoch of reionization through detection of the global 21-cm
  signal}},\ }\href {https://doi.org/10.1007/s10686-018-9584-3} {\bibfield
  {journal} {\bibinfo  {journal} {Experimental Astronomy}\ }\textbf {\bibinfo
  {volume} {45}},\ \bibinfo {pages} {269} (\bibinfo {year} {2018})},\ \Eprint
  {https://arxiv.org/abs/1710.01101} {arXiv:1710.01101 [astro-ph.IM]}
  \BibitemShut {NoStop}%
\bibitem [{\citenamefont {{de Lera Acedo}}(2019)}]{deLera:2019}%
  \BibitemOpen
  \bibfield  {author} {\bibinfo {author} {\bibfnamefont {E.}~\bibnamefont {{de
  Lera Acedo}}},\ }\bibfield  {title} {\bibinfo {title} {Reach: Radio
  experiment for the analysis of cosmic hydrogen},\ }in\ \href
  {https://doi.org/10.1109/ICEAA.2019.8879199} {\emph {\bibinfo {booktitle}
  {2019 International Conference on Electromagnetics in Advanced Applications
  (ICEAA)}}}\ (\bibinfo {year} {2019})\ pp.\ \bibinfo {pages}
  {0626--0629}\BibitemShut {NoStop}%
\bibitem [{\citenamefont {Parsons}\ \emph {et~al.}(2019)\citenamefont {Parsons}
  \emph {et~al.}}]{2019BAAS...51g.241P}%
  \BibitemOpen
  \bibfield  {author} {\bibinfo {author} {\bibfnamefont {A.}~\bibnamefont
  {Parsons}} \emph {et~al.},\ }\bibfield  {title} {\bibinfo {title} {{A Roadmap
  for Astrophysics and Cosmology with High-Redshift 21 cm Intensity Mapping}},\
  }in\ \href@noop {} {\emph {\bibinfo {booktitle} {Bulletin of the American
  Astronomical Society}}},\ Vol.~\bibinfo {volume} {51}\ (\bibinfo {year}
  {2019})\ p.\ \bibinfo {pages} {241},\ \Eprint
  {https://arxiv.org/abs/1907.06440} {arXiv:1907.06440 [astro-ph.IM]}
  \BibitemShut {NoStop}%
\bibitem [{\citenamefont {Eastwood}\ \emph {et~al.}(2019)\citenamefont
  {Eastwood} \emph {et~al.}}]{2019AJ....158...84E}%
  \BibitemOpen
  \bibfield  {author} {\bibinfo {author} {\bibfnamefont {M.~W.}\ \bibnamefont
  {Eastwood}} \emph {et~al.},\ }\bibfield  {title} {\bibinfo {title} {{The 21
  cm Power Spectrum from the Cosmic Dawn: First Results from the OVRO-LWA}},\
  }\href {https://doi.org/10.3847/1538-3881/ab2629} {\bibfield  {journal}
  {\bibinfo  {journal} {Astron. J.}\ }\textbf {\bibinfo {volume} {158}},\
  \bibinfo {eid} {84} (\bibinfo {year} {2019})},\ \Eprint
  {https://arxiv.org/abs/1906.08943} {arXiv:1906.08943 [astro-ph.CO]}
  \BibitemShut {NoStop}%
\bibitem [{\citenamefont {Mellema}\ \emph {et~al.}(2013)\citenamefont {Mellema}
  \emph {et~al.}}]{Mellema:2012ht}%
  \BibitemOpen
  \bibfield  {author} {\bibinfo {author} {\bibfnamefont {G.}~\bibnamefont
  {Mellema}} \emph {et~al.},\ }\bibfield  {title} {\bibinfo {title}
  {{Reionization and the Cosmic Dawn with the Square Kilometre Array}},\ }\href
  {https://doi.org/10.1007/s10686-013-9334-5} {\bibfield  {journal} {\bibinfo
  {journal} {Exper. Astron.}\ }\textbf {\bibinfo {volume} {36}},\ \bibinfo
  {pages} {235} (\bibinfo {year} {2013})},\ \Eprint
  {https://arxiv.org/abs/1210.0197} {arXiv:1210.0197 [astro-ph.CO]}
  \BibitemShut {NoStop}%
\bibitem [{\citenamefont {Visbal}\ \emph {et~al.}(2015)\citenamefont {Visbal},
  \citenamefont {Haiman},\ and\ \citenamefont {Bryan}}]{Visbal:2015sca}%
  \BibitemOpen
  \bibfield  {author} {\bibinfo {author} {\bibfnamefont {E.}~\bibnamefont
  {Visbal}}, \bibinfo {author} {\bibfnamefont {Z.}~\bibnamefont {Haiman}},\
  and\ \bibinfo {author} {\bibfnamefont {G.~L.}\ \bibnamefont {Bryan}},\
  }\bibfield  {title} {\bibinfo {title} {{Looking for Population III stars with
  He II line intensity mapping}},\ }\href
  {https://doi.org/10.1093/mnras/stv785} {\bibfield  {journal} {\bibinfo
  {journal} {Mon. Not. Roy. Astron. Soc.}\ }\textbf {\bibinfo {volume} {450}},\
  \bibinfo {pages} {2506} (\bibinfo {year} {2015})},\ \Eprint
  {https://arxiv.org/abs/1501.03177} {arXiv:1501.03177 [astro-ph.CO]}
  \BibitemShut {NoStop}%
\bibitem [{\citenamefont {{Bagla}}\ and\ \citenamefont
  {{Loeb}}(2009)}]{2009arXiv0905.1698B}%
  \BibitemOpen
  \bibfield  {author} {\bibinfo {author} {\bibfnamefont {J.~S.}\ \bibnamefont
  {{Bagla}}}\ and\ \bibinfo {author} {\bibfnamefont {A.}~\bibnamefont
  {{Loeb}}},\ }\bibfield  {title} {\bibinfo {title} {{The hyperfine transition
  of 3He+ as a probe of the intergalactic medium}},\ }\href@noop {} {\bibfield
  {journal} {\bibinfo  {journal} {arXiv e-prints}\ ,\ \bibinfo {eid}
  {arXiv:0905.1698}} (\bibinfo {year} {2009})},\ \Eprint
  {https://arxiv.org/abs/0905.1698} {arXiv:0905.1698 [astro-ph.CO]}
  \BibitemShut {NoStop}%
\bibitem [{\citenamefont {{McQuinn}}\ and\ \citenamefont
  {{Switzer}}(2009)}]{2009PhRvD..80f3010M}%
  \BibitemOpen
  \bibfield  {author} {\bibinfo {author} {\bibfnamefont {M.}~\bibnamefont
  {{McQuinn}}}\ and\ \bibinfo {author} {\bibfnamefont {E.~R.}\ \bibnamefont
  {{Switzer}}},\ }\bibfield  {title} {\bibinfo {title} {{Redshifted
  intergalactic He+3 8.7 GHz hyperfine absorption}},\ }\href
  {https://doi.org/10.1103/PhysRevD.80.063010} {\bibfield  {journal} {\bibinfo
  {journal} {Phys. Rev. D}\ }\textbf {\bibinfo {volume} {80}},\ \bibinfo {eid}
  {063010} (\bibinfo {year} {2009})},\ \Eprint
  {https://arxiv.org/abs/0905.1715} {arXiv:0905.1715 [astro-ph.CO]}
  \BibitemShut {NoStop}%
\bibitem [{\citenamefont {Khullar}\ \emph {et~al.}(2020)\citenamefont
  {Khullar}, \citenamefont {Ma}, \citenamefont {Busch}, \citenamefont {Ciardi},
  \citenamefont {Eide},\ and\ \citenamefont
  {Kakiichi}}]{10.1093/mnras/staa1951}%
  \BibitemOpen
  \bibfield  {author} {\bibinfo {author} {\bibfnamefont {S.}~\bibnamefont
  {Khullar}}, \bibinfo {author} {\bibfnamefont {Q.}~\bibnamefont {Ma}},
  \bibinfo {author} {\bibfnamefont {P.}~\bibnamefont {Busch}}, \bibinfo
  {author} {\bibfnamefont {B.}~\bibnamefont {Ciardi}}, \bibinfo {author}
  {\bibfnamefont {M.~B.}\ \bibnamefont {Eide}},\ and\ \bibinfo {author}
  {\bibfnamefont {K.}~\bibnamefont {Kakiichi}},\ }\bibfield  {title} {\bibinfo
  {title} {{Probing the high-z IGM with the hyperfine transition of 3He+}},\
  }\href {https://doi.org/10.1093/mnras/staa1951} {\bibfield  {journal}
  {\bibinfo  {journal} {Monthly Notices of the Royal Astronomical Society}\
  }\textbf {\bibinfo {volume} {497}},\ \bibinfo {pages} {572} (\bibinfo {year}
  {2020})},\ \Eprint
  {https://arxiv.org/abs/https://academic.oup.com/mnras/article-pdf/497/1/572/33527980/staa1951.pdf}
  {https://academic.oup.com/mnras/article-pdf/497/1/572/33527980/staa1951.pdf}
  \BibitemShut {NoStop}%
\bibitem [{\citenamefont {{Gong}}\ \emph {et~al.}(2013)\citenamefont {{Gong}},
  \citenamefont {{Cooray}},\ and\ \citenamefont
  {{Santos}}}]{2013ApJ...768..130G}%
  \BibitemOpen
  \bibfield  {author} {\bibinfo {author} {\bibfnamefont {Y.}~\bibnamefont
  {{Gong}}}, \bibinfo {author} {\bibfnamefont {A.}~\bibnamefont {{Cooray}}},\
  and\ \bibinfo {author} {\bibfnamefont {M.~G.}\ \bibnamefont {{Santos}}},\
  }\bibfield  {title} {\bibinfo {title} {{Probing the Pre-reionization Epoch
  with Molecular Hydrogen Intensity Mapping}},\ }\href
  {https://doi.org/10.1088/0004-637X/768/2/130} {\bibfield  {journal} {\bibinfo
   {journal} {Astrophys. J.}\ }\textbf {\bibinfo {volume} {768}},\ \bibinfo
  {eid} {130} (\bibinfo {year} {2013})},\ \Eprint
  {https://arxiv.org/abs/1212.2964} {arXiv:1212.2964 [astro-ph.CO]}
  \BibitemShut {NoStop}%
\bibitem [{\citenamefont {Kosenko}\ and\ \citenamefont
  {Ivanchik}(2018)}]{Kosenko_2018}%
  \BibitemOpen
  \bibfield  {author} {\bibinfo {author} {\bibfnamefont {D.~N.}\ \bibnamefont
  {Kosenko}}\ and\ \bibinfo {author} {\bibfnamefont {A.~V.}\ \bibnamefont
  {Ivanchik}},\ }\bibfield  {title} {\bibinfo {title} {The influence of
  baryon-photon ratio on 21 and 92 cm brightness temperature},\ }\href
  {https://doi.org/10.1088/1742-6596/1038/1/012011} {\bibfield  {journal}
  {\bibinfo  {journal} {Journal of Physics: Conference Series}\ }\textbf
  {\bibinfo {volume} {1038}},\ \bibinfo {pages} {012011} (\bibinfo {year}
  {2018})}\BibitemShut {NoStop}%
\bibitem [{\citenamefont {{Sigurdson}}\ and\ \citenamefont
  {{Furlanetto}}(2006)}]{2006PhRvL..97i1301S}%
  \BibitemOpen
  \bibfield  {author} {\bibinfo {author} {\bibfnamefont {K.}~\bibnamefont
  {{Sigurdson}}}\ and\ \bibinfo {author} {\bibfnamefont {S.~R.}\ \bibnamefont
  {{Furlanetto}}},\ }\bibfield  {title} {\bibinfo {title} {{Measuring the
  Primordial Deuterium Abundance during the Cosmic Dark Ages}},\ }\href
  {https://doi.org/10.1103/PhysRevLett.97.091301} {\bibfield  {journal}
  {\bibinfo  {journal} {Phys. Rev. Lett.}\ }\textbf {\bibinfo {volume} {97}},\
  \bibinfo {eid} {091301} (\bibinfo {year} {2006})},\ \Eprint
  {https://arxiv.org/abs/astro-ph/0505173} {arXiv:astro-ph/0505173 [astro-ph]}
  \BibitemShut {NoStop}%
\bibitem [{\citenamefont {Kovetz}\ \emph {et~al.}(2017)\citenamefont {Kovetz}
  \emph {et~al.}}]{Kovetz:2017agg}%
  \BibitemOpen
  \bibfield  {author} {\bibinfo {author} {\bibfnamefont {E.~D.}\ \bibnamefont
  {Kovetz}} \emph {et~al.},\ }\bibfield  {title} {\bibinfo {title}
  {{Line-Intensity Mapping: 2017 Status Report}},\ }\href@noop {} {\  (\bibinfo
  {year} {2017})},\ \Eprint {https://arxiv.org/abs/1709.09066}
  {arXiv:1709.09066 [astro-ph.CO]} \BibitemShut {NoStop}%
\bibitem [{\citenamefont {Kovetz}\ \emph {et~al.}(2019)\citenamefont {Kovetz}
  \emph {et~al.}}]{Kovetz:2019uss}%
  \BibitemOpen
  \bibfield  {author} {\bibinfo {author} {\bibfnamefont {E.~D.}\ \bibnamefont
  {Kovetz}} \emph {et~al.},\ }\bibfield  {title} {\bibinfo {title}
  {{Astrophysics and Cosmology with Line-Intensity Mapping}},\ }\href@noop {}
  {\  (\bibinfo {year} {2019})},\ \Eprint {https://arxiv.org/abs/1903.04496}
  {arXiv:1903.04496 [astro-ph.CO]} \BibitemShut {NoStop}%
\bibitem [{\citenamefont {Chang}\ \emph {et~al.}(2019)\citenamefont {Chang}
  \emph {et~al.}}]{Chang:2019xgc}%
  \BibitemOpen
  \bibfield  {author} {\bibinfo {author} {\bibfnamefont {T.-C.}\ \bibnamefont
  {Chang}} \emph {et~al.},\ }\bibfield  {title} {\bibinfo {title} {{Tomography
  of the Cosmic Dawn and Reionization Eras with Multiple Tracers}},\
  }\href@noop {} {\  (\bibinfo {year} {2019})},\ \Eprint
  {https://arxiv.org/abs/1903.11744} {arXiv:1903.11744 [astro-ph.CO]}
  \BibitemShut {NoStop}%
\bibitem [{\citenamefont {{Liu}}\ and\ \citenamefont
  {{Shaw}}(2020)}]{2020PASP..132f2001L}%
  \BibitemOpen
  \bibfield  {author} {\bibinfo {author} {\bibfnamefont {A.}~\bibnamefont
  {{Liu}}}\ and\ \bibinfo {author} {\bibfnamefont {J.~R.}\ \bibnamefont
  {{Shaw}}},\ }\bibfield  {title} {\bibinfo {title} {{Data Analysis for
  Precision 21 cm Cosmology}},\ }\href
  {https://doi.org/10.1088/1538-3873/ab5bfd} {\bibfield  {journal} {\bibinfo
  {journal} {Publ. Astron. Soc. Pac.}\ }\textbf {\bibinfo {volume} {132}},\
  \bibinfo {eid} {062001} (\bibinfo {year} {2020})},\ \Eprint
  {https://arxiv.org/abs/1907.08211} {arXiv:1907.08211 [astro-ph.IM]}
  \BibitemShut {NoStop}%
\bibitem [{\citenamefont {Moroi}\ \emph {et~al.}(2018)\citenamefont {Moroi},
  \citenamefont {Nakayama},\ and\ \citenamefont {Tang}}]{Moroi:2018vci}%
  \BibitemOpen
  \bibfield  {author} {\bibinfo {author} {\bibfnamefont {T.}~\bibnamefont
  {Moroi}}, \bibinfo {author} {\bibfnamefont {K.}~\bibnamefont {Nakayama}},\
  and\ \bibinfo {author} {\bibfnamefont {Y.}~\bibnamefont {Tang}},\ }\bibfield
  {title} {\bibinfo {title} {{Axion-photon conversion and effects on 21 cm
  observation}},\ }\href {https://doi.org/10.1016/j.physletb.2018.07.002}
  {\bibfield  {journal} {\bibinfo  {journal} {Phys. Lett. B}\ }\textbf
  {\bibinfo {volume} {783}},\ \bibinfo {pages} {301} (\bibinfo {year}
  {2018})},\ \Eprint {https://arxiv.org/abs/1804.10378} {arXiv:1804.10378
  [hep-ph]} \BibitemShut {NoStop}%
\bibitem [{\citenamefont {Pospelov}\ \emph {et~al.}(2018)\citenamefont
  {Pospelov}, \citenamefont {Pradler}, \citenamefont {Ruderman},\ and\
  \citenamefont {Urbano}}]{Pospelov:2018kdh}%
  \BibitemOpen
  \bibfield  {author} {\bibinfo {author} {\bibfnamefont {M.}~\bibnamefont
  {Pospelov}}, \bibinfo {author} {\bibfnamefont {J.}~\bibnamefont {Pradler}},
  \bibinfo {author} {\bibfnamefont {J.~T.}\ \bibnamefont {Ruderman}},\ and\
  \bibinfo {author} {\bibfnamefont {A.}~\bibnamefont {Urbano}},\ }\bibfield
  {title} {\bibinfo {title} {{Room for New Physics in the Rayleigh-Jeans Tail
  of the Cosmic Microwave Background}},\ }\href
  {https://doi.org/10.1103/PhysRevLett.121.031103} {\bibfield  {journal}
  {\bibinfo  {journal} {Phys. Rev. Lett.}\ }\textbf {\bibinfo {volume} {121}},\
  \bibinfo {pages} {031103} (\bibinfo {year} {2018})},\ \Eprint
  {https://arxiv.org/abs/1803.07048} {arXiv:1803.07048 [hep-ph]} \BibitemShut
  {NoStop}%
\bibitem [{\citenamefont {Choi}\ \emph {et~al.}(2020)\citenamefont {Choi},
  \citenamefont {Seong},\ and\ \citenamefont {Yun}}]{Choi:2019jwx}%
  \BibitemOpen
  \bibfield  {author} {\bibinfo {author} {\bibfnamefont {K.}~\bibnamefont
  {Choi}}, \bibinfo {author} {\bibfnamefont {H.}~\bibnamefont {Seong}},\ and\
  \bibinfo {author} {\bibfnamefont {S.}~\bibnamefont {Yun}},\ }\bibfield
  {title} {\bibinfo {title} {{Axion-photon-dark photon oscillation and its
  implication for 21 cm observation}},\ }\href
  {https://doi.org/10.1103/PhysRevD.102.075024} {\bibfield  {journal} {\bibinfo
   {journal} {Phys. Rev. D}\ }\textbf {\bibinfo {volume} {102}},\ \bibinfo
  {pages} {075024} (\bibinfo {year} {2020})},\ \Eprint
  {https://arxiv.org/abs/1911.00532} {arXiv:1911.00532 [hep-ph]} \BibitemShut
  {NoStop}%
\bibitem [{\citenamefont {Brandenberger}\ \emph
  {et~al.}(2019{\natexlab{a}})\citenamefont {Brandenberger}, \citenamefont
  {Cyr},\ and\ \citenamefont {Schaeffer}}]{Brandenberger:2018dfj}%
  \BibitemOpen
  \bibfield  {author} {\bibinfo {author} {\bibfnamefont {R.}~\bibnamefont
  {Brandenberger}}, \bibinfo {author} {\bibfnamefont {B.}~\bibnamefont {Cyr}},\
  and\ \bibinfo {author} {\bibfnamefont {T.}~\bibnamefont {Schaeffer}},\
  }\bibfield  {title} {\bibinfo {title} {{On the Possible Enhancement of the
  Global $21$-cm Signal at Reionization from the Decay of Cosmic String
  Cusps}},\ }\href {https://doi.org/10.1088/1475-7516/2019/04/020} {\bibfield
  {journal} {\bibinfo  {journal} {JCAP}\ }\textbf {\bibinfo {volume} {04}},\
  \bibinfo {pages} {020}},\ \Eprint {https://arxiv.org/abs/1810.03219}
  {arXiv:1810.03219 [astro-ph.CO]} \BibitemShut {NoStop}%
\bibitem [{\citenamefont {Brandenberger}\ \emph
  {et~al.}(2019{\natexlab{b}})\citenamefont {Brandenberger}, \citenamefont
  {Cyr},\ and\ \citenamefont {Shi}}]{Brandenberger:2019lfm}%
  \BibitemOpen
  \bibfield  {author} {\bibinfo {author} {\bibfnamefont {R.}~\bibnamefont
  {Brandenberger}}, \bibinfo {author} {\bibfnamefont {B.}~\bibnamefont {Cyr}},\
  and\ \bibinfo {author} {\bibfnamefont {R.}~\bibnamefont {Shi}},\ }\bibfield
  {title} {\bibinfo {title} {{Constraints on Superconducting Cosmic Strings
  from the Global $21$-cm Signal before Reionization}},\ }\href
  {https://doi.org/10.1088/1475-7516/2019/09/009} {\bibfield  {journal}
  {\bibinfo  {journal} {JCAP}\ }\textbf {\bibinfo {volume} {09}},\ \bibinfo
  {pages} {009}},\ \Eprint {https://arxiv.org/abs/1902.08282} {arXiv:1902.08282
  [astro-ph.CO]} \BibitemShut {NoStop}%
\bibitem [{\citenamefont {Johns}\ and\ \citenamefont
  {Koren}(2020)}]{companion}%
  \BibitemOpen
  \bibfield  {author} {\bibinfo {author} {\bibfnamefont {L.}~\bibnamefont
  {Johns}}\ and\ \bibinfo {author} {\bibfnamefont {S.}~\bibnamefont {Koren}},\
  }\bibfield  {title} {\bibinfo {title} {Hydrogen mixing as a novel mechanism
  for colder baryons in 21 cm cosmology},\ }\href@noop {} {\  (\bibinfo {year}
  {2020})}\BibitemShut {NoStop}%
\bibitem [{\citenamefont {Lee}\ and\ \citenamefont {Yang}(1956)}]{Lee:1956qn}%
  \BibitemOpen
  \bibfield  {author} {\bibinfo {author} {\bibfnamefont {T.}~\bibnamefont
  {Lee}}\ and\ \bibinfo {author} {\bibfnamefont {C.-N.}\ \bibnamefont {Yang}},\
  }\bibfield  {title} {\bibinfo {title} {{Question of Parity Conservation in
  Weak Interactions}},\ }\href {https://doi.org/10.1103/PhysRev.104.254}
  {\bibfield  {journal} {\bibinfo  {journal} {Phys. Rev.}\ }\textbf {\bibinfo
  {volume} {104}},\ \bibinfo {pages} {254} (\bibinfo {year}
  {1956})}\BibitemShut {NoStop}%
\bibitem [{\citenamefont {Okun}(2007)}]{Okun:2006eb}%
  \BibitemOpen
  \bibfield  {author} {\bibinfo {author} {\bibfnamefont {L.}~\bibnamefont
  {Okun}},\ }\bibfield  {title} {\bibinfo {title} {{Mirror particles and mirror
  matter: 50 years of speculations and search}},\ }\href
  {https://doi.org/10.1070/PU2007v050n04ABEH006227} {\bibfield  {journal}
  {\bibinfo  {journal} {Phys. Usp.}\ }\textbf {\bibinfo {volume} {50}},\
  \bibinfo {pages} {380} (\bibinfo {year} {2007})},\ \Eprint
  {https://arxiv.org/abs/hep-ph/0606202} {arXiv:hep-ph/0606202} \BibitemShut
  {NoStop}%
\bibitem [{\citenamefont {Blinnikov}\ and\ \citenamefont
  {Khlopov}(1982)}]{Blinnikov:1982eh}%
  \BibitemOpen
  \bibfield  {author} {\bibinfo {author} {\bibfnamefont {S.}~\bibnamefont
  {Blinnikov}}\ and\ \bibinfo {author} {\bibfnamefont {M.}~\bibnamefont
  {Khlopov}},\ }\bibfield  {title} {\bibinfo {title} {{On Possible Effects of
  'Mirror' Particles}},\ }\href@noop {} {\bibfield  {journal} {\bibinfo
  {journal} {Sov. J. Nucl. Phys.}\ }\textbf {\bibinfo {volume} {36}},\ \bibinfo
  {pages} {472} (\bibinfo {year} {1982})}\BibitemShut {NoStop}%
\bibitem [{\citenamefont {Gross}\ \emph {et~al.}(1985)\citenamefont {Gross},
  \citenamefont {Harvey}, \citenamefont {Martinec},\ and\ \citenamefont
  {Rohm}}]{Gross:1984dd}%
  \BibitemOpen
  \bibfield  {author} {\bibinfo {author} {\bibfnamefont {D.~J.}\ \bibnamefont
  {Gross}}, \bibinfo {author} {\bibfnamefont {J.~A.}\ \bibnamefont {Harvey}},
  \bibinfo {author} {\bibfnamefont {E.~J.}\ \bibnamefont {Martinec}},\ and\
  \bibinfo {author} {\bibfnamefont {R.}~\bibnamefont {Rohm}},\ }\bibfield
  {title} {\bibinfo {title} {{The Heterotic String}},\ }\href
  {https://doi.org/10.1103/PhysRevLett.54.502} {\bibfield  {journal} {\bibinfo
  {journal} {Phys. Rev. Lett.}\ }\textbf {\bibinfo {volume} {54}},\ \bibinfo
  {pages} {502} (\bibinfo {year} {1985})}\BibitemShut {NoStop}%
\bibitem [{\citenamefont {Foot}\ \emph {et~al.}(1991)\citenamefont {Foot},
  \citenamefont {Lew},\ and\ \citenamefont {Volkas}}]{Foot:1991bp}%
  \BibitemOpen
  \bibfield  {author} {\bibinfo {author} {\bibfnamefont {R.}~\bibnamefont
  {Foot}}, \bibinfo {author} {\bibfnamefont {H.}~\bibnamefont {Lew}},\ and\
  \bibinfo {author} {\bibfnamefont {R.}~\bibnamefont {Volkas}},\ }\bibfield
  {title} {\bibinfo {title} {{A Model with fundamental improper space-time
  symmetries}},\ }\href {https://doi.org/10.1016/0370-2693(91)91013-L}
  {\bibfield  {journal} {\bibinfo  {journal} {Phys. Lett. B}\ }\textbf
  {\bibinfo {volume} {272}},\ \bibinfo {pages} {67} (\bibinfo {year}
  {1991})}\BibitemShut {NoStop}%
\bibitem [{\citenamefont {Chacko}\ \emph {et~al.}(2006)\citenamefont {Chacko},
  \citenamefont {Goh},\ and\ \citenamefont {Harnik}}]{Chacko:2005pe}%
  \BibitemOpen
  \bibfield  {author} {\bibinfo {author} {\bibfnamefont {Z.}~\bibnamefont
  {Chacko}}, \bibinfo {author} {\bibfnamefont {H.-S.}\ \bibnamefont {Goh}},\
  and\ \bibinfo {author} {\bibfnamefont {R.}~\bibnamefont {Harnik}},\
  }\bibfield  {title} {\bibinfo {title} {{The Twin Higgs: Natural electroweak
  breaking from mirror symmetry}},\ }\href
  {https://doi.org/10.1103/PhysRevLett.96.231802} {\bibfield  {journal}
  {\bibinfo  {journal} {Phys. Rev. Lett.}\ }\textbf {\bibinfo {volume} {96}},\
  \bibinfo {pages} {231802} (\bibinfo {year} {2006})},\ \Eprint
  {https://arxiv.org/abs/hep-ph/0506256} {arXiv:hep-ph/0506256} \BibitemShut
  {NoStop}%
\bibitem [{\citenamefont {McKeen}\ \emph {et~al.}(2020)\citenamefont {McKeen},
  \citenamefont {Pospelov},\ and\ \citenamefont
  {Raj}}]{PhysRevLett.125.231803}%
  \BibitemOpen
  \bibfield  {author} {\bibinfo {author} {\bibfnamefont {D.}~\bibnamefont
  {McKeen}}, \bibinfo {author} {\bibfnamefont {M.}~\bibnamefont {Pospelov}},\
  and\ \bibinfo {author} {\bibfnamefont {N.}~\bibnamefont {Raj}},\ }\bibfield
  {title} {\bibinfo {title} {Hydrogen portal to exotic radioactivity},\ }\href
  {https://doi.org/10.1103/PhysRevLett.125.231803} {\bibfield  {journal}
  {\bibinfo  {journal} {Phys. Rev. Lett.}\ }\textbf {\bibinfo {volume} {125}},\
  \bibinfo {pages} {231803} (\bibinfo {year} {2020})}\BibitemShut {NoStop}%
\bibitem [{\citenamefont {Pritchard}\ and\ \citenamefont
  {Loeb}(2012)}]{Pritchard_2012}%
  \BibitemOpen
  \bibfield  {author} {\bibinfo {author} {\bibfnamefont {J.~R.}\ \bibnamefont
  {Pritchard}}\ and\ \bibinfo {author} {\bibfnamefont {A.}~\bibnamefont
  {Loeb}},\ }\bibfield  {title} {\bibinfo {title} {21 cm cosmology in the 21st
  century},\ }\href {https://doi.org/10.1088/0034-4885/75/8/086901} {\bibfield
  {journal} {\bibinfo  {journal} {Reports on Progress in Physics}\ }\textbf
  {\bibinfo {volume} {75}},\ \bibinfo {pages} {086901} (\bibinfo {year}
  {2012})}\BibitemShut {NoStop}%
\bibitem [{\citenamefont {{Field}}(1958)}]{4065250}%
  \BibitemOpen
  \bibfield  {author} {\bibinfo {author} {\bibfnamefont {G.~B.}\ \bibnamefont
  {{Field}}},\ }\bibfield  {title} {\bibinfo {title} {Excitation of the
  hydrogen 21-cm line},\ }\href {https://doi.org/10.1109/JRPROC.1958.286741}
  {\bibfield  {journal} {\bibinfo  {journal} {Proceedings of the IRE}\ }\textbf
  {\bibinfo {volume} {46}},\ \bibinfo {pages} {240} (\bibinfo {year}
  {1958})}\BibitemShut {NoStop}%
\bibitem [{\citenamefont {Wouthuysen}(1952)}]{wouthuysen1952excitation}%
  \BibitemOpen
  \bibfield  {author} {\bibinfo {author} {\bibfnamefont {S.~A.}\ \bibnamefont
  {Wouthuysen}},\ }\bibfield  {title} {\bibinfo {title} {On the excitation
  mechanism of the 21-cm (radio-frequency) interstellar hydrogen emission
  line.},\ }\href@noop {} {\bibfield  {journal} {\bibinfo  {journal} {Astron.
  J.}\ }\textbf {\bibinfo {volume} {57}},\ \bibinfo {pages} {31} (\bibinfo
  {year} {1952})}\BibitemShut {NoStop}%
\bibitem [{\citenamefont {{Barkana}}\ and\ \citenamefont
  {{Loeb}}(2001)}]{Barkana:2001avi}%
  \BibitemOpen
  \bibfield  {author} {\bibinfo {author} {\bibfnamefont {R.}~\bibnamefont
  {{Barkana}}}\ and\ \bibinfo {author} {\bibfnamefont {A.}~\bibnamefont
  {{Loeb}}},\ }\bibfield  {title} {\bibinfo {title} {{In the beginning: the
  first sources of light and the reionization of the universe}},\ }\href
  {https://doi.org/10.1016/S0370-1573(01)00019-9} {\bibfield  {journal}
  {\bibinfo  {journal} {Phys. Rept.}\ }\textbf {\bibinfo {volume} {349}},\
  \bibinfo {pages} {125} (\bibinfo {year} {2001})},\ \Eprint
  {https://arxiv.org/abs/astro-ph/0010468} {arXiv:astro-ph/0010468 [astro-ph]}
  \BibitemShut {NoStop}%
\bibitem [{\citenamefont {Loeb}\ \emph {et~al.}(2008)\citenamefont {Loeb},
  \citenamefont {Ferrara},\ and\ \citenamefont {Ellis}}]{loeb2008first}%
  \BibitemOpen
  \bibfield  {author} {\bibinfo {author} {\bibfnamefont {A.}~\bibnamefont
  {Loeb}}, \bibinfo {author} {\bibfnamefont {A.}~\bibnamefont {Ferrara}},\ and\
  \bibinfo {author} {\bibfnamefont {R.~S.}\ \bibnamefont {Ellis}},\ }\href@noop
  {} {\emph {\bibinfo {title} {First light in the universe}}}\ (\bibinfo
  {publisher} {Springer},\ \bibinfo {year} {2008})\BibitemShut {NoStop}%
\bibitem [{\citenamefont {Kainulainen}(1990)}]{KAINULAINEN1990191}%
  \BibitemOpen
  \bibfield  {author} {\bibinfo {author} {\bibfnamefont {K.}~\bibnamefont
  {Kainulainen}},\ }\bibfield  {title} {\bibinfo {title} {Light singlet
  neutrinos and the primordial nucleosynthesis},\ }\href
  {https://doi.org/https://doi.org/10.1016/0370-2693(90)90054-A} {\bibfield
  {journal} {\bibinfo  {journal} {Physics Letters B}\ }\textbf {\bibinfo
  {volume} {244}},\ \bibinfo {pages} {191 } (\bibinfo {year}
  {1990})}\BibitemShut {NoStop}%
\bibitem [{\citenamefont {Cline}(1992)}]{PhysRevLett.68.3137}%
  \BibitemOpen
  \bibfield  {author} {\bibinfo {author} {\bibfnamefont {J.~M.}\ \bibnamefont
  {Cline}},\ }\bibfield  {title} {\bibinfo {title} {Constraints on almost-dirac
  neutrinos from neutrino-antineutrino oscillations},\ }\href
  {https://doi.org/10.1103/PhysRevLett.68.3137} {\bibfield  {journal} {\bibinfo
   {journal} {Phys. Rev. Lett.}\ }\textbf {\bibinfo {volume} {68}},\ \bibinfo
  {pages} {3137} (\bibinfo {year} {1992})}\BibitemShut {NoStop}%
\bibitem [{\citenamefont {Dodelson}\ and\ \citenamefont
  {Widrow}(1994)}]{PhysRevLett.72.17}%
  \BibitemOpen
  \bibfield  {author} {\bibinfo {author} {\bibfnamefont {S.}~\bibnamefont
  {Dodelson}}\ and\ \bibinfo {author} {\bibfnamefont {L.~M.}\ \bibnamefont
  {Widrow}},\ }\bibfield  {title} {\bibinfo {title} {Sterile neutrinos as dark
  matter},\ }\href {https://doi.org/10.1103/PhysRevLett.72.17} {\bibfield
  {journal} {\bibinfo  {journal} {Phys. Rev. Lett.}\ }\textbf {\bibinfo
  {volume} {72}},\ \bibinfo {pages} {17} (\bibinfo {year} {1994})}\BibitemShut
  {NoStop}%
\bibitem [{\citenamefont {Lee}\ \emph {et~al.}(2000)\citenamefont {Lee},
  \citenamefont {Volkas},\ and\ \citenamefont {Wong}}]{PhysRevD.62.093025}%
  \BibitemOpen
  \bibfield  {author} {\bibinfo {author} {\bibfnamefont {K.~S.~M.}\
  \bibnamefont {Lee}}, \bibinfo {author} {\bibfnamefont {R.~R.}\ \bibnamefont
  {Volkas}},\ and\ \bibinfo {author} {\bibfnamefont {Y.~Y.~Y.}\ \bibnamefont
  {Wong}},\ }\bibfield  {title} {\bibinfo {title} {Further studies on relic
  neutrino asymmetry generation. ii. a rigorous treatment of repopulation in
  the adiabatic limit},\ }\href {https://doi.org/10.1103/PhysRevD.62.093025}
  {\bibfield  {journal} {\bibinfo  {journal} {Phys. Rev. D}\ }\textbf {\bibinfo
  {volume} {62}},\ \bibinfo {pages} {093025} (\bibinfo {year}
  {2000})}\BibitemShut {NoStop}%
\bibitem [{\citenamefont {Johns}(2019)}]{PhysRevD.100.083536}%
  \BibitemOpen
  \bibfield  {author} {\bibinfo {author} {\bibfnamefont {L.}~\bibnamefont
  {Johns}},\ }\bibfield  {title} {\bibinfo {title} {Derivation of the sterile
  neutrino boltzmann equation from quantum kinetics},\ }\href
  {https://doi.org/10.1103/PhysRevD.100.083536} {\bibfield  {journal} {\bibinfo
   {journal} {Phys. Rev. D}\ }\textbf {\bibinfo {volume} {100}},\ \bibinfo
  {pages} {083536} (\bibinfo {year} {2019})}\BibitemShut {NoStop}%
\bibitem [{\citenamefont {Kaplan}(2005)}]{Kaplan:2005es}%
  \BibitemOpen
  \bibfield  {author} {\bibinfo {author} {\bibfnamefont {D.~B.}\ \bibnamefont
  {Kaplan}},\ }\bibfield  {title} {\bibinfo {title} {{Five lectures on
  effective field theory}}\ }(\bibinfo {year} {2005})\ \Eprint
  {https://arxiv.org/abs/nucl-th/0510023} {arXiv:nucl-th/0510023} \BibitemShut
  {NoStop}%
\bibitem [{\citenamefont {Subramanian}(2019)}]{Subramanian:2019jyd}%
  \BibitemOpen
  \bibfield  {author} {\bibinfo {author} {\bibfnamefont {K.}~\bibnamefont
  {Subramanian}},\ }\bibfield  {title} {\bibinfo {title} {{From primordial seed
  magnetic fields to the galactic dynamo}},\ }\href
  {https://doi.org/10.3390/galaxies7020047} {\bibfield  {journal} {\bibinfo
  {journal} {Galaxies}\ }\textbf {\bibinfo {volume} {7}},\ \bibinfo {pages}
  {47} (\bibinfo {year} {2019})},\ \Eprint {https://arxiv.org/abs/1903.03744}
  {arXiv:1903.03744 [astro-ph.CO]} \BibitemShut {NoStop}%
\bibitem [{\citenamefont {Gopal}\ and\ \citenamefont
  {Sethi}(2005)}]{10.1111/j.1365-2966.2005.09442.x}%
  \BibitemOpen
  \bibfield  {author} {\bibinfo {author} {\bibfnamefont {R.}~\bibnamefont
  {Gopal}}\ and\ \bibinfo {author} {\bibfnamefont {S.~K.}\ \bibnamefont
  {Sethi}},\ }\bibfield  {title} {\bibinfo {title} {{Generation of magnetic
  field in the pre-recombination era}},\ }\href
  {https://doi.org/10.1111/j.1365-2966.2005.09442.x} {\bibfield  {journal}
  {\bibinfo  {journal} {Monthly Notices of the Royal Astronomical Society}\
  }\textbf {\bibinfo {volume} {363}},\ \bibinfo {pages} {521} (\bibinfo {year}
  {2005})},\ \Eprint
  {https://arxiv.org/abs/https://academic.oup.com/mnras/article-pdf/363/2/521/3912327/363-2-521.pdf}
  {https://academic.oup.com/mnras/article-pdf/363/2/521/3912327/363-2-521.pdf}
  \BibitemShut {NoStop}%
\bibitem [{\citenamefont {Matarrese}\ \emph {et~al.}(2005)\citenamefont
  {Matarrese}, \citenamefont {Mollerach}, \citenamefont {Notari},\ and\
  \citenamefont {Riotto}}]{PhysRevD.71.043502}%
  \BibitemOpen
  \bibfield  {author} {\bibinfo {author} {\bibfnamefont {S.}~\bibnamefont
  {Matarrese}}, \bibinfo {author} {\bibfnamefont {S.}~\bibnamefont
  {Mollerach}}, \bibinfo {author} {\bibfnamefont {A.}~\bibnamefont {Notari}},\
  and\ \bibinfo {author} {\bibfnamefont {A.}~\bibnamefont {Riotto}},\
  }\bibfield  {title} {\bibinfo {title} {Large-scale magnetic fields from
  density perturbations},\ }\href {https://doi.org/10.1103/PhysRevD.71.043502}
  {\bibfield  {journal} {\bibinfo  {journal} {Phys. Rev. D}\ }\textbf {\bibinfo
  {volume} {71}},\ \bibinfo {pages} {043502} (\bibinfo {year}
  {2005})}\BibitemShut {NoStop}%
\bibitem [{\citenamefont {Kulsrud}\ \emph {et~al.}(1997)\citenamefont
  {Kulsrud}, \citenamefont {Cen}, \citenamefont {Ostriker},\ and\ \citenamefont
  {Ryu}}]{Kulsrud_1997}%
  \BibitemOpen
  \bibfield  {author} {\bibinfo {author} {\bibfnamefont {R.~M.}\ \bibnamefont
  {Kulsrud}}, \bibinfo {author} {\bibfnamefont {R.}~\bibnamefont {Cen}},
  \bibinfo {author} {\bibfnamefont {J.~P.}\ \bibnamefont {Ostriker}},\ and\
  \bibinfo {author} {\bibfnamefont {D.}~\bibnamefont {Ryu}},\ }\bibfield
  {title} {\bibinfo {title} {The protogalactic origin for cosmic magnetic
  fields},\ }\href {https://doi.org/10.1086/303987} {\bibfield  {journal}
  {\bibinfo  {journal} {The Astrophysical Journal}\ }\textbf {\bibinfo {volume}
  {480}},\ \bibinfo {pages} {481} (\bibinfo {year} {1997})}\BibitemShut
  {NoStop}%
\bibitem [{\citenamefont {Gnedin}\ \emph {et~al.}(2000)\citenamefont {Gnedin},
  \citenamefont {Ferrara},\ and\ \citenamefont {Zweibel}}]{Gnedin_2000}%
  \BibitemOpen
  \bibfield  {author} {\bibinfo {author} {\bibfnamefont {N.~Y.}\ \bibnamefont
  {Gnedin}}, \bibinfo {author} {\bibfnamefont {A.}~\bibnamefont {Ferrara}},\
  and\ \bibinfo {author} {\bibfnamefont {E.~G.}\ \bibnamefont {Zweibel}},\
  }\bibfield  {title} {\bibinfo {title} {Generation of the primordial magnetic
  fields during cosmological reionization},\ }\href
  {https://doi.org/10.1086/309272} {\bibfield  {journal} {\bibinfo  {journal}
  {The Astrophysical Journal}\ }\textbf {\bibinfo {volume} {539}},\ \bibinfo
  {pages} {505} (\bibinfo {year} {2000})}\BibitemShut {NoStop}%
\bibitem [{\citenamefont {{Subramanian}}\ \emph {et~al.}(1994)\citenamefont
  {{Subramanian}}, \citenamefont {{Narasimha}},\ and\ \citenamefont
  {{Chitre}}}]{1994MNRAS.271L..15S}%
  \BibitemOpen
  \bibfield  {author} {\bibinfo {author} {\bibfnamefont {K.}~\bibnamefont
  {{Subramanian}}}, \bibinfo {author} {\bibfnamefont {D.}~\bibnamefont
  {{Narasimha}}},\ and\ \bibinfo {author} {\bibfnamefont {S.~M.}\ \bibnamefont
  {{Chitre}}},\ }\bibfield  {title} {\bibinfo {title} {{Thermal generation of
  cosmological seed magnetic fields in ionization fronts}},\ }\href
  {https://doi.org/10.1093/mnras/271.1.L15} {\bibfield  {journal} {\bibinfo
  {journal} {Mon. Not. Roy. Astron. Soc.}\ }\textbf {\bibinfo {volume} {271}},\
  \bibinfo {pages} {L15} (\bibinfo {year} {1994})}\BibitemShut {NoStop}%
\bibitem [{\citenamefont {Takahashi}\ \emph {et~al.}(2005)\citenamefont
  {Takahashi}, \citenamefont {Ichiki}, \citenamefont {Ohno},\ and\
  \citenamefont {Hanayama}}]{PhysRevLett.95.121301}%
  \BibitemOpen
  \bibfield  {author} {\bibinfo {author} {\bibfnamefont {K.}~\bibnamefont
  {Takahashi}}, \bibinfo {author} {\bibfnamefont {K.}~\bibnamefont {Ichiki}},
  \bibinfo {author} {\bibfnamefont {H.}~\bibnamefont {Ohno}},\ and\ \bibinfo
  {author} {\bibfnamefont {H.}~\bibnamefont {Hanayama}},\ }\bibfield  {title}
  {\bibinfo {title} {Magnetic field generation from cosmological
  perturbations},\ }\href {https://doi.org/10.1103/PhysRevLett.95.121301}
  {\bibfield  {journal} {\bibinfo  {journal} {Phys. Rev. Lett.}\ }\textbf
  {\bibinfo {volume} {95}},\ \bibinfo {pages} {121301} (\bibinfo {year}
  {2005})}\BibitemShut {NoStop}%
\bibitem [{\citenamefont {Kobayashi}\ \emph {et~al.}(2007)\citenamefont
  {Kobayashi}, \citenamefont {Maartens}, \citenamefont {Shiromizu},\ and\
  \citenamefont {Takahashi}}]{PhysRevD.75.103501}%
  \BibitemOpen
  \bibfield  {author} {\bibinfo {author} {\bibfnamefont {T.}~\bibnamefont
  {Kobayashi}}, \bibinfo {author} {\bibfnamefont {R.}~\bibnamefont {Maartens}},
  \bibinfo {author} {\bibfnamefont {T.}~\bibnamefont {Shiromizu}},\ and\
  \bibinfo {author} {\bibfnamefont {K.}~\bibnamefont {Takahashi}},\ }\bibfield
  {title} {\bibinfo {title} {Cosmological magnetic fields from nonlinear
  effects},\ }\href {https://doi.org/10.1103/PhysRevD.75.103501} {\bibfield
  {journal} {\bibinfo  {journal} {Phys. Rev. D}\ }\textbf {\bibinfo {volume}
  {75}},\ \bibinfo {pages} {103501} (\bibinfo {year} {2007})}\BibitemShut
  {NoStop}%
\bibitem [{\citenamefont {Durrive}\ \emph {et~al.}(2017)\citenamefont
  {Durrive}, \citenamefont {Tashiro}, \citenamefont {Langer},\ and\
  \citenamefont {Sugiyama}}]{10.1093/mnras/stx2007}%
  \BibitemOpen
  \bibfield  {author} {\bibinfo {author} {\bibfnamefont {J.-B.}\ \bibnamefont
  {Durrive}}, \bibinfo {author} {\bibfnamefont {H.}~\bibnamefont {Tashiro}},
  \bibinfo {author} {\bibfnamefont {M.}~\bibnamefont {Langer}},\ and\ \bibinfo
  {author} {\bibfnamefont {N.}~\bibnamefont {Sugiyama}},\ }\bibfield  {title}
  {\bibinfo {title} {{Mean energy density of photogenerated magnetic fields
  throughout the Epoch of Reionization}},\ }\href
  {https://doi.org/10.1093/mnras/stx2007} {\bibfield  {journal} {\bibinfo
  {journal} {Monthly Notices of the Royal Astronomical Society}\ }\textbf
  {\bibinfo {volume} {472}},\ \bibinfo {pages} {1649} (\bibinfo {year}
  {2017})},\ \Eprint
  {https://arxiv.org/abs/https://academic.oup.com/mnras/article-pdf/472/2/1649/19917665/stx2007.pdf}
  {https://academic.oup.com/mnras/article-pdf/472/2/1649/19917665/stx2007.pdf}
  \BibitemShut {NoStop}%
\bibitem [{\citenamefont {Durrive}\ and\ \citenamefont
  {Langer}(2015)}]{10.1093/mnras/stv1578}%
  \BibitemOpen
  \bibfield  {author} {\bibinfo {author} {\bibfnamefont {J.-B.}\ \bibnamefont
  {Durrive}}\ and\ \bibinfo {author} {\bibfnamefont {M.}~\bibnamefont
  {Langer}},\ }\bibfield  {title} {\bibinfo {title} {{Intergalactic
  magnetogenesis at Cosmic Dawn by photoionization}},\ }\href
  {https://doi.org/10.1093/mnras/stv1578} {\bibfield  {journal} {\bibinfo
  {journal} {Monthly Notices of the Royal Astronomical Society}\ }\textbf
  {\bibinfo {volume} {453}},\ \bibinfo {pages} {345} (\bibinfo {year}
  {2015})},\ \Eprint
  {https://arxiv.org/abs/https://academic.oup.com/mnras/article-pdf/453/1/345/4913470/stv1578.pdf}
  {https://academic.oup.com/mnras/article-pdf/453/1/345/4913470/stv1578.pdf}
  \BibitemShut {NoStop}%
\bibitem [{\citenamefont {Mu{\~n}oz}\ \emph {et~al.}(2015)\citenamefont
  {Mu{\~n}oz}, \citenamefont {Kovetz},\ and\ \citenamefont
  {Ali-Ha{\"\i}moud}}]{Munoz:2015bca}%
  \BibitemOpen
  \bibfield  {author} {\bibinfo {author} {\bibfnamefont {J.~B.}\ \bibnamefont
  {Mu{\~n}oz}}, \bibinfo {author} {\bibfnamefont {E.~D.}\ \bibnamefont
  {Kovetz}},\ and\ \bibinfo {author} {\bibfnamefont {Y.}~\bibnamefont
  {Ali-Ha{\"\i}moud}},\ }\bibfield  {title} {\bibinfo {title} {{Heating of
  Baryons due to Scattering with Dark Matter During the Dark Ages}},\ }\href
  {https://doi.org/10.1103/PhysRevD.92.083528} {\bibfield  {journal} {\bibinfo
  {journal} {Phys. Rev. D}\ }\textbf {\bibinfo {volume} {92}},\ \bibinfo
  {pages} {083528} (\bibinfo {year} {2015})},\ \Eprint
  {https://arxiv.org/abs/1509.00029} {arXiv:1509.00029 [astro-ph.CO]}
  \BibitemShut {NoStop}%
\bibitem [{\citenamefont {Mu\~noz}\ and\ \citenamefont
  {Loeb}(2017)}]{Munoz:2017qpy}%
  \BibitemOpen
  \bibfield  {author} {\bibinfo {author} {\bibfnamefont {J.~B.}\ \bibnamefont
  {Mu\~noz}}\ and\ \bibinfo {author} {\bibfnamefont {A.}~\bibnamefont {Loeb}},\
  }\bibfield  {title} {\bibinfo {title} {{Constraints on Dark Matter-Baryon
  Scattering from the Temperature Evolution of the Intergalactic Medium}},\
  }\href {https://doi.org/10.1088/1475-7516/2017/11/043} {\bibfield  {journal}
  {\bibinfo  {journal} {JCAP}\ }\textbf {\bibinfo {volume} {11}},\ \bibinfo
  {pages} {043}},\ \Eprint {https://arxiv.org/abs/1708.08923} {arXiv:1708.08923
  [astro-ph.CO]} \BibitemShut {NoStop}%
\bibitem [{\citenamefont {Ballesteros}\ \emph {et~al.}(2020)\citenamefont
  {Ballesteros}, \citenamefont {Garcia},\ and\ \citenamefont
  {Pierre}}]{Ballesteros:2020adh}%
  \BibitemOpen
  \bibfield  {author} {\bibinfo {author} {\bibfnamefont {G.}~\bibnamefont
  {Ballesteros}}, \bibinfo {author} {\bibfnamefont {M.~A.}\ \bibnamefont
  {Garcia}},\ and\ \bibinfo {author} {\bibfnamefont {M.}~\bibnamefont
  {Pierre}},\ }\bibfield  {title} {\bibinfo {title} {{How warm are non-thermal
  relics? Lyman-$\alpha$ bounds on out-of-equilibrium dark matter}},\
  }\href@noop {} {\  (\bibinfo {year} {2020})},\ \Eprint
  {https://arxiv.org/abs/2011.13458} {arXiv:2011.13458 [hep-ph]} \BibitemShut
  {NoStop}%
\bibitem [{\citenamefont {Chikashige}\ \emph {et~al.}(1981)\citenamefont
  {Chikashige}, \citenamefont {Mohapatra},\ and\ \citenamefont
  {Peccei}}]{Chikashige:1980ui}%
  \BibitemOpen
  \bibfield  {author} {\bibinfo {author} {\bibfnamefont {Y.}~\bibnamefont
  {Chikashige}}, \bibinfo {author} {\bibfnamefont {R.~N.}\ \bibnamefont
  {Mohapatra}},\ and\ \bibinfo {author} {\bibfnamefont {R.}~\bibnamefont
  {Peccei}},\ }\bibfield  {title} {\bibinfo {title} {{Are There Real Goldstone
  Bosons Associated with Broken Lepton Number?}},\ }\href
  {https://doi.org/10.1016/0370-2693(81)90011-3} {\bibfield  {journal}
  {\bibinfo  {journal} {Phys. Lett. B}\ }\textbf {\bibinfo {volume} {98}},\
  \bibinfo {pages} {265} (\bibinfo {year} {1981})}\BibitemShut {NoStop}%
\bibitem [{\citenamefont {{Kuz'min}}(1970)}]{1970JETPL..12..228K}%
  \BibitemOpen
  \bibfield  {author} {\bibinfo {author} {\bibfnamefont {V.~A.}\ \bibnamefont
  {{Kuz'min}}},\ }\bibfield  {title} {\bibinfo {title} {{CP-noninvariance and
  baryon asymmetry of the universe.}},\ }\href@noop {} {\bibfield  {journal}
  {\bibinfo  {journal} {Soviet Journal of Experimental and Theoretical Physics
  Letters}\ }\textbf {\bibinfo {volume} {12}},\ \bibinfo {pages} {228}
  (\bibinfo {year} {1970})}\BibitemShut {NoStop}%
\bibitem [{\citenamefont {Feinberg}\ \emph {et~al.}(1978)\citenamefont
  {Feinberg}, \citenamefont {Goldhaber},\ and\ \citenamefont
  {Steigman}}]{Feinberg:1978sd}%
  \BibitemOpen
  \bibfield  {author} {\bibinfo {author} {\bibfnamefont {G.}~\bibnamefont
  {Feinberg}}, \bibinfo {author} {\bibfnamefont {M.}~\bibnamefont
  {Goldhaber}},\ and\ \bibinfo {author} {\bibfnamefont {G.}~\bibnamefont
  {Steigman}},\ }\bibfield  {title} {\bibinfo {title} {{Multiplicative Baryon
  Number Conservation and the Oscillation of Hydrogen Into Anti-hydrogen}},\
  }\href {https://doi.org/10.1103/PhysRevD.18.1602} {\bibfield  {journal}
  {\bibinfo  {journal} {Phys. Rev. D}\ }\textbf {\bibinfo {volume} {18}},\
  \bibinfo {pages} {1602} (\bibinfo {year} {1978})}\BibitemShut {NoStop}%
\bibitem [{\citenamefont {Misra}\ and\ \citenamefont
  {Sarkar}(1983)}]{Misra:1982mg}%
  \BibitemOpen
  \bibfield  {author} {\bibinfo {author} {\bibfnamefont {S.}~\bibnamefont
  {Misra}}\ and\ \bibinfo {author} {\bibfnamefont {U.}~\bibnamefont {Sarkar}},\
  }\bibfield  {title} {\bibinfo {title} {{$n \bar{n}$ Oscillation, $H \bar{H}$
  Oscillation and the Double Proton Decay: Are They Suppressed by Wave Function
  Effects?}},\ }\href {https://doi.org/10.1103/PhysRevD.28.249} {\bibfield
  {journal} {\bibinfo  {journal} {Phys. Rev. D}\ }\textbf {\bibinfo {volume}
  {28}},\ \bibinfo {pages} {249} (\bibinfo {year} {1983})}\BibitemShut
  {NoStop}%
\bibitem [{\citenamefont {Arnellos}\ and\ \citenamefont
  {Marciano}(1982)}]{Arnellos:1982nt}%
  \BibitemOpen
  \bibfield  {author} {\bibinfo {author} {\bibfnamefont {L.}~\bibnamefont
  {Arnellos}}\ and\ \bibinfo {author} {\bibfnamefont {W.~J.}\ \bibnamefont
  {Marciano}},\ }\bibfield  {title} {\bibinfo {title} {{Hydrogen -
  Anti-hydrogen Oscillations, Double Proton Decay and Grand Unified
  Theories}},\ }\href {https://doi.org/10.1103/PhysRevLett.48.1708} {\bibfield
  {journal} {\bibinfo  {journal} {Phys. Rev. Lett.}\ }\textbf {\bibinfo
  {volume} {48}},\ \bibinfo {pages} {1708} (\bibinfo {year}
  {1982})}\BibitemShut {NoStop}%
\bibitem [{\citenamefont {Mohapatra}\ and\ \citenamefont
  {Senjanovic}(1982)}]{Mohapatra:1982aj}%
  \BibitemOpen
  \bibfield  {author} {\bibinfo {author} {\bibfnamefont {R.~N.}\ \bibnamefont
  {Mohapatra}}\ and\ \bibinfo {author} {\bibfnamefont {G.}~\bibnamefont
  {Senjanovic}},\ }\bibfield  {title} {\bibinfo {title} {{Hydrogen -
  Anti-hydrogen Oscillations and Spontaneously Broken Global $B-L$ Symmetry}},\
  }\href {https://doi.org/10.1103/PhysRevLett.49.7} {\bibfield  {journal}
  {\bibinfo  {journal} {Phys. Rev. Lett.}\ }\textbf {\bibinfo {volume} {49}},\
  \bibinfo {pages} {7} (\bibinfo {year} {1982})}\BibitemShut {NoStop}%
\bibitem [{\citenamefont {Caswell}\ \emph {et~al.}(1983)\citenamefont
  {Caswell}, \citenamefont {Milutinovic},\ and\ \citenamefont
  {Senjanovic}}]{Caswell:1982qs}%
  \BibitemOpen
  \bibfield  {author} {\bibinfo {author} {\bibfnamefont {W.~E.}\ \bibnamefont
  {Caswell}}, \bibinfo {author} {\bibfnamefont {J.}~\bibnamefont
  {Milutinovic}},\ and\ \bibinfo {author} {\bibfnamefont {G.}~\bibnamefont
  {Senjanovic}},\ }\bibfield  {title} {\bibinfo {title} {{Matter - Antimatter
  Transition Operators: A Manual for Modeling}},\ }\href
  {https://doi.org/10.1016/0370-2693(83)91585-X} {\bibfield  {journal}
  {\bibinfo  {journal} {Phys. Lett. B}\ }\textbf {\bibinfo {volume} {122}},\
  \bibinfo {pages} {373} (\bibinfo {year} {1983})}\BibitemShut {NoStop}%
\bibitem [{\citenamefont {Mohapatra}(2009)}]{Mohapatra:2009wp}%
  \BibitemOpen
  \bibfield  {author} {\bibinfo {author} {\bibfnamefont {R.}~\bibnamefont
  {Mohapatra}},\ }\bibfield  {title} {\bibinfo {title} {{Neutron-Anti-Neutron
  Oscillation: Theory and Phenomenology}},\ }\href
  {https://doi.org/10.1088/0954-3899/36/10/104006} {\bibfield  {journal}
  {\bibinfo  {journal} {J. Phys. G}\ }\textbf {\bibinfo {volume} {36}},\
  \bibinfo {pages} {104006} (\bibinfo {year} {2009})},\ \Eprint
  {https://arxiv.org/abs/0902.0834} {arXiv:0902.0834 [hep-ph]} \BibitemShut
  {NoStop}%
\bibitem [{\citenamefont {Phillips}\ \emph {et~al.}(2016)\citenamefont
  {Phillips} \emph {et~al.}}]{Phillips:2014fgb}%
  \BibitemOpen
  \bibfield  {author} {\bibinfo {author} {\bibfnamefont {I.}~\bibnamefont
  {Phillips}, \bibfnamefont {D.G.}} \emph {et~al.},\ }\bibfield  {title}
  {\bibinfo {title} {{Neutron-Antineutron Oscillations: Theoretical Status and
  Experimental Prospects}},\ }\href
  {https://doi.org/10.1016/j.physrep.2015.11.001} {\bibfield  {journal}
  {\bibinfo  {journal} {Phys. Rept.}\ }\textbf {\bibinfo {volume} {612}},\
  \bibinfo {pages} {1} (\bibinfo {year} {2016})},\ \Eprint
  {https://arxiv.org/abs/1410.1100} {arXiv:1410.1100 [hep-ex]} \BibitemShut
  {NoStop}%
\bibitem [{\citenamefont {Grossman}\ \emph {et~al.}(2018)\citenamefont
  {Grossman}, \citenamefont {Ng},\ and\ \citenamefont
  {Ray}}]{Grossman:2018rdg}%
  \BibitemOpen
  \bibfield  {author} {\bibinfo {author} {\bibfnamefont {Y.}~\bibnamefont
  {Grossman}}, \bibinfo {author} {\bibfnamefont {W.~H.}\ \bibnamefont {Ng}},\
  and\ \bibinfo {author} {\bibfnamefont {S.}~\bibnamefont {Ray}},\ }\bibfield
  {title} {\bibinfo {title} {{Revisiting the bounds on hydrogen-antihydrogen
  oscillations from diffuse $\gamma$-ray surveys}},\ }\href
  {https://doi.org/10.1103/PhysRevD.98.035020} {\bibfield  {journal} {\bibinfo
  {journal} {Phys. Rev. D}\ }\textbf {\bibinfo {volume} {98}},\ \bibinfo
  {pages} {035020} (\bibinfo {year} {2018})},\ \Eprint
  {https://arxiv.org/abs/1806.08233} {arXiv:1806.08233 [hep-ph]} \BibitemShut
  {NoStop}%
\bibitem [{\citenamefont {Berezhiani}\ and\ \citenamefont
  {Bento}(2006{\natexlab{a}})}]{Berezhiani:2005hv}%
  \BibitemOpen
  \bibfield  {author} {\bibinfo {author} {\bibfnamefont {Z.}~\bibnamefont
  {Berezhiani}}\ and\ \bibinfo {author} {\bibfnamefont {L.}~\bibnamefont
  {Bento}},\ }\bibfield  {title} {\bibinfo {title} {{Neutron - mirror neutron
  oscillations: How fast might they be?}},\ }\href
  {https://doi.org/10.1103/PhysRevLett.96.081801} {\bibfield  {journal}
  {\bibinfo  {journal} {Phys. Rev. Lett.}\ }\textbf {\bibinfo {volume} {96}},\
  \bibinfo {pages} {081801} (\bibinfo {year} {2006}{\natexlab{a}})},\ \Eprint
  {https://arxiv.org/abs/hep-ph/0507031} {arXiv:hep-ph/0507031} \BibitemShut
  {NoStop}%
\bibitem [{\citenamefont {Berezhiani}\ and\ \citenamefont
  {Bento}(2006{\natexlab{b}})}]{Berezhiani:2006je}%
  \BibitemOpen
  \bibfield  {author} {\bibinfo {author} {\bibfnamefont {Z.}~\bibnamefont
  {Berezhiani}}\ and\ \bibinfo {author} {\bibfnamefont {L.}~\bibnamefont
  {Bento}},\ }\bibfield  {title} {\bibinfo {title} {{Fast neutron: Mirror
  neutron oscillation and ultra high energy cosmic rays}},\ }\href
  {https://doi.org/10.1016/j.physletb.2006.03.008} {\bibfield  {journal}
  {\bibinfo  {journal} {Phys. Lett. B}\ }\textbf {\bibinfo {volume} {635}},\
  \bibinfo {pages} {253} (\bibinfo {year} {2006}{\natexlab{b}})},\ \Eprint
  {https://arxiv.org/abs/hep-ph/0602227} {arXiv:hep-ph/0602227} \BibitemShut
  {NoStop}%
\bibitem [{\citenamefont {Berezhiani}(2009)}]{Berezhiani:2008bc}%
  \BibitemOpen
  \bibfield  {author} {\bibinfo {author} {\bibfnamefont {Z.}~\bibnamefont
  {Berezhiani}},\ }\bibfield  {title} {\bibinfo {title} {{More about neutron -
  mirror neutron oscillation}},\ }\href
  {https://doi.org/10.1140/epjc/s10052-009-1165-1} {\bibfield  {journal}
  {\bibinfo  {journal} {Eur. Phys. J. C}\ }\textbf {\bibinfo {volume} {64}},\
  \bibinfo {pages} {421} (\bibinfo {year} {2009})},\ \Eprint
  {https://arxiv.org/abs/0804.2088} {arXiv:0804.2088 [hep-ph]} \BibitemShut
  {NoStop}%
\bibitem [{\citenamefont {Berezhiani}\ \emph {et~al.}(2017)\citenamefont
  {Berezhiani}, \citenamefont {Frost}, \citenamefont {Kamyshkov}, \citenamefont
  {Rybolt},\ and\ \citenamefont {Varriano}}]{Berezhiani:2017azg}%
  \BibitemOpen
  \bibfield  {author} {\bibinfo {author} {\bibfnamefont {Z.}~\bibnamefont
  {Berezhiani}}, \bibinfo {author} {\bibfnamefont {M.}~\bibnamefont {Frost}},
  \bibinfo {author} {\bibfnamefont {Y.}~\bibnamefont {Kamyshkov}}, \bibinfo
  {author} {\bibfnamefont {B.}~\bibnamefont {Rybolt}},\ and\ \bibinfo {author}
  {\bibfnamefont {L.}~\bibnamefont {Varriano}},\ }\bibfield  {title} {\bibinfo
  {title} {{Neutron Disappearance and Regeneration from Mirror State}},\ }\href
  {https://doi.org/10.1103/PhysRevD.96.035039} {\bibfield  {journal} {\bibinfo
  {journal} {Phys. Rev. D}\ }\textbf {\bibinfo {volume} {96}},\ \bibinfo
  {pages} {035039} (\bibinfo {year} {2017})},\ \Eprint
  {https://arxiv.org/abs/1703.06735} {arXiv:1703.06735 [hep-ex]} \BibitemShut
  {NoStop}%
\bibitem [{\citenamefont {Berezhiani}\ \emph {et~al.}(2018)\citenamefont
  {Berezhiani}, \citenamefont {Biondi}, \citenamefont {Geltenbort},
  \citenamefont {Krasnoshchekova}, \citenamefont {Varlamov}, \citenamefont
  {Vassiljev},\ and\ \citenamefont {Zherebtsov}}]{Berezhiani:2017jkn}%
  \BibitemOpen
  \bibfield  {author} {\bibinfo {author} {\bibfnamefont {Z.}~\bibnamefont
  {Berezhiani}}, \bibinfo {author} {\bibfnamefont {R.}~\bibnamefont {Biondi}},
  \bibinfo {author} {\bibfnamefont {P.}~\bibnamefont {Geltenbort}}, \bibinfo
  {author} {\bibfnamefont {I.}~\bibnamefont {Krasnoshchekova}}, \bibinfo
  {author} {\bibfnamefont {V.}~\bibnamefont {Varlamov}}, \bibinfo {author}
  {\bibfnamefont {A.}~\bibnamefont {Vassiljev}},\ and\ \bibinfo {author}
  {\bibfnamefont {O.}~\bibnamefont {Zherebtsov}},\ }\bibfield  {title}
  {\bibinfo {title} {{New experimental limits on neutron - mirror neutron
  oscillations in the presence of mirror magnetic field}},\ }\href
  {https://doi.org/10.1140/epjc/s10052-018-6189-y} {\bibfield  {journal}
  {\bibinfo  {journal} {Eur. Phys. J. C}\ }\textbf {\bibinfo {volume} {78}},\
  \bibinfo {pages} {717} (\bibinfo {year} {2018})},\ \Eprint
  {https://arxiv.org/abs/1712.05761} {arXiv:1712.05761 [hep-ex]} \BibitemShut
  {NoStop}%
\bibitem [{\citenamefont {Berezhiani}(2019)}]{Berezhiani:2018eds}%
  \BibitemOpen
  \bibfield  {author} {\bibinfo {author} {\bibfnamefont {Z.}~\bibnamefont
  {Berezhiani}},\ }\bibfield  {title} {\bibinfo {title} {{Neutron lifetime
  puzzle and neutron\textendash{}mirror neutron oscillation}},\ }\href
  {https://doi.org/10.1140/epjc/s10052-019-6995-x} {\bibfield  {journal}
  {\bibinfo  {journal} {Eur. Phys. J. C}\ }\textbf {\bibinfo {volume} {79}},\
  \bibinfo {pages} {484} (\bibinfo {year} {2019})},\ \Eprint
  {https://arxiv.org/abs/1807.07906} {arXiv:1807.07906 [hep-ph]} \BibitemShut
  {NoStop}%
\bibitem [{\citenamefont {Ban}\ \emph {et~al.}(2007)\citenamefont {Ban} \emph
  {et~al.}}]{Ban:2007tp}%
  \BibitemOpen
  \bibfield  {author} {\bibinfo {author} {\bibfnamefont {G.}~\bibnamefont
  {Ban}} \emph {et~al.},\ }\bibfield  {title} {\bibinfo {title} {{A Direct
  experimental limit on neutron: Mirror neutron oscillations}},\ }\href
  {https://doi.org/10.1103/PhysRevLett.99.161603} {\bibfield  {journal}
  {\bibinfo  {journal} {Phys. Rev. Lett.}\ }\textbf {\bibinfo {volume} {99}},\
  \bibinfo {pages} {161603} (\bibinfo {year} {2007})},\ \Eprint
  {https://arxiv.org/abs/0705.2336} {arXiv:0705.2336 [nucl-ex]} \BibitemShut
  {NoStop}%
\bibitem [{\citenamefont {Altarev}\ \emph {et~al.}(2009)\citenamefont {Altarev}
  \emph {et~al.}}]{Altarev:2009tg}%
  \BibitemOpen
  \bibfield  {author} {\bibinfo {author} {\bibfnamefont {I.}~\bibnamefont
  {Altarev}} \emph {et~al.},\ }\bibfield  {title} {\bibinfo {title} {{Neutron
  to Mirror-Neutron Oscillations in the Presence of Mirror Magnetic Fields}},\
  }\href {https://doi.org/10.1103/PhysRevD.80.032003} {\bibfield  {journal}
  {\bibinfo  {journal} {Phys. Rev. D}\ }\textbf {\bibinfo {volume} {80}},\
  \bibinfo {pages} {032003} (\bibinfo {year} {2009})},\ \Eprint
  {https://arxiv.org/abs/0905.4208} {arXiv:0905.4208 [nucl-ex]} \BibitemShut
  {NoStop}%
\bibitem [{\citenamefont {Serebrov}\ \emph {et~al.}(2008)\citenamefont
  {Serebrov} \emph {et~al.}}]{Serebrov:2007gw}%
  \BibitemOpen
  \bibfield  {author} {\bibinfo {author} {\bibfnamefont {A.}~\bibnamefont
  {Serebrov}} \emph {et~al.},\ }\bibfield  {title} {\bibinfo {title}
  {{Experimental search for neutron: Mirror neutron oscillations using storage
  of ultracold neutrons}},\ }\href
  {https://doi.org/10.1016/j.physletb.2008.04.014} {\bibfield  {journal}
  {\bibinfo  {journal} {Phys. Lett. B}\ }\textbf {\bibinfo {volume} {663}},\
  \bibinfo {pages} {181} (\bibinfo {year} {2008})},\ \Eprint
  {https://arxiv.org/abs/0706.3600} {arXiv:0706.3600 [nucl-ex]} \BibitemShut
  {NoStop}%
\bibitem [{\citenamefont {Mohapatra}\ and\ \citenamefont
  {Nussinov}(2018)}]{Mohapatra:2017lqw}%
  \BibitemOpen
  \bibfield  {author} {\bibinfo {author} {\bibfnamefont {R.~N.}\ \bibnamefont
  {Mohapatra}}\ and\ \bibinfo {author} {\bibfnamefont {S.}~\bibnamefont
  {Nussinov}},\ }\bibfield  {title} {\bibinfo {title} {{Constraints on Mirror
  Models of Dark Matter from Observable Neutron-Mirror Neutron Oscillation}},\
  }\href {https://doi.org/10.1016/j.physletb.2017.11.022} {\bibfield  {journal}
  {\bibinfo  {journal} {Phys. Lett. B}\ }\textbf {\bibinfo {volume} {776}},\
  \bibinfo {pages} {22} (\bibinfo {year} {2018})},\ \Eprint
  {https://arxiv.org/abs/1709.01637} {arXiv:1709.01637 [hep-ph]} \BibitemShut
  {NoStop}%
\bibitem [{\citenamefont {Burdman}\ \emph {et~al.}(2015)\citenamefont
  {Burdman}, \citenamefont {Chacko}, \citenamefont {Harnik}, \citenamefont
  {de~Lima},\ and\ \citenamefont {Verhaaren}}]{Burdman:2014zta}%
  \BibitemOpen
  \bibfield  {author} {\bibinfo {author} {\bibfnamefont {G.}~\bibnamefont
  {Burdman}}, \bibinfo {author} {\bibfnamefont {Z.}~\bibnamefont {Chacko}},
  \bibinfo {author} {\bibfnamefont {R.}~\bibnamefont {Harnik}}, \bibinfo
  {author} {\bibfnamefont {L.}~\bibnamefont {de~Lima}},\ and\ \bibinfo {author}
  {\bibfnamefont {C.~B.}\ \bibnamefont {Verhaaren}},\ }\bibfield  {title}
  {\bibinfo {title} {{Colorless Top Partners, a 125 GeV Higgs, and the Limits
  on Naturalness}},\ }\href {https://doi.org/10.1103/PhysRevD.91.055007}
  {\bibfield  {journal} {\bibinfo  {journal} {Phys. Rev. D}\ }\textbf {\bibinfo
  {volume} {91}},\ \bibinfo {pages} {055007} (\bibinfo {year} {2015})},\
  \Eprint {https://arxiv.org/abs/1411.3310} {arXiv:1411.3310 [hep-ph]}
  \BibitemShut {NoStop}%
\bibitem [{\citenamefont {Craig}\ \emph {et~al.}(2015)\citenamefont {Craig},
  \citenamefont {Katz}, \citenamefont {Strassler},\ and\ \citenamefont
  {Sundrum}}]{Craig:2015pha}%
  \BibitemOpen
  \bibfield  {author} {\bibinfo {author} {\bibfnamefont {N.}~\bibnamefont
  {Craig}}, \bibinfo {author} {\bibfnamefont {A.}~\bibnamefont {Katz}},
  \bibinfo {author} {\bibfnamefont {M.}~\bibnamefont {Strassler}},\ and\
  \bibinfo {author} {\bibfnamefont {R.}~\bibnamefont {Sundrum}},\ }\bibfield
  {title} {\bibinfo {title} {{Naturalness in the Dark at the LHC}},\ }\href
  {https://doi.org/10.1007/JHEP07(2015)105} {\bibfield  {journal} {\bibinfo
  {journal} {JHEP}\ }\textbf {\bibinfo {volume} {07}},\ \bibinfo {pages}
  {105}},\ \Eprint {https://arxiv.org/abs/1501.05310} {arXiv:1501.05310
  [hep-ph]} \BibitemShut {NoStop}%
\bibitem [{\citenamefont {Cs{\'a}ki}\ \emph {et~al.}(2019)\citenamefont
  {Cs{\'a}ki}, \citenamefont {Guan}, \citenamefont {Ma},\ and\ \citenamefont
  {Shu}}]{Csaki:2019qgb}%
  \BibitemOpen
  \bibfield  {author} {\bibinfo {author} {\bibfnamefont {C.}~\bibnamefont
  {Cs{\'a}ki}}, \bibinfo {author} {\bibfnamefont {C.-S.}\ \bibnamefont {Guan}},
  \bibinfo {author} {\bibfnamefont {T.}~\bibnamefont {Ma}},\ and\ \bibinfo
  {author} {\bibfnamefont {J.}~\bibnamefont {Shu}},\ }\bibfield  {title}
  {\bibinfo {title} {{Twin Higgs with Exact $Z_2$}},\ }\href@noop {} {\
  (\bibinfo {year} {2019})},\ \Eprint {https://arxiv.org/abs/1910.14085}
  {arXiv:1910.14085 [hep-ph]} \BibitemShut {NoStop}%
\bibitem [{\citenamefont {Craig}\ \emph {et~al.}(2017)\citenamefont {Craig},
  \citenamefont {Koren},\ and\ \citenamefont {Trott}}]{Craig:2016lyx}%
  \BibitemOpen
  \bibfield  {author} {\bibinfo {author} {\bibfnamefont {N.}~\bibnamefont
  {Craig}}, \bibinfo {author} {\bibfnamefont {S.}~\bibnamefont {Koren}},\ and\
  \bibinfo {author} {\bibfnamefont {T.}~\bibnamefont {Trott}},\ }\bibfield
  {title} {\bibinfo {title} {{Cosmological Signals of a Mirror Twin Higgs}},\
  }\href {https://doi.org/10.1007/JHEP05(2017)038} {\bibfield  {journal}
  {\bibinfo  {journal} {JHEP}\ }\textbf {\bibinfo {volume} {05}},\ \bibinfo
  {pages} {038}},\ \Eprint {https://arxiv.org/abs/1611.07977} {arXiv:1611.07977
  [hep-ph]} \BibitemShut {NoStop}%
\bibitem [{\citenamefont {Koren}\ and\ \citenamefont
  {McGehee}(2020)}]{Koren:2019iuv}%
  \BibitemOpen
  \bibfield  {author} {\bibinfo {author} {\bibfnamefont {S.}~\bibnamefont
  {Koren}}\ and\ \bibinfo {author} {\bibfnamefont {R.}~\bibnamefont
  {McGehee}},\ }\bibfield  {title} {\bibinfo {title} {{Freezing-in twin dark
  matter}},\ }\href {https://doi.org/10.1103/PhysRevD.101.055024} {\bibfield
  {journal} {\bibinfo  {journal} {Phys. Rev. D}\ }\textbf {\bibinfo {volume}
  {101}},\ \bibinfo {pages} {055024} (\bibinfo {year} {2020})},\ \Eprint
  {https://arxiv.org/abs/1908.03559} {arXiv:1908.03559 [hep-ph]} \BibitemShut
  {NoStop}%
\bibitem [{\citenamefont {Chacko}\ \emph {et~al.}(2017)\citenamefont {Chacko},
  \citenamefont {Craig}, \citenamefont {Fox},\ and\ \citenamefont
  {Harnik}}]{Chacko:2016hvu}%
  \BibitemOpen
  \bibfield  {author} {\bibinfo {author} {\bibfnamefont {Z.}~\bibnamefont
  {Chacko}}, \bibinfo {author} {\bibfnamefont {N.}~\bibnamefont {Craig}},
  \bibinfo {author} {\bibfnamefont {P.~J.}\ \bibnamefont {Fox}},\ and\ \bibinfo
  {author} {\bibfnamefont {R.}~\bibnamefont {Harnik}},\ }\bibfield  {title}
  {\bibinfo {title} {{Cosmology in Mirror Twin Higgs and Neutrino Masses}},\
  }\href {https://doi.org/10.1007/JHEP07(2017)023} {\bibfield  {journal}
  {\bibinfo  {journal} {JHEP}\ }\textbf {\bibinfo {volume} {07}},\ \bibinfo
  {pages} {023}},\ \Eprint {https://arxiv.org/abs/1611.07975} {arXiv:1611.07975
  [hep-ph]} \BibitemShut {NoStop}%
\bibitem [{\citenamefont {Koren}(2020)}]{Koren:2020biu}%
  \BibitemOpen
  \bibfield  {author} {\bibinfo {author} {\bibfnamefont {S.}~\bibnamefont
  {Koren}},\ }\bibfield  {title} {\bibinfo {title} {{The Hierarchy Problem:
  From the Fundamentals to the Frontiers}},\ }\href@noop {} {\bibfield
  {journal} {\bibinfo  {journal} {PhD thesis}\ } (\bibinfo {year} {2020})},\
  \Eprint {https://arxiv.org/abs/2009.11870} {arXiv:2009.11870 [hep-ph]}
  \BibitemShut {NoStop}%
\bibitem [{\citenamefont {Frampton}(1992)}]{Frampton:1991ay}%
  \BibitemOpen
  \bibfield  {author} {\bibinfo {author} {\bibfnamefont {P.~H.}\ \bibnamefont
  {Frampton}},\ }\bibfield  {title} {\bibinfo {title} {{Light leptoquarks as
  possible signature of strong electroweak unification}},\ }\href
  {https://doi.org/10.1142/S0217732392000525} {\bibfield  {journal} {\bibinfo
  {journal} {Mod. Phys. Lett. A}\ }\textbf {\bibinfo {volume} {7}},\ \bibinfo
  {pages} {559} (\bibinfo {year} {1992})}\BibitemShut {NoStop}%
\bibitem [{\citenamefont {Hall}\ and\ \citenamefont
  {Suzuki}(1984)}]{Hall:1983id}%
  \BibitemOpen
  \bibfield  {author} {\bibinfo {author} {\bibfnamefont {L.~J.}\ \bibnamefont
  {Hall}}\ and\ \bibinfo {author} {\bibfnamefont {M.}~\bibnamefont {Suzuki}},\
  }\bibfield  {title} {\bibinfo {title} {{Explicit R-Parity Breaking in
  Supersymmetric Models}},\ }\href
  {https://doi.org/10.1016/0550-3213(84)90513-3} {\bibfield  {journal}
  {\bibinfo  {journal} {Nucl. Phys. B}\ }\textbf {\bibinfo {volume} {231}},\
  \bibinfo {pages} {419} (\bibinfo {year} {1984})}\BibitemShut {NoStop}%
\bibitem [{\citenamefont {Zwirner}(1983)}]{Zwirner:1984is}%
  \BibitemOpen
  \bibfield  {author} {\bibinfo {author} {\bibfnamefont {F.}~\bibnamefont
  {Zwirner}},\ }\bibfield  {title} {\bibinfo {title} {{Observable Delta B=2
  Transitions Without Nucleon Decay in a Minimal Supersymmetric Extension of
  the Standard Model}},\ }\href {https://doi.org/10.1016/0370-2693(83)90230-7}
  {\bibfield  {journal} {\bibinfo  {journal} {Phys. Lett. B}\ }\textbf
  {\bibinfo {volume} {132}},\ \bibinfo {pages} {103} (\bibinfo {year}
  {1983})}\BibitemShut {NoStop}%
\bibitem [{\citenamefont {Dawson}(1985)}]{Dawson:1985vr}%
  \BibitemOpen
  \bibfield  {author} {\bibinfo {author} {\bibfnamefont {S.}~\bibnamefont
  {Dawson}},\ }\bibfield  {title} {\bibinfo {title} {{R-Parity Breaking in
  Supersymmetric Theories}},\ }\href
  {https://doi.org/10.1016/0550-3213(85)90577-2} {\bibfield  {journal}
  {\bibinfo  {journal} {Nucl. Phys. B}\ }\textbf {\bibinfo {volume} {261}},\
  \bibinfo {pages} {297} (\bibinfo {year} {1985})}\BibitemShut {NoStop}%
\bibitem [{\citenamefont {Schrempp}\ and\ \citenamefont
  {Schrempp}(1985)}]{Schrempp:1984nj}%
  \BibitemOpen
  \bibfield  {author} {\bibinfo {author} {\bibfnamefont {B.}~\bibnamefont
  {Schrempp}}\ and\ \bibinfo {author} {\bibfnamefont {F.}~\bibnamefont
  {Schrempp}},\ }\bibfield  {title} {\bibinfo {title} {{Light Leptoquarks}},\
  }\href {https://doi.org/10.1016/0370-2693(85)91450-9} {\bibfield  {journal}
  {\bibinfo  {journal} {Phys. Lett. B}\ }\textbf {\bibinfo {volume} {153}},\
  \bibinfo {pages} {101} (\bibinfo {year} {1985})}\BibitemShut {NoStop}%
\bibitem [{\citenamefont {Gripaios}\ \emph {et~al.}(2015)\citenamefont
  {Gripaios}, \citenamefont {Nardecchia},\ and\ \citenamefont
  {Renner}}]{Gripaios:2014tna}%
  \BibitemOpen
  \bibfield  {author} {\bibinfo {author} {\bibfnamefont {B.}~\bibnamefont
  {Gripaios}}, \bibinfo {author} {\bibfnamefont {M.}~\bibnamefont
  {Nardecchia}},\ and\ \bibinfo {author} {\bibfnamefont {S.}~\bibnamefont
  {Renner}},\ }\bibfield  {title} {\bibinfo {title} {{Composite leptoquarks and
  anomalies in $B$-meson decays}},\ }\href
  {https://doi.org/10.1007/JHEP05(2015)006} {\bibfield  {journal} {\bibinfo
  {journal} {JHEP}\ }\textbf {\bibinfo {volume} {05}},\ \bibinfo {pages}
  {006}},\ \Eprint {https://arxiv.org/abs/1412.1791} {arXiv:1412.1791 [hep-ph]}
  \BibitemShut {NoStop}%
\bibitem [{\citenamefont {de~Blas}\ \emph {et~al.}(2018)\citenamefont
  {de~Blas}, \citenamefont {Criado}, \citenamefont {Perez-Victoria},\ and\
  \citenamefont {Santiago}}]{deBlas:2017xtg}%
  \BibitemOpen
  \bibfield  {author} {\bibinfo {author} {\bibfnamefont {J.}~\bibnamefont
  {de~Blas}}, \bibinfo {author} {\bibfnamefont {J.}~\bibnamefont {Criado}},
  \bibinfo {author} {\bibfnamefont {M.}~\bibnamefont {Perez-Victoria}},\ and\
  \bibinfo {author} {\bibfnamefont {J.}~\bibnamefont {Santiago}},\ }\bibfield
  {title} {\bibinfo {title} {{Effective description of general extensions of
  the Standard Model: the complete tree-level dictionary}},\ }\href
  {https://doi.org/10.1007/JHEP03(2018)109} {\bibfield  {journal} {\bibinfo
  {journal} {JHEP}\ }\textbf {\bibinfo {volume} {03}},\ \bibinfo {pages}
  {109}},\ \Eprint {https://arxiv.org/abs/1711.10391} {arXiv:1711.10391
  [hep-ph]} \BibitemShut {NoStop}%
\bibitem [{\citenamefont {Aad}\ \emph {et~al.}(2020{\natexlab{a}})\citenamefont
  {Aad} \emph {et~al.}}]{Aad:2020iuy}%
  \BibitemOpen
  \bibfield  {author} {\bibinfo {author} {\bibfnamefont {G.}~\bibnamefont
  {Aad}} \emph {et~al.} (\bibinfo {collaboration} {ATLAS}),\ }\bibfield
  {title} {\bibinfo {title} {{Search for pairs of scalar leptoquarks decaying
  into quarks and electrons or muons in $\sqrt{s}=13$ TeV pp collisions with
  the ATLAS detector}},\ }\href@noop {} {\  (\bibinfo {year}
  {2020}{\natexlab{a}})},\ \Eprint {https://arxiv.org/abs/2006.05872}
  {arXiv:2006.05872 [hep-ex]} \BibitemShut {NoStop}%
\bibitem [{\citenamefont {Sirunyan}\ \emph {et~al.}(2019)\citenamefont
  {Sirunyan} \emph {et~al.}}]{Sirunyan:2018btu}%
  \BibitemOpen
  \bibfield  {author} {\bibinfo {author} {\bibfnamefont {A.~M.}\ \bibnamefont
  {Sirunyan}} \emph {et~al.} (\bibinfo {collaboration} {CMS}),\ }\bibfield
  {title} {\bibinfo {title} {{Search for pair production of first-generation
  scalar leptoquarks at $\sqrt{s} =$ 13 TeV}},\ }\href
  {https://doi.org/10.1103/PhysRevD.99.052002} {\bibfield  {journal} {\bibinfo
  {journal} {Phys. Rev. D}\ }\textbf {\bibinfo {volume} {99}},\ \bibinfo
  {pages} {052002} (\bibinfo {year} {2019})},\ \Eprint
  {https://arxiv.org/abs/1811.01197} {arXiv:1811.01197 [hep-ex]} \BibitemShut
  {NoStop}%
\bibitem [{\citenamefont {Aaboud}\ \emph {et~al.}(2019)\citenamefont {Aaboud}
  \emph {et~al.}}]{Aaboud:2019jcc}%
  \BibitemOpen
  \bibfield  {author} {\bibinfo {author} {\bibfnamefont {M.}~\bibnamefont
  {Aaboud}} \emph {et~al.} (\bibinfo {collaboration} {ATLAS}),\ }\bibfield
  {title} {\bibinfo {title} {{Searches for scalar leptoquarks and differential
  cross-section measurements in dilepton-dijet events in proton-proton
  collisions at a centre-of-mass energy of $\sqrt{s}$ = 13 TeV with the ATLAS
  experiment}},\ }\href {https://doi.org/10.1140/epjc/s10052-019-7181-x}
  {\bibfield  {journal} {\bibinfo  {journal} {Eur. Phys. J. C}\ }\textbf
  {\bibinfo {volume} {79}},\ \bibinfo {pages} {733} (\bibinfo {year} {2019})},\
  \Eprint {https://arxiv.org/abs/1902.00377} {arXiv:1902.00377 [hep-ex]}
  \BibitemShut {NoStop}%
\bibitem [{\citenamefont {Greljo}\ and\ \citenamefont
  {Selimovic}(2020)}]{Greljo:2020nud}%
  \BibitemOpen
  \bibfield  {author} {\bibinfo {author} {\bibfnamefont {A.}~\bibnamefont
  {Greljo}}\ and\ \bibinfo {author} {\bibfnamefont {N.}~\bibnamefont
  {Selimovic}},\ }\bibfield  {title} {\bibinfo {title} {{Lepton-Quark Fusion at
  Hadron Colliders, precisely}},\ }\href@noop {} {\  (\bibinfo {year}
  {2020})},\ \Eprint {https://arxiv.org/abs/2012.02092} {arXiv:2012.02092
  [hep-ph]} \BibitemShut {NoStop}%
\bibitem [{\citenamefont {Aad}\ \emph {et~al.}(2020{\natexlab{b}})\citenamefont
  {Aad} \emph {et~al.}}]{Aad:2019hjw}%
  \BibitemOpen
  \bibfield  {author} {\bibinfo {author} {\bibfnamefont {G.}~\bibnamefont
  {Aad}} \emph {et~al.} (\bibinfo {collaboration} {ATLAS}),\ }\bibfield
  {title} {\bibinfo {title} {{Search for new resonances in mass distributions
  of jet pairs using 139 fb$^{-1}$ of $pp$ collisions at $\sqrt{s}=13$ TeV with
  the ATLAS detector}},\ }\href {https://doi.org/10.1007/JHEP03(2020)145}
  {\bibfield  {journal} {\bibinfo  {journal} {JHEP}\ }\textbf {\bibinfo
  {volume} {03}},\ \bibinfo {pages} {145}},\ \Eprint
  {https://arxiv.org/abs/1910.08447} {arXiv:1910.08447 [hep-ex]} \BibitemShut
  {NoStop}%
\bibitem [{\citenamefont {Sirunyan}\ \emph {et~al.}(2018)\citenamefont
  {Sirunyan} \emph {et~al.}}]{Sirunyan:2018xlo}%
  \BibitemOpen
  \bibfield  {author} {\bibinfo {author} {\bibfnamefont {A.~M.}\ \bibnamefont
  {Sirunyan}} \emph {et~al.} (\bibinfo {collaboration} {CMS}),\ }\bibfield
  {title} {\bibinfo {title} {{Search for narrow and broad dijet resonances in
  proton-proton collisions at $ \sqrt{s}=13 $ TeV and constraints on dark
  matter mediators and other new particles}},\ }\href
  {https://doi.org/10.1007/JHEP08(2018)130} {\bibfield  {journal} {\bibinfo
  {journal} {JHEP}\ }\textbf {\bibinfo {volume} {08}},\ \bibinfo {pages}
  {130}},\ \Eprint {https://arxiv.org/abs/1806.00843} {arXiv:1806.00843
  [hep-ex]} \BibitemShut {NoStop}%
\bibitem [{\citenamefont {Ciarcelluti}(2010)}]{Ciarcelluti:2010zz}%
  \BibitemOpen
  \bibfield  {author} {\bibinfo {author} {\bibfnamefont {P.}~\bibnamefont
  {Ciarcelluti}},\ }\bibfield  {title} {\bibinfo {title} {{Cosmology with
  mirror dark matter}},\ }\href {https://doi.org/10.1142/S0218271810018438}
  {\bibfield  {journal} {\bibinfo  {journal} {Int. J. Mod. Phys. D}\ }\textbf
  {\bibinfo {volume} {19}},\ \bibinfo {pages} {2151} (\bibinfo {year}
  {2010})},\ \Eprint {https://arxiv.org/abs/1102.5530} {arXiv:1102.5530
  [astro-ph.CO]} \BibitemShut {NoStop}%
\bibitem [{\citenamefont {Foot}(2014)}]{Foot:2014mia}%
  \BibitemOpen
  \bibfield  {author} {\bibinfo {author} {\bibfnamefont {R.}~\bibnamefont
  {Foot}},\ }\bibfield  {title} {\bibinfo {title} {{Mirror dark matter:
  Cosmology, galaxy structure and direct detection}},\ }\href
  {https://doi.org/10.1142/S0217751X14300130} {\bibfield  {journal} {\bibinfo
  {journal} {Int. J. Mod. Phys. A}\ }\textbf {\bibinfo {volume} {29}},\
  \bibinfo {pages} {1430013} (\bibinfo {year} {2014})},\ \Eprint
  {https://arxiv.org/abs/1401.3965} {arXiv:1401.3965 [astro-ph.CO]}
  \BibitemShut {NoStop}%
\bibitem [{\citenamefont {Kolb}\ \emph {et~al.}(1985)\citenamefont {Kolb},
  \citenamefont {Seckel},\ and\ \citenamefont {Turner}}]{Kolb:1985bf}%
  \BibitemOpen
  \bibfield  {author} {\bibinfo {author} {\bibfnamefont {E.~W.}\ \bibnamefont
  {Kolb}}, \bibinfo {author} {\bibfnamefont {D.}~\bibnamefont {Seckel}},\ and\
  \bibinfo {author} {\bibfnamefont {M.~S.}\ \bibnamefont {Turner}},\ }\bibfield
   {title} {\bibinfo {title} {{The Shadow World}},\ }\href
  {https://doi.org/10.1038/314415a0} {\bibfield  {journal} {\bibinfo  {journal}
  {Nature}\ }\textbf {\bibinfo {volume} {314}},\ \bibinfo {pages} {415}
  (\bibinfo {year} {1985})}\BibitemShut {NoStop}%
\bibitem [{\citenamefont {Hodges}(1993)}]{Hodges:1993yb}%
  \BibitemOpen
  \bibfield  {author} {\bibinfo {author} {\bibfnamefont {H.}~\bibnamefont
  {Hodges}},\ }\bibfield  {title} {\bibinfo {title} {{Mirror baryons as the
  dark matter}},\ }\href {https://doi.org/10.1103/PhysRevD.47.456} {\bibfield
  {journal} {\bibinfo  {journal} {Phys. Rev. D}\ }\textbf {\bibinfo {volume}
  {47}},\ \bibinfo {pages} {456} (\bibinfo {year} {1993})}\BibitemShut
  {NoStop}%
\bibitem [{\citenamefont {Berezinsky}\ and\ \citenamefont
  {Vilenkin}(2000)}]{Berezinsky:1999az}%
  \BibitemOpen
  \bibfield  {author} {\bibinfo {author} {\bibfnamefont {V.}~\bibnamefont
  {Berezinsky}}\ and\ \bibinfo {author} {\bibfnamefont {A.}~\bibnamefont
  {Vilenkin}},\ }\bibfield  {title} {\bibinfo {title} {{Ultrahigh-energy
  neutrinos from hidden sector topological defects}},\ }\href
  {https://doi.org/10.1103/PhysRevD.62.083512} {\bibfield  {journal} {\bibinfo
  {journal} {Phys. Rev. D}\ }\textbf {\bibinfo {volume} {62}},\ \bibinfo
  {pages} {083512} (\bibinfo {year} {2000})},\ \Eprint
  {https://arxiv.org/abs/hep-ph/9908257} {arXiv:hep-ph/9908257} \BibitemShut
  {NoStop}%
\bibitem [{\citenamefont {Roux}\ and\ \citenamefont
  {Cline}(2020)}]{Roux:2020wkp}%
  \BibitemOpen
  \bibfield  {author} {\bibinfo {author} {\bibfnamefont {J.-S.}\ \bibnamefont
  {Roux}}\ and\ \bibinfo {author} {\bibfnamefont {J.~M.}\ \bibnamefont
  {Cline}},\ }\bibfield  {title} {\bibinfo {title} {{Constraining galactic
  structures of mirror dark matter}},\ }\href
  {https://doi.org/10.1103/PhysRevD.102.063518} {\bibfield  {journal} {\bibinfo
   {journal} {Phys. Rev. D}\ }\textbf {\bibinfo {volume} {102}},\ \bibinfo
  {pages} {063518} (\bibinfo {year} {2020})},\ \Eprint
  {https://arxiv.org/abs/2001.11504} {arXiv:2001.11504 [astro-ph.CO]}
  \BibitemShut {NoStop}%
\bibitem [{\citenamefont {Aghanim}\ \emph {et~al.}(2020)\citenamefont {Aghanim}
  \emph {et~al.}}]{Aghanim:2018eyx}%
  \BibitemOpen
  \bibfield  {author} {\bibinfo {author} {\bibfnamefont {N.}~\bibnamefont
  {Aghanim}} \emph {et~al.} (\bibinfo {collaboration} {Planck}),\ }\bibfield
  {title} {\bibinfo {title} {{Planck 2018 results. VI. Cosmological
  parameters}},\ }\href {https://doi.org/10.1051/0004-6361/201833910}
  {\bibfield  {journal} {\bibinfo  {journal} {Astron. Astrophys.}\ }\textbf
  {\bibinfo {volume} {641}},\ \bibinfo {pages} {A6} (\bibinfo {year} {2020})},\
  \Eprint {https://arxiv.org/abs/1807.06209} {arXiv:1807.06209 [astro-ph.CO]}
  \BibitemShut {NoStop}%
\bibitem [{\citenamefont {Khlopov}\ \emph {et~al.}(1991)\citenamefont
  {Khlopov}, \citenamefont {Beskin}, \citenamefont {Bochkarev}, \citenamefont
  {Pustylnik},\ and\ \citenamefont {Pustylnik}}]{Khlopov:1989fj}%
  \BibitemOpen
  \bibfield  {author} {\bibinfo {author} {\bibfnamefont {M.}~\bibnamefont
  {Khlopov}}, \bibinfo {author} {\bibfnamefont {G.}~\bibnamefont {Beskin}},
  \bibinfo {author} {\bibfnamefont {N.}~\bibnamefont {Bochkarev}}, \bibinfo
  {author} {\bibfnamefont {L.}~\bibnamefont {Pustylnik}},\ and\ \bibinfo
  {author} {\bibfnamefont {S.}~\bibnamefont {Pustylnik}},\ }\bibfield  {title}
  {\bibinfo {title} {{Observational Physics of Mirror World}},\ }\href@noop {}
  {\bibfield  {journal} {\bibinfo  {journal} {Sov. Astron.}\ }\textbf {\bibinfo
  {volume} {35}},\ \bibinfo {pages} {21} (\bibinfo {year} {1991})}\BibitemShut
  {NoStop}%
\bibitem [{\citenamefont {Berezhiani}\ \emph
  {et~al.}(2001{\natexlab{a}})\citenamefont {Berezhiani}, \citenamefont
  {Comelli},\ and\ \citenamefont {Villante}}]{Berezhiani:2000gw}%
  \BibitemOpen
  \bibfield  {author} {\bibinfo {author} {\bibfnamefont {Z.}~\bibnamefont
  {Berezhiani}}, \bibinfo {author} {\bibfnamefont {D.}~\bibnamefont
  {Comelli}},\ and\ \bibinfo {author} {\bibfnamefont {F.~L.}\ \bibnamefont
  {Villante}},\ }\bibfield  {title} {\bibinfo {title} {{The Early mirror
  universe: Inflation, baryogenesis, nucleosynthesis and dark matter}},\ }\href
  {https://doi.org/10.1016/S0370-2693(01)00217-9} {\bibfield  {journal}
  {\bibinfo  {journal} {Phys. Lett. B}\ }\textbf {\bibinfo {volume} {503}},\
  \bibinfo {pages} {362} (\bibinfo {year} {2001}{\natexlab{a}})},\ \Eprint
  {https://arxiv.org/abs/hep-ph/0008105} {arXiv:hep-ph/0008105} \BibitemShut
  {NoStop}%
\bibitem [{\citenamefont {Berezhiani}\ \emph
  {et~al.}(2001{\natexlab{b}})\citenamefont {Berezhiani}, \citenamefont
  {Gianfagna},\ and\ \citenamefont {Giannotti}}]{Berezhiani:2000gh}%
  \BibitemOpen
  \bibfield  {author} {\bibinfo {author} {\bibfnamefont {Z.}~\bibnamefont
  {Berezhiani}}, \bibinfo {author} {\bibfnamefont {L.}~\bibnamefont
  {Gianfagna}},\ and\ \bibinfo {author} {\bibfnamefont {M.}~\bibnamefont
  {Giannotti}},\ }\bibfield  {title} {\bibinfo {title} {{Strong CP problem and
  mirror world: The Weinberg-Wilczek axion revisited}},\ }\href
  {https://doi.org/10.1016/S0370-2693(00)01392-7} {\bibfield  {journal}
  {\bibinfo  {journal} {Phys. Lett. B}\ }\textbf {\bibinfo {volume} {500}},\
  \bibinfo {pages} {286} (\bibinfo {year} {2001}{\natexlab{b}})},\ \Eprint
  {https://arxiv.org/abs/hep-ph/0009290} {arXiv:hep-ph/0009290} \BibitemShut
  {NoStop}%
\bibitem [{\citenamefont {Berezhiani}\ \emph {et~al.}(2005)\citenamefont
  {Berezhiani}, \citenamefont {Ciarcelluti}, \citenamefont {Comelli},\ and\
  \citenamefont {Villante}}]{Berezhiani:2003wj}%
  \BibitemOpen
  \bibfield  {author} {\bibinfo {author} {\bibfnamefont {Z.}~\bibnamefont
  {Berezhiani}}, \bibinfo {author} {\bibfnamefont {P.}~\bibnamefont
  {Ciarcelluti}}, \bibinfo {author} {\bibfnamefont {D.}~\bibnamefont
  {Comelli}},\ and\ \bibinfo {author} {\bibfnamefont {F.~L.}\ \bibnamefont
  {Villante}},\ }\bibfield  {title} {\bibinfo {title} {{Structure formation
  with mirror dark matter: CMB and LSS}},\ }\href
  {https://doi.org/10.1142/S0218271805005165} {\bibfield  {journal} {\bibinfo
  {journal} {Int. J. Mod. Phys. D}\ }\textbf {\bibinfo {volume} {14}},\
  \bibinfo {pages} {107} (\bibinfo {year} {2005})},\ \Eprint
  {https://arxiv.org/abs/astro-ph/0312605} {arXiv:astro-ph/0312605}
  \BibitemShut {NoStop}%
\bibitem [{\citenamefont {Ciarcelluti}(2003)}]{Ciarcelluti:2003wm}%
  \BibitemOpen
  \bibfield  {author} {\bibinfo {author} {\bibfnamefont {P.}~\bibnamefont
  {Ciarcelluti}},\ }\emph {\bibinfo {title} {{Cosmology of the mirror
  universe}}},\ \href@noop {} {\bibinfo {type} {Other thesis}} (\bibinfo {year}
  {2003}),\ \Eprint {https://arxiv.org/abs/astro-ph/0312607}
  {arXiv:astro-ph/0312607} \BibitemShut {NoStop}%
\bibitem [{\citenamefont {Ignatiev}\ and\ \citenamefont
  {Volkas}(2003)}]{Ignatiev:2003js}%
  \BibitemOpen
  \bibfield  {author} {\bibinfo {author} {\bibfnamefont {A.}~\bibnamefont
  {Ignatiev}}\ and\ \bibinfo {author} {\bibfnamefont {R.}~\bibnamefont
  {Volkas}},\ }\bibfield  {title} {\bibinfo {title} {{Mirror dark matter and
  large scale structure}},\ }\href {https://doi.org/10.1103/PhysRevD.68.023518}
  {\bibfield  {journal} {\bibinfo  {journal} {Phys. Rev. D}\ }\textbf {\bibinfo
  {volume} {68}},\ \bibinfo {pages} {023518} (\bibinfo {year} {2003})},\
  \Eprint {https://arxiv.org/abs/hep-ph/0304260} {arXiv:hep-ph/0304260}
  \BibitemShut {NoStop}%
\bibitem [{\citenamefont
  {Ciarcelluti}(2005{\natexlab{a}})}]{Ciarcelluti:2004ik}%
  \BibitemOpen
  \bibfield  {author} {\bibinfo {author} {\bibfnamefont {P.}~\bibnamefont
  {Ciarcelluti}},\ }\bibfield  {title} {\bibinfo {title} {{Cosmology with
  mirror dark matter. 1. Linear evolution of perturbations}},\ }\href
  {https://doi.org/10.1142/S0218271805006213} {\bibfield  {journal} {\bibinfo
  {journal} {Int. J. Mod. Phys. D}\ }\textbf {\bibinfo {volume} {14}},\
  \bibinfo {pages} {187} (\bibinfo {year} {2005}{\natexlab{a}})},\ \Eprint
  {https://arxiv.org/abs/astro-ph/0409630} {arXiv:astro-ph/0409630}
  \BibitemShut {NoStop}%
\bibitem [{\citenamefont
  {Ciarcelluti}(2005{\natexlab{b}})}]{Ciarcelluti:2004ip}%
  \BibitemOpen
  \bibfield  {author} {\bibinfo {author} {\bibfnamefont {P.}~\bibnamefont
  {Ciarcelluti}},\ }\bibfield  {title} {\bibinfo {title} {{Cosmology with
  mirror dark matter. 2. Cosmic microwave background and large scale
  structure}},\ }\href {https://doi.org/10.1142/S0218271805006225} {\bibfield
  {journal} {\bibinfo  {journal} {Int. J. Mod. Phys. D}\ }\textbf {\bibinfo
  {volume} {14}},\ \bibinfo {pages} {223} (\bibinfo {year}
  {2005}{\natexlab{b}})},\ \Eprint {https://arxiv.org/abs/astro-ph/0409633}
  {arXiv:astro-ph/0409633} \BibitemShut {NoStop}%
\bibitem [{\citenamefont {Ciarcelluti}\ and\ \citenamefont
  {Foot}(2009)}]{Ciarcelluti:2008qk}%
  \BibitemOpen
  \bibfield  {author} {\bibinfo {author} {\bibfnamefont {P.}~\bibnamefont
  {Ciarcelluti}}\ and\ \bibinfo {author} {\bibfnamefont {R.}~\bibnamefont
  {Foot}},\ }\bibfield  {title} {\bibinfo {title} {{Early Universe cosmology in
  the light of the mirror dark matter interpretation of the DAMA/Libra
  signal}},\ }\href {https://doi.org/10.1016/j.physletb.2009.07.052} {\bibfield
   {journal} {\bibinfo  {journal} {Phys. Lett. B}\ }\textbf {\bibinfo {volume}
  {679}},\ \bibinfo {pages} {278} (\bibinfo {year} {2009})},\ \Eprint
  {https://arxiv.org/abs/0809.4438} {arXiv:0809.4438 [astro-ph]} \BibitemShut
  {NoStop}%
\bibitem [{\citenamefont {Foot}(2012)}]{Foot:2011ve}%
  \BibitemOpen
  \bibfield  {author} {\bibinfo {author} {\bibfnamefont {R.}~\bibnamefont
  {Foot}},\ }\bibfield  {title} {\bibinfo {title} {{Mirror dark matter
  cosmology - predictions for $N_{eff} [CMB]$ and $N_{eff} [BBN]$}},\ }\href
  {https://doi.org/10.1016/j.physletb.2012.04.023} {\bibfield  {journal}
  {\bibinfo  {journal} {Phys. Lett. B}\ }\textbf {\bibinfo {volume} {711}},\
  \bibinfo {pages} {238} (\bibinfo {year} {2012})},\ \Eprint
  {https://arxiv.org/abs/1111.6366} {arXiv:1111.6366 [astro-ph.CO]}
  \BibitemShut {NoStop}%
\bibitem [{\citenamefont {Dine}\ and\ \citenamefont
  {Fischler}(1983)}]{Dine:1982ah}%
  \BibitemOpen
  \bibfield  {author} {\bibinfo {author} {\bibfnamefont {M.}~\bibnamefont
  {Dine}}\ and\ \bibinfo {author} {\bibfnamefont {W.}~\bibnamefont
  {Fischler}},\ }\bibfield  {title} {\bibinfo {title} {{The Not So Harmless
  Axion}},\ }\href {https://doi.org/10.1016/0370-2693(83)90639-1} {\bibfield
  {journal} {\bibinfo  {journal} {Phys. Lett. B}\ }\textbf {\bibinfo {volume}
  {120}},\ \bibinfo {pages} {137} (\bibinfo {year} {1983})}\BibitemShut
  {NoStop}%
\bibitem [{\citenamefont {Abbott}\ and\ \citenamefont
  {Sikivie}(1983)}]{Abbott:1982af}%
  \BibitemOpen
  \bibfield  {author} {\bibinfo {author} {\bibfnamefont {L.}~\bibnamefont
  {Abbott}}\ and\ \bibinfo {author} {\bibfnamefont {P.}~\bibnamefont
  {Sikivie}},\ }\bibfield  {title} {\bibinfo {title} {{A Cosmological Bound on
  the Invisible Axion}},\ }\href {https://doi.org/10.1016/0370-2693(83)90638-X}
  {\bibfield  {journal} {\bibinfo  {journal} {Phys. Lett. B}\ }\textbf
  {\bibinfo {volume} {120}},\ \bibinfo {pages} {133} (\bibinfo {year}
  {1983})}\BibitemShut {NoStop}%
\bibitem [{\citenamefont {Preskill}\ \emph {et~al.}(1983)\citenamefont
  {Preskill}, \citenamefont {Wise},\ and\ \citenamefont
  {Wilczek}}]{Preskill:1982cy}%
  \BibitemOpen
  \bibfield  {author} {\bibinfo {author} {\bibfnamefont {J.}~\bibnamefont
  {Preskill}}, \bibinfo {author} {\bibfnamefont {M.~B.}\ \bibnamefont {Wise}},\
  and\ \bibinfo {author} {\bibfnamefont {F.}~\bibnamefont {Wilczek}},\
  }\bibfield  {title} {\bibinfo {title} {{Cosmology of the Invisible Axion}},\
  }\href {https://doi.org/10.1016/0370-2693(83)90637-8} {\bibfield  {journal}
  {\bibinfo  {journal} {Phys. Lett. B}\ }\textbf {\bibinfo {volume} {120}},\
  \bibinfo {pages} {127} (\bibinfo {year} {1983})}\BibitemShut {NoStop}%
\bibitem [{\citenamefont {Co}\ \emph {et~al.}(2020)\citenamefont {Co},
  \citenamefont {Hall},\ and\ \citenamefont {Harigaya}}]{Co:2019jts}%
  \BibitemOpen
  \bibfield  {author} {\bibinfo {author} {\bibfnamefont {R.~T.}\ \bibnamefont
  {Co}}, \bibinfo {author} {\bibfnamefont {L.~J.}\ \bibnamefont {Hall}},\ and\
  \bibinfo {author} {\bibfnamefont {K.}~\bibnamefont {Harigaya}},\ }\bibfield
  {title} {\bibinfo {title} {{Axion Kinetic Misalignment Mechanism}},\ }\href
  {https://doi.org/10.1103/PhysRevLett.124.251802} {\bibfield  {journal}
  {\bibinfo  {journal} {Phys. Rev. Lett.}\ }\textbf {\bibinfo {volume} {124}},\
  \bibinfo {pages} {251802} (\bibinfo {year} {2020})},\ \Eprint
  {https://arxiv.org/abs/1910.14152} {arXiv:1910.14152 [hep-ph]} \BibitemShut
  {NoStop}%
\bibitem [{\citenamefont {Houston}\ \emph {et~al.}(2018)\citenamefont
  {Houston}, \citenamefont {Li}, \citenamefont {Li}, \citenamefont {Yang},\
  and\ \citenamefont {Zhang}}]{Houston:2018vrf}%
  \BibitemOpen
  \bibfield  {author} {\bibinfo {author} {\bibfnamefont {N.}~\bibnamefont
  {Houston}}, \bibinfo {author} {\bibfnamefont {C.}~\bibnamefont {Li}},
  \bibinfo {author} {\bibfnamefont {T.}~\bibnamefont {Li}}, \bibinfo {author}
  {\bibfnamefont {Q.}~\bibnamefont {Yang}},\ and\ \bibinfo {author}
  {\bibfnamefont {X.}~\bibnamefont {Zhang}},\ }\bibfield  {title} {\bibinfo
  {title} {{Natural Explanation for 21 cm Absorption Signals via Axion-Induced
  Cooling}},\ }\href {https://doi.org/10.1103/PhysRevLett.121.111301}
  {\bibfield  {journal} {\bibinfo  {journal} {Phys. Rev. Lett.}\ }\textbf
  {\bibinfo {volume} {121}},\ \bibinfo {pages} {111301} (\bibinfo {year}
  {2018})},\ \Eprint {https://arxiv.org/abs/1805.04426} {arXiv:1805.04426
  [hep-ph]} \BibitemShut {NoStop}%
\bibitem [{\citenamefont {Li}\ \emph {et~al.}(2018)\citenamefont {Li},
  \citenamefont {Houston}, \citenamefont {Li}, \citenamefont {Yang},\ and\
  \citenamefont {Zhang}}]{Houston:2018vbk}%
  \BibitemOpen
  \bibfield  {author} {\bibinfo {author} {\bibfnamefont {C.}~\bibnamefont
  {Li}}, \bibinfo {author} {\bibfnamefont {N.}~\bibnamefont {Houston}},
  \bibinfo {author} {\bibfnamefont {T.}~\bibnamefont {Li}}, \bibinfo {author}
  {\bibfnamefont {Q.}~\bibnamefont {Yang}},\ and\ \bibinfo {author}
  {\bibfnamefont {X.}~\bibnamefont {Zhang}},\ }\bibfield  {title} {\bibinfo
  {title} {{A detailed exploration of the EDGES 21 cm absorption anomaly and
  axion-induced cooling}},\ }\href@noop {} {\  (\bibinfo {year} {2018})},\
  \Eprint {https://arxiv.org/abs/1812.03931} {arXiv:1812.03931 [hep-ph]}
  \BibitemShut {NoStop}%
\bibitem [{\citenamefont {Sikivie}(2019)}]{Sikivie:2018tml}%
  \BibitemOpen
  \bibfield  {author} {\bibinfo {author} {\bibfnamefont {P.}~\bibnamefont
  {Sikivie}},\ }\bibfield  {title} {\bibinfo {title} {{Axion dark matter and
  the 21-cm signal}},\ }\href {https://doi.org/10.1016/j.dark.2019.100289}
  {\bibfield  {journal} {\bibinfo  {journal} {Phys. Dark Univ.}\ }\textbf
  {\bibinfo {volume} {24}},\ \bibinfo {pages} {100289} (\bibinfo {year}
  {2019})},\ \Eprint {https://arxiv.org/abs/1805.05577} {arXiv:1805.05577
  [astro-ph.CO]} \BibitemShut {NoStop}%
\bibitem [{\citenamefont {Lawson}\ and\ \citenamefont
  {Zhitnitsky}(2018)}]{Lawson:2018qkc}%
  \BibitemOpen
  \bibfield  {author} {\bibinfo {author} {\bibfnamefont {K.}~\bibnamefont
  {Lawson}}\ and\ \bibinfo {author} {\bibfnamefont {A.}~\bibnamefont
  {Zhitnitsky}},\ }\bibfield  {title} {\bibinfo {title} {{The 21 cm absorption
  line and the axion quark nugget dark matter model}},\ }\href
  {https://doi.org/10.1016/j.dark.2019.100295} {\bibfield  {journal} {\bibinfo
  {journal} {Phys. Dark Univ.}\ }\textbf {\bibinfo {volume} {100295}},\
  \bibinfo {pages} {2019} (\bibinfo {year} {2018})},\ \Eprint
  {https://arxiv.org/abs/1804.07340} {arXiv:1804.07340 [hep-ph]} \BibitemShut
  {NoStop}%
\bibitem [{\citenamefont {Lambiase}\ and\ \citenamefont
  {Mohanty}(2020)}]{Lambiase:2018lhs}%
  \BibitemOpen
  \bibfield  {author} {\bibinfo {author} {\bibfnamefont {G.}~\bibnamefont
  {Lambiase}}\ and\ \bibinfo {author} {\bibfnamefont {S.}~\bibnamefont
  {Mohanty}},\ }\bibfield  {title} {\bibinfo {title} {{Hydrogen spin
  oscillations in a background of axions and the 21-cm brightness
  temperature}},\ }\href {https://doi.org/10.1093/mnras/staa1070} {\bibfield
  {journal} {\bibinfo  {journal} {Mon. Not. Roy. Astron. Soc.}\ }\textbf
  {\bibinfo {volume} {494}},\ \bibinfo {pages} {5961} (\bibinfo {year}
  {2020})},\ \Eprint {https://arxiv.org/abs/1804.05318} {arXiv:1804.05318
  [hep-ph]} \BibitemShut {NoStop}%
\bibitem [{\citenamefont {Tucker-Smith}\ and\ \citenamefont
  {Weiner}(2001)}]{TuckerSmith:2001hy}%
  \BibitemOpen
  \bibfield  {author} {\bibinfo {author} {\bibfnamefont {D.}~\bibnamefont
  {Tucker-Smith}}\ and\ \bibinfo {author} {\bibfnamefont {N.}~\bibnamefont
  {Weiner}},\ }\bibfield  {title} {\bibinfo {title} {{Inelastic dark matter}},\
  }\href {https://doi.org/10.1103/PhysRevD.64.043502} {\bibfield  {journal}
  {\bibinfo  {journal} {Phys. Rev. D}\ }\textbf {\bibinfo {volume} {64}},\
  \bibinfo {pages} {043502} (\bibinfo {year} {2001})},\ \Eprint
  {https://arxiv.org/abs/hep-ph/0101138} {arXiv:hep-ph/0101138} \BibitemShut
  {NoStop}%
\bibitem [{\citenamefont {Finkbeiner}\ and\ \citenamefont
  {Weiner}(2007)}]{Finkbeiner:2007kk}%
  \BibitemOpen
  \bibfield  {author} {\bibinfo {author} {\bibfnamefont {D.~P.}\ \bibnamefont
  {Finkbeiner}}\ and\ \bibinfo {author} {\bibfnamefont {N.}~\bibnamefont
  {Weiner}},\ }\bibfield  {title} {\bibinfo {title} {{Exciting Dark Matter and
  the INTEGRAL/SPI 511 keV signal}},\ }\href
  {https://doi.org/10.1103/PhysRevD.76.083519} {\bibfield  {journal} {\bibinfo
  {journal} {Phys. Rev. D}\ }\textbf {\bibinfo {volume} {76}},\ \bibinfo
  {pages} {083519} (\bibinfo {year} {2007})},\ \Eprint
  {https://arxiv.org/abs/astro-ph/0702587} {arXiv:astro-ph/0702587}
  \BibitemShut {NoStop}%
\bibitem [{\citenamefont {Chang}\ \emph {et~al.}(2009)\citenamefont {Chang},
  \citenamefont {Kribs}, \citenamefont {Tucker-Smith},\ and\ \citenamefont
  {Weiner}}]{Chang:2008gd}%
  \BibitemOpen
  \bibfield  {author} {\bibinfo {author} {\bibfnamefont {S.}~\bibnamefont
  {Chang}}, \bibinfo {author} {\bibfnamefont {G.~D.}\ \bibnamefont {Kribs}},
  \bibinfo {author} {\bibfnamefont {D.}~\bibnamefont {Tucker-Smith}},\ and\
  \bibinfo {author} {\bibfnamefont {N.}~\bibnamefont {Weiner}},\ }\bibfield
  {title} {\bibinfo {title} {{Inelastic Dark Matter in Light of DAMA/LIBRA}},\
  }\href {https://doi.org/10.1103/PhysRevD.79.043513} {\bibfield  {journal}
  {\bibinfo  {journal} {Phys. Rev. D}\ }\textbf {\bibinfo {volume} {79}},\
  \bibinfo {pages} {043513} (\bibinfo {year} {2009})},\ \Eprint
  {https://arxiv.org/abs/0807.2250} {arXiv:0807.2250 [hep-ph]} \BibitemShut
  {NoStop}%
\bibitem [{\citenamefont {Khlopov}\ and\ \citenamefont
  {Kouvaris}(2008)}]{Khlopov:2008ty}%
  \BibitemOpen
  \bibfield  {author} {\bibinfo {author} {\bibfnamefont {M.~Y.}\ \bibnamefont
  {Khlopov}}\ and\ \bibinfo {author} {\bibfnamefont {C.}~\bibnamefont
  {Kouvaris}},\ }\bibfield  {title} {\bibinfo {title} {{Composite dark matter
  from a model with composite Higgs boson}},\ }\href
  {https://doi.org/10.1103/PhysRevD.78.065040} {\bibfield  {journal} {\bibinfo
  {journal} {Phys. Rev. D}\ }\textbf {\bibinfo {volume} {78}},\ \bibinfo
  {pages} {065040} (\bibinfo {year} {2008})},\ \Eprint
  {https://arxiv.org/abs/0806.1191} {arXiv:0806.1191 [astro-ph]} \BibitemShut
  {NoStop}%
\bibitem [{\citenamefont {Batell}\ \emph {et~al.}(2009)\citenamefont {Batell},
  \citenamefont {Pospelov},\ and\ \citenamefont {Ritz}}]{Batell:2009vb}%
  \BibitemOpen
  \bibfield  {author} {\bibinfo {author} {\bibfnamefont {B.}~\bibnamefont
  {Batell}}, \bibinfo {author} {\bibfnamefont {M.}~\bibnamefont {Pospelov}},\
  and\ \bibinfo {author} {\bibfnamefont {A.}~\bibnamefont {Ritz}},\ }\bibfield
  {title} {\bibinfo {title} {{Direct Detection of Multi-component Secluded
  WIMPs}},\ }\href {https://doi.org/10.1103/PhysRevD.79.115019} {\bibfield
  {journal} {\bibinfo  {journal} {Phys. Rev. D}\ }\textbf {\bibinfo {volume}
  {79}},\ \bibinfo {pages} {115019} (\bibinfo {year} {2009})},\ \Eprint
  {https://arxiv.org/abs/0903.3396} {arXiv:0903.3396 [hep-ph]} \BibitemShut
  {NoStop}%
\bibitem [{\citenamefont {Graham}\ \emph {et~al.}(2010)\citenamefont {Graham},
  \citenamefont {Harnik}, \citenamefont {Rajendran},\ and\ \citenamefont
  {Saraswat}}]{Graham:2010ca}%
  \BibitemOpen
  \bibfield  {author} {\bibinfo {author} {\bibfnamefont {P.~W.}\ \bibnamefont
  {Graham}}, \bibinfo {author} {\bibfnamefont {R.}~\bibnamefont {Harnik}},
  \bibinfo {author} {\bibfnamefont {S.}~\bibnamefont {Rajendran}},\ and\
  \bibinfo {author} {\bibfnamefont {P.}~\bibnamefont {Saraswat}},\ }\bibfield
  {title} {\bibinfo {title} {{Exothermic Dark Matter}},\ }\href
  {https://doi.org/10.1103/PhysRevD.82.063512} {\bibfield  {journal} {\bibinfo
  {journal} {Phys. Rev. D}\ }\textbf {\bibinfo {volume} {82}},\ \bibinfo
  {pages} {063512} (\bibinfo {year} {2010})},\ \Eprint
  {https://arxiv.org/abs/1004.0937} {arXiv:1004.0937 [hep-ph]} \BibitemShut
  {NoStop}%
\bibitem [{\citenamefont {Spier Moreira~Alves}\ \emph
  {et~al.}(2010)\citenamefont {Spier Moreira~Alves}, \citenamefont {Behbahani},
  \citenamefont {Schuster},\ and\ \citenamefont {Wacker}}]{Alves:2010dd}%
  \BibitemOpen
  \bibfield  {author} {\bibinfo {author} {\bibfnamefont {D.}~\bibnamefont
  {Spier Moreira~Alves}}, \bibinfo {author} {\bibfnamefont {S.~R.}\
  \bibnamefont {Behbahani}}, \bibinfo {author} {\bibfnamefont {P.}~\bibnamefont
  {Schuster}},\ and\ \bibinfo {author} {\bibfnamefont {J.~G.}\ \bibnamefont
  {Wacker}},\ }\bibfield  {title} {\bibinfo {title} {{The Cosmology of
  Composite Inelastic Dark Matter}},\ }\href
  {https://doi.org/10.1007/JHEP06(2010)113} {\bibfield  {journal} {\bibinfo
  {journal} {JHEP}\ }\textbf {\bibinfo {volume} {06}},\ \bibinfo {pages}
  {113}},\ \Eprint {https://arxiv.org/abs/1003.4729} {arXiv:1003.4729 [hep-ph]}
  \BibitemShut {NoStop}%
\bibitem [{\citenamefont {McCullough}\ and\ \citenamefont
  {Randall}(2013)}]{McCullough:2013jma}%
  \BibitemOpen
  \bibfield  {author} {\bibinfo {author} {\bibfnamefont {M.}~\bibnamefont
  {McCullough}}\ and\ \bibinfo {author} {\bibfnamefont {L.}~\bibnamefont
  {Randall}},\ }\bibfield  {title} {\bibinfo {title} {{Exothermic Double-Disk
  Dark Matter}},\ }\href {https://doi.org/10.1088/1475-7516/2013/10/058}
  {\bibfield  {journal} {\bibinfo  {journal} {JCAP}\ }\textbf {\bibinfo
  {volume} {10}},\ \bibinfo {pages} {058}},\ \Eprint
  {https://arxiv.org/abs/1307.4095} {arXiv:1307.4095 [hep-ph]} \BibitemShut
  {NoStop}%
\bibitem [{\citenamefont {Barello}\ \emph {et~al.}(2014)\citenamefont
  {Barello}, \citenamefont {Chang},\ and\ \citenamefont
  {Newby}}]{Barello:2014uda}%
  \BibitemOpen
  \bibfield  {author} {\bibinfo {author} {\bibfnamefont {G.}~\bibnamefont
  {Barello}}, \bibinfo {author} {\bibfnamefont {S.}~\bibnamefont {Chang}},\
  and\ \bibinfo {author} {\bibfnamefont {C.~A.}\ \bibnamefont {Newby}},\
  }\bibfield  {title} {\bibinfo {title} {{A Model Independent Approach to
  Inelastic Dark Matter Scattering}},\ }\href
  {https://doi.org/10.1103/PhysRevD.90.094027} {\bibfield  {journal} {\bibinfo
  {journal} {Phys. Rev. D}\ }\textbf {\bibinfo {volume} {90}},\ \bibinfo
  {pages} {094027} (\bibinfo {year} {2014})},\ \Eprint
  {https://arxiv.org/abs/1409.0536} {arXiv:1409.0536 [hep-ph]} \BibitemShut
  {NoStop}%
\bibitem [{\citenamefont {Blennow}\ \emph {et~al.}(2017)\citenamefont
  {Blennow}, \citenamefont {Clementz},\ and\ \citenamefont
  {Herrero-Garcia}}]{Blennow:2016gde}%
  \BibitemOpen
  \bibfield  {author} {\bibinfo {author} {\bibfnamefont {M.}~\bibnamefont
  {Blennow}}, \bibinfo {author} {\bibfnamefont {S.}~\bibnamefont {Clementz}},\
  and\ \bibinfo {author} {\bibfnamefont {J.}~\bibnamefont {Herrero-Garcia}},\
  }\bibfield  {title} {\bibinfo {title} {{Self-interacting inelastic dark
  matter: A viable solution to the small scale structure problems}},\ }\href
  {https://doi.org/10.1088/1475-7516/2017/03/048} {\bibfield  {journal}
  {\bibinfo  {journal} {JCAP}\ }\textbf {\bibinfo {volume} {03}},\ \bibinfo
  {pages} {048}},\ \Eprint {https://arxiv.org/abs/1612.06681} {arXiv:1612.06681
  [hep-ph]} \BibitemShut {NoStop}%
\bibitem [{\citenamefont {Peebles}(1968)}]{peebles1968recombination}%
  \BibitemOpen
  \bibfield  {author} {\bibinfo {author} {\bibfnamefont {P.~J.~E.}\
  \bibnamefont {Peebles}},\ }\bibfield  {title} {\bibinfo {title}
  {Recombination of the primeval plasma},\ }\href@noop {} {\bibfield  {journal}
  {\bibinfo  {journal} {Astrophys. J.}\ }\textbf {\bibinfo {volume} {153}},\
  \bibinfo {pages} {1} (\bibinfo {year} {1968})}\BibitemShut {NoStop}%
\bibitem [{\citenamefont {Zeldovich}\ \emph {et~al.}(1969)\citenamefont
  {Zeldovich}, \citenamefont {Kurt}, \citenamefont {Sunyaev} \emph
  {et~al.}}]{zeldovich1969recombination}%
  \BibitemOpen
  \bibfield  {author} {\bibinfo {author} {\bibfnamefont {Y.~B.}\ \bibnamefont
  {Zeldovich}}, \bibinfo {author} {\bibfnamefont {V.}~\bibnamefont {Kurt}},
  \bibinfo {author} {\bibfnamefont {R.}~\bibnamefont {Sunyaev}}, \emph
  {et~al.},\ }\bibfield  {title} {\bibinfo {title} {Recombination of hydrogen
  in the hot model of the universe},\ }\href@noop {} {\bibfield  {journal}
  {\bibinfo  {journal} {Sov. Phys. JETP}\ }\textbf {\bibinfo {volume} {28}},\
  \bibinfo {pages} {146} (\bibinfo {year} {1969})}\BibitemShut {NoStop}%
\bibitem [{\citenamefont {Hu}\ and\ \citenamefont
  {Sugiyama}(1995)}]{Hu:1994jd}%
  \BibitemOpen
  \bibfield  {author} {\bibinfo {author} {\bibfnamefont {W.}~\bibnamefont
  {Hu}}\ and\ \bibinfo {author} {\bibfnamefont {N.}~\bibnamefont {Sugiyama}},\
  }\bibfield  {title} {\bibinfo {title} {{Toward understanding CMB anisotropies
  and their implications}},\ }\href {https://doi.org/10.1103/PhysRevD.51.2599}
  {\bibfield  {journal} {\bibinfo  {journal} {Phys. Rev. D}\ }\textbf {\bibinfo
  {volume} {51}},\ \bibinfo {pages} {2599} (\bibinfo {year} {1995})},\ \Eprint
  {https://arxiv.org/abs/astro-ph/9411008} {arXiv:astro-ph/9411008}
  \BibitemShut {NoStop}%
\bibitem [{\citenamefont {Zaldarriaga}\ and\ \citenamefont
  {Harari}(1995)}]{Zaldarriaga:1995gi}%
  \BibitemOpen
  \bibfield  {author} {\bibinfo {author} {\bibfnamefont {M.}~\bibnamefont
  {Zaldarriaga}}\ and\ \bibinfo {author} {\bibfnamefont {D.~D.}\ \bibnamefont
  {Harari}},\ }\bibfield  {title} {\bibinfo {title} {{Analytic approach to the
  polarization of the cosmic microwave background in flat and open
  universes}},\ }\href {https://doi.org/10.1103/PhysRevD.52.3276} {\bibfield
  {journal} {\bibinfo  {journal} {Phys. Rev. D}\ }\textbf {\bibinfo {volume}
  {52}},\ \bibinfo {pages} {3276} (\bibinfo {year} {1995})},\ \Eprint
  {https://arxiv.org/abs/astro-ph/9504085} {arXiv:astro-ph/9504085}
  \BibitemShut {NoStop}%
\bibitem [{\citenamefont {Hadzhiyska}\ and\ \citenamefont
  {Spergel}(2019)}]{Hadzhiyska:2018mwh}%
  \BibitemOpen
  \bibfield  {author} {\bibinfo {author} {\bibfnamefont {B.}~\bibnamefont
  {Hadzhiyska}}\ and\ \bibinfo {author} {\bibfnamefont {D.~N.}\ \bibnamefont
  {Spergel}},\ }\bibfield  {title} {\bibinfo {title} {{Measuring the Duration
  of Last Scattering}},\ }\href {https://doi.org/10.1103/PhysRevD.99.043537}
  {\bibfield  {journal} {\bibinfo  {journal} {Phys. Rev. D}\ }\textbf {\bibinfo
  {volume} {99}},\ \bibinfo {pages} {043537} (\bibinfo {year} {2019})},\
  \Eprint {https://arxiv.org/abs/1808.04083} {arXiv:1808.04083 [astro-ph.CO]}
  \BibitemShut {NoStop}%
\bibitem [{\citenamefont {Hoyle}(1953)}]{hoyle1953fragmentation}%
  \BibitemOpen
  \bibfield  {author} {\bibinfo {author} {\bibfnamefont {F.}~\bibnamefont
  {Hoyle}},\ }\bibfield  {title} {\bibinfo {title} {On the fragmentation of gas
  clouds into galaxies and stars.},\ }\href@noop {} {\bibfield  {journal}
  {\bibinfo  {journal} {Astrophys. J.}\ }\textbf {\bibinfo {volume} {118}},\
  \bibinfo {pages} {513} (\bibinfo {year} {1953})}\BibitemShut {NoStop}%
\bibitem [{\citenamefont {{Rees}}\ and\ \citenamefont
  {{Ostriker}}(1977)}]{1977MNRAS.179..541R}%
  \BibitemOpen
  \bibfield  {author} {\bibinfo {author} {\bibfnamefont {M.~J.}\ \bibnamefont
  {{Rees}}}\ and\ \bibinfo {author} {\bibfnamefont {J.~P.}\ \bibnamefont
  {{Ostriker}}},\ }\bibfield  {title} {\bibinfo {title} {{Cooling, dynamics and
  fragmentation of massive gas clouds: clues to the masses and radii of
  galaxies and clusters.}},\ }\href {https://doi.org/10.1093/mnras/179.4.541}
  {\bibfield  {journal} {\bibinfo  {journal} {Mon. Not. Roy. Astron. Soc.}\
  }\textbf {\bibinfo {volume} {179}},\ \bibinfo {pages} {541} (\bibinfo {year}
  {1977})}\BibitemShut {NoStop}%
\bibitem [{\citenamefont {{McDowell}}(1961)}]{1961Obs....81..240M}%
  \BibitemOpen
  \bibfield  {author} {\bibinfo {author} {\bibfnamefont {M.~R.~C.}\
  \bibnamefont {{McDowell}}},\ }\bibfield  {title} {\bibinfo {title} {{On the
  formation of H2 in H I regions}},\ }\href@noop {} {\bibfield  {journal}
  {\bibinfo  {journal} {The Observatory}\ }\textbf {\bibinfo {volume} {81}},\
  \bibinfo {pages} {240} (\bibinfo {year} {1961})}\BibitemShut {NoStop}%
\bibitem [{\citenamefont {{Peebles}}\ and\ \citenamefont
  {{Dicke}}(1968)}]{1968ApJ...154..891P}%
  \BibitemOpen
  \bibfield  {author} {\bibinfo {author} {\bibfnamefont {P.~J.~E.}\
  \bibnamefont {{Peebles}}}\ and\ \bibinfo {author} {\bibfnamefont {R.~H.}\
  \bibnamefont {{Dicke}}},\ }\bibfield  {title} {\bibinfo {title} {{Origin of
  the Globular Star Clusters}},\ }\href {https://doi.org/10.1086/149811}
  {\bibfield  {journal} {\bibinfo  {journal} {Astrophys. J.}\ }\textbf
  {\bibinfo {volume} {154}},\ \bibinfo {pages} {891} (\bibinfo {year}
  {1968})}\BibitemShut {NoStop}%
\bibitem [{\citenamefont {{Shaw}}\ \emph {et~al.}(2005)\citenamefont {{Shaw}},
  \citenamefont {{Ferland}}, \citenamefont {{Abel}}, \citenamefont
  {{Stancil}},\ and\ \citenamefont {{van Hoof}}}]{2005ApJ...624..794S}%
  \BibitemOpen
  \bibfield  {author} {\bibinfo {author} {\bibfnamefont {G.}~\bibnamefont
  {{Shaw}}}, \bibinfo {author} {\bibfnamefont {G.~J.}\ \bibnamefont
  {{Ferland}}}, \bibinfo {author} {\bibfnamefont {N.~P.}\ \bibnamefont
  {{Abel}}}, \bibinfo {author} {\bibfnamefont {P.~C.}\ \bibnamefont
  {{Stancil}}},\ and\ \bibinfo {author} {\bibfnamefont {P.~A.~M.}\ \bibnamefont
  {{van Hoof}}},\ }\bibfield  {title} {\bibinfo {title} {{Molecular Hydrogen in
  Star-forming Regions: Implementation of its Microphysics in CLOUDY}},\ }\href
  {https://doi.org/10.1086/429215} {\bibfield  {journal} {\bibinfo  {journal}
  {Astrophys. J.}\ }\textbf {\bibinfo {volume} {624}},\ \bibinfo {pages} {794}
  (\bibinfo {year} {2005})},\ \Eprint {https://arxiv.org/abs/astro-ph/0501485}
  {arXiv:astro-ph/0501485 [astro-ph]} \BibitemShut {NoStop}%
\bibitem [{\citenamefont {{Tegmark}}\ \emph {et~al.}(1997)\citenamefont
  {{Tegmark}}, \citenamefont {{Silk}}, \citenamefont {{Rees}}, \citenamefont
  {{Blanchard}}, \citenamefont {{Abel}},\ and\ \citenamefont
  {{Palla}}}]{1997ApJ...474....1T}%
  \BibitemOpen
  \bibfield  {author} {\bibinfo {author} {\bibfnamefont {M.}~\bibnamefont
  {{Tegmark}}}, \bibinfo {author} {\bibfnamefont {J.}~\bibnamefont {{Silk}}},
  \bibinfo {author} {\bibfnamefont {M.~J.}\ \bibnamefont {{Rees}}}, \bibinfo
  {author} {\bibfnamefont {A.}~\bibnamefont {{Blanchard}}}, \bibinfo {author}
  {\bibfnamefont {T.}~\bibnamefont {{Abel}}},\ and\ \bibinfo {author}
  {\bibfnamefont {F.}~\bibnamefont {{Palla}}},\ }\bibfield  {title} {\bibinfo
  {title} {{How Small Were the First Cosmological Objects?}},\ }\href
  {https://doi.org/10.1086/303434} {\bibfield  {journal} {\bibinfo  {journal}
  {Astrophys. J.}\ }\textbf {\bibinfo {volume} {474}},\ \bibinfo {pages} {1}
  (\bibinfo {year} {1997})},\ \Eprint {https://arxiv.org/abs/astro-ph/9603007}
  {arXiv:astro-ph/9603007 [astro-ph]} \BibitemShut {NoStop}%
\bibitem [{\citenamefont {{Bromm}}\ \emph {et~al.}(2001)\citenamefont
  {{Bromm}}, \citenamefont {{Ferrara}}, \citenamefont {{Coppi}},\ and\
  \citenamefont {{Larson}}}]{2001MNRAS.328..969B}%
  \BibitemOpen
  \bibfield  {author} {\bibinfo {author} {\bibfnamefont {V.}~\bibnamefont
  {{Bromm}}}, \bibinfo {author} {\bibfnamefont {A.}~\bibnamefont {{Ferrara}}},
  \bibinfo {author} {\bibfnamefont {P.~S.}\ \bibnamefont {{Coppi}}},\ and\
  \bibinfo {author} {\bibfnamefont {R.~B.}\ \bibnamefont {{Larson}}},\
  }\bibfield  {title} {\bibinfo {title} {{The fragmentation of pre-enriched
  primordial objects}},\ }\href
  {https://doi.org/10.1046/j.1365-8711.2001.04915.x} {\bibfield  {journal}
  {\bibinfo  {journal} {Mon. Not. Roy. Astron. Soc.}\ }\textbf {\bibinfo
  {volume} {328}},\ \bibinfo {pages} {969} (\bibinfo {year} {2001})},\ \Eprint
  {https://arxiv.org/abs/astro-ph/0104271} {arXiv:astro-ph/0104271 [astro-ph]}
  \BibitemShut {NoStop}%
\bibitem [{\citenamefont {{O'Leary}}\ and\ \citenamefont
  {{McQuinn}}(2012)}]{2012ApJ...760....4O}%
  \BibitemOpen
  \bibfield  {author} {\bibinfo {author} {\bibfnamefont {R.~M.}\ \bibnamefont
  {{O'Leary}}}\ and\ \bibinfo {author} {\bibfnamefont {M.}~\bibnamefont
  {{McQuinn}}},\ }\bibfield  {title} {\bibinfo {title} {{The Formation of the
  First Cosmic Structures and the Physics of the z
  \raisebox{-0.5ex}\textasciitilde 20 Universe}},\ }\href
  {https://doi.org/10.1088/0004-637X/760/1/4} {\bibfield  {journal} {\bibinfo
  {journal} {Astrophys. J.}\ }\textbf {\bibinfo {volume} {760}},\ \bibinfo
  {eid} {4} (\bibinfo {year} {2012})},\ \Eprint
  {https://arxiv.org/abs/1204.1344} {arXiv:1204.1344 [astro-ph.CO]}
  \BibitemShut {NoStop}%
\bibitem [{\citenamefont {{Shapiro}}\ and\ \citenamefont
  {{Kang}}(1987)}]{Shapiro:1987kan}%
  \BibitemOpen
  \bibfield  {author} {\bibinfo {author} {\bibfnamefont {P.~R.}\ \bibnamefont
  {{Shapiro}}}\ and\ \bibinfo {author} {\bibfnamefont {H.}~\bibnamefont
  {{Kang}}},\ }\bibfield  {title} {\bibinfo {title} {{Hydrogen Molecules and
  the Radiative Cooling of Pregalactic Shocks}},\ }\href
  {https://doi.org/10.1086/165350} {\bibfield  {journal} {\bibinfo  {journal}
  {Astrophys. J.}\ }\textbf {\bibinfo {volume} {318}},\ \bibinfo {pages} {32}
  (\bibinfo {year} {1987})}\BibitemShut {NoStop}%
\bibitem [{\citenamefont {{Oh}}\ and\ \citenamefont
  {{Haiman}}(2002)}]{Oh2002:Hai}%
  \BibitemOpen
  \bibfield  {author} {\bibinfo {author} {\bibfnamefont {S.~P.}\ \bibnamefont
  {{Oh}}}\ and\ \bibinfo {author} {\bibfnamefont {Z.}~\bibnamefont
  {{Haiman}}},\ }\bibfield  {title} {\bibinfo {title} {{Second-Generation
  Objects in the Universe: Radiative Cooling and Collapse of Halos with Virial
  Temperatures above {}10$^{4}$ K}},\ }\href {https://doi.org/10.1086/339393}
  {\bibfield  {journal} {\bibinfo  {journal} {Astrophys. J.}\ }\textbf
  {\bibinfo {volume} {569}},\ \bibinfo {pages} {558} (\bibinfo {year}
  {2002})},\ \Eprint {https://arxiv.org/abs/astro-ph/0108071}
  {arXiv:astro-ph/0108071 [astro-ph]} \BibitemShut {NoStop}%
\bibitem [{\citenamefont {{Johnson}}\ and\ \citenamefont
  {{Bromm}}(2006)}]{Johnson2006:Bro}%
  \BibitemOpen
  \bibfield  {author} {\bibinfo {author} {\bibfnamefont {J.~L.}\ \bibnamefont
  {{Johnson}}}\ and\ \bibinfo {author} {\bibfnamefont {V.}~\bibnamefont
  {{Bromm}}},\ }\bibfield  {title} {\bibinfo {title} {{The cooling of
  shock-compressed primordial gas}},\ }\href
  {https://doi.org/10.1111/j.1365-2966.2005.09846.x} {\bibfield  {journal}
  {\bibinfo  {journal} {Mon. Not. Roy. Astron. Soc.}\ }\textbf {\bibinfo
  {volume} {366}},\ \bibinfo {pages} {247} (\bibinfo {year} {2006})},\ \Eprint
  {https://arxiv.org/abs/astro-ph/0505304} {arXiv:astro-ph/0505304 [astro-ph]}
  \BibitemShut {NoStop}%
\bibitem [{\citenamefont {D'Amico}\ \emph {et~al.}(2018)\citenamefont
  {D'Amico}, \citenamefont {Panci}, \citenamefont {Lupi}, \citenamefont
  {Bovino},\ and\ \citenamefont {Silk}}]{DAmico:2017lqj}%
  \BibitemOpen
  \bibfield  {author} {\bibinfo {author} {\bibfnamefont {G.}~\bibnamefont
  {D'Amico}}, \bibinfo {author} {\bibfnamefont {P.}~\bibnamefont {Panci}},
  \bibinfo {author} {\bibfnamefont {A.}~\bibnamefont {Lupi}}, \bibinfo {author}
  {\bibfnamefont {S.}~\bibnamefont {Bovino}},\ and\ \bibinfo {author}
  {\bibfnamefont {J.}~\bibnamefont {Silk}},\ }\bibfield  {title} {\bibinfo
  {title} {{Massive Black Holes from Dissipative Dark Matter}},\ }\href
  {https://doi.org/10.1093/mnras/stx2419} {\bibfield  {journal} {\bibinfo
  {journal} {Mon. Not. Roy. Astron. Soc.}\ }\textbf {\bibinfo {volume} {473}},\
  \bibinfo {pages} {328} (\bibinfo {year} {2018})},\ \Eprint
  {https://arxiv.org/abs/1707.03419} {arXiv:1707.03419 [astro-ph.CO]}
  \BibitemShut {NoStop}%
\bibitem [{\citenamefont {Latif}\ \emph {et~al.}(2019)\citenamefont {Latif},
  \citenamefont {Lupi}, \citenamefont {Schleicher}, \citenamefont {D'Amico},
  \citenamefont {Panci},\ and\ \citenamefont {Bovino}}]{Latif:2018kqv}%
  \BibitemOpen
  \bibfield  {author} {\bibinfo {author} {\bibfnamefont {M.}~\bibnamefont
  {Latif}}, \bibinfo {author} {\bibfnamefont {A.}~\bibnamefont {Lupi}},
  \bibinfo {author} {\bibfnamefont {D.}~\bibnamefont {Schleicher}}, \bibinfo
  {author} {\bibfnamefont {G.}~\bibnamefont {D'Amico}}, \bibinfo {author}
  {\bibfnamefont {P.}~\bibnamefont {Panci}},\ and\ \bibinfo {author}
  {\bibfnamefont {S.}~\bibnamefont {Bovino}},\ }\bibfield  {title} {\bibinfo
  {title} {{Black hole formation in the context of dissipative dark matter}},\
  }\href {https://doi.org/10.1093/mnras/stz608} {\bibfield  {journal} {\bibinfo
   {journal} {Mon. Not. Roy. Astron. Soc.}\ }\textbf {\bibinfo {volume}
  {485}},\ \bibinfo {pages} {3352} (\bibinfo {year} {2019})},\ \Eprint
  {https://arxiv.org/abs/1812.03104} {arXiv:1812.03104 [astro-ph.CO]}
  \BibitemShut {NoStop}%
\bibitem [{\citenamefont {Pollack}\ \emph {et~al.}(2015)\citenamefont
  {Pollack}, \citenamefont {Spergel},\ and\ \citenamefont
  {Steinhardt}}]{Pollack:2014rja}%
  \BibitemOpen
  \bibfield  {author} {\bibinfo {author} {\bibfnamefont {J.}~\bibnamefont
  {Pollack}}, \bibinfo {author} {\bibfnamefont {D.~N.}\ \bibnamefont
  {Spergel}},\ and\ \bibinfo {author} {\bibfnamefont {P.~J.}\ \bibnamefont
  {Steinhardt}},\ }\bibfield  {title} {\bibinfo {title} {{Supermassive Black
  Holes from Ultra-Strongly Self-Interacting Dark Matter}},\ }\href
  {https://doi.org/10.1088/0004-637X/804/2/131} {\bibfield  {journal} {\bibinfo
   {journal} {Astrophys. J.}\ }\textbf {\bibinfo {volume} {804}},\ \bibinfo
  {pages} {131} (\bibinfo {year} {2015})},\ \Eprint
  {https://arxiv.org/abs/1501.00017} {arXiv:1501.00017 [astro-ph.CO]}
  \BibitemShut {NoStop}%
\bibitem [{\citenamefont {Choquette}\ \emph {et~al.}(2019)\citenamefont
  {Choquette}, \citenamefont {Cline},\ and\ \citenamefont
  {Cornell}}]{Choquette:2018lvq}%
  \BibitemOpen
  \bibfield  {author} {\bibinfo {author} {\bibfnamefont {J.}~\bibnamefont
  {Choquette}}, \bibinfo {author} {\bibfnamefont {J.~M.}\ \bibnamefont
  {Cline}},\ and\ \bibinfo {author} {\bibfnamefont {J.~M.}\ \bibnamefont
  {Cornell}},\ }\bibfield  {title} {\bibinfo {title} {{Early formation of
  supermassive black holes via dark matter self-interactions}},\ }\href
  {https://doi.org/10.1088/1475-7516/2019/07/036} {\bibfield  {journal}
  {\bibinfo  {journal} {JCAP}\ }\textbf {\bibinfo {volume} {07}},\ \bibinfo
  {pages} {036}},\ \Eprint {https://arxiv.org/abs/1812.05088} {arXiv:1812.05088
  [astro-ph.CO]} \BibitemShut {NoStop}%
\bibitem [{\citenamefont {Mohapatra}\ and\ \citenamefont
  {Teplitz}(1997)}]{Mohapatra:1996yy}%
  \BibitemOpen
  \bibfield  {author} {\bibinfo {author} {\bibfnamefont {R.}~\bibnamefont
  {Mohapatra}}\ and\ \bibinfo {author} {\bibfnamefont {V.~L.}\ \bibnamefont
  {Teplitz}},\ }\bibfield  {title} {\bibinfo {title} {{Structures in the mirror
  universe}},\ }\href {https://doi.org/10.1086/303762} {\bibfield  {journal}
  {\bibinfo  {journal} {Astrophys. J.}\ }\textbf {\bibinfo {volume} {478}},\
  \bibinfo {pages} {29} (\bibinfo {year} {1997})},\ \Eprint
  {https://arxiv.org/abs/astro-ph/9603049} {arXiv:astro-ph/9603049}
  \BibitemShut {NoStop}%
\bibitem [{\citenamefont {Foot}(1999)}]{Foot:1999hm}%
  \BibitemOpen
  \bibfield  {author} {\bibinfo {author} {\bibfnamefont {R.}~\bibnamefont
  {Foot}},\ }\bibfield  {title} {\bibinfo {title} {{Have mirror stars been
  observed?}},\ }\href {https://doi.org/10.1016/S0370-2693(99)00230-0}
  {\bibfield  {journal} {\bibinfo  {journal} {Phys. Lett. B}\ }\textbf
  {\bibinfo {volume} {452}},\ \bibinfo {pages} {83} (\bibinfo {year} {1999})},\
  \Eprint {https://arxiv.org/abs/astro-ph/9902065} {arXiv:astro-ph/9902065}
  \BibitemShut {NoStop}%
\bibitem [{\citenamefont {Mohapatra}\ and\ \citenamefont
  {Teplitz}(1999)}]{Mohapatra:1999ih}%
  \BibitemOpen
  \bibfield  {author} {\bibinfo {author} {\bibfnamefont {R.~N.}\ \bibnamefont
  {Mohapatra}}\ and\ \bibinfo {author} {\bibfnamefont {V.~L.}\ \bibnamefont
  {Teplitz}},\ }\bibfield  {title} {\bibinfo {title} {{Mirror matter MACHOs}},\
  }\href {https://doi.org/10.1016/S0370-2693(99)00789-3} {\bibfield  {journal}
  {\bibinfo  {journal} {Phys. Lett. B}\ }\textbf {\bibinfo {volume} {462}},\
  \bibinfo {pages} {302} (\bibinfo {year} {1999})},\ \Eprint
  {https://arxiv.org/abs/astro-ph/9902085} {arXiv:astro-ph/9902085}
  \BibitemShut {NoStop}%
\bibitem [{\citenamefont {Foot}\ \emph {et~al.}(2001)\citenamefont {Foot},
  \citenamefont {Ignatiev},\ and\ \citenamefont {Volkas}}]{Foot:2000vy}%
  \BibitemOpen
  \bibfield  {author} {\bibinfo {author} {\bibfnamefont {R.}~\bibnamefont
  {Foot}}, \bibinfo {author} {\bibfnamefont {A.}~\bibnamefont {Ignatiev}},\
  and\ \bibinfo {author} {\bibfnamefont {R.}~\bibnamefont {Volkas}},\
  }\bibfield  {title} {\bibinfo {title} {{Physics of mirror photons}},\ }\href
  {https://doi.org/10.1016/S0370-2693(01)00228-3} {\bibfield  {journal}
  {\bibinfo  {journal} {Phys. Lett. B}\ }\textbf {\bibinfo {volume} {503}},\
  \bibinfo {pages} {355} (\bibinfo {year} {2001})},\ \Eprint
  {https://arxiv.org/abs/astro-ph/0011156} {arXiv:astro-ph/0011156}
  \BibitemShut {NoStop}%
\bibitem [{\citenamefont {Mohapatra}\ and\ \citenamefont
  {Teplitz}(2000)}]{Mohapatra:2000qx}%
  \BibitemOpen
  \bibfield  {author} {\bibinfo {author} {\bibfnamefont {R.~N.}\ \bibnamefont
  {Mohapatra}}\ and\ \bibinfo {author} {\bibfnamefont {V.~L.}\ \bibnamefont
  {Teplitz}},\ }\bibfield  {title} {\bibinfo {title} {{Mirror dark matter and
  galaxy core densities of galaxies}},\ }\href
  {https://doi.org/10.1103/PhysRevD.62.063506} {\bibfield  {journal} {\bibinfo
  {journal} {Phys. Rev. D}\ }\textbf {\bibinfo {volume} {62}},\ \bibinfo
  {pages} {063506} (\bibinfo {year} {2000})},\ \Eprint
  {https://arxiv.org/abs/astro-ph/0001362} {arXiv:astro-ph/0001362}
  \BibitemShut {NoStop}%
\bibitem [{\citenamefont {Mohapatra}\ \emph {et~al.}(2002)\citenamefont
  {Mohapatra}, \citenamefont {Nussinov},\ and\ \citenamefont
  {Teplitz}}]{Mohapatra:2001sx}%
  \BibitemOpen
  \bibfield  {author} {\bibinfo {author} {\bibfnamefont {R.}~\bibnamefont
  {Mohapatra}}, \bibinfo {author} {\bibfnamefont {S.}~\bibnamefont
  {Nussinov}},\ and\ \bibinfo {author} {\bibfnamefont {V.}~\bibnamefont
  {Teplitz}},\ }\bibfield  {title} {\bibinfo {title} {{Mirror matter as
  selfinteracting dark matter}},\ }\href
  {https://doi.org/10.1103/PhysRevD.66.063002} {\bibfield  {journal} {\bibinfo
  {journal} {Phys. Rev. D}\ }\textbf {\bibinfo {volume} {66}},\ \bibinfo
  {pages} {063002} (\bibinfo {year} {2002})},\ \Eprint
  {https://arxiv.org/abs/hep-ph/0111381} {arXiv:hep-ph/0111381} \BibitemShut
  {NoStop}%
\bibitem [{\citenamefont {Foot}\ \emph {et~al.}(2000)\citenamefont {Foot},
  \citenamefont {Lew},\ and\ \citenamefont {Volkas}}]{Foot:2000tp}%
  \BibitemOpen
  \bibfield  {author} {\bibinfo {author} {\bibfnamefont {R.}~\bibnamefont
  {Foot}}, \bibinfo {author} {\bibfnamefont {H.}~\bibnamefont {Lew}},\ and\
  \bibinfo {author} {\bibfnamefont {R.}~\bibnamefont {Volkas}},\ }\bibfield
  {title} {\bibinfo {title} {{Unbroken versus broken mirror world: A Tale of
  two vacua}},\ }\href {https://doi.org/10.1088/1126-6708/2000/07/032}
  {\bibfield  {journal} {\bibinfo  {journal} {JHEP}\ }\textbf {\bibinfo
  {volume} {07}},\ \bibinfo {pages} {032}},\ \Eprint
  {https://arxiv.org/abs/hep-ph/0006027} {arXiv:hep-ph/0006027} \BibitemShut
  {NoStop}%
\bibitem [{\citenamefont {Berezhiani}(2004)}]{Berezhiani:2003xm}%
  \BibitemOpen
  \bibfield  {author} {\bibinfo {author} {\bibfnamefont {Z.}~\bibnamefont
  {Berezhiani}},\ }\bibfield  {title} {\bibinfo {title} {{Mirror world and its
  cosmological consequences}},\ }\href
  {https://doi.org/10.1142/S0217751X04020075} {\bibfield  {journal} {\bibinfo
  {journal} {Int. J. Mod. Phys. A}\ }\textbf {\bibinfo {volume} {19}},\
  \bibinfo {pages} {3775} (\bibinfo {year} {2004})},\ \Eprint
  {https://arxiv.org/abs/hep-ph/0312335} {arXiv:hep-ph/0312335} \BibitemShut
  {NoStop}%
\bibitem [{\citenamefont {Berezhiani}\ \emph {et~al.}(2006)\citenamefont
  {Berezhiani}, \citenamefont {Cassisi}, \citenamefont {Ciarcelluti},\ and\
  \citenamefont {Pietrinferni}}]{Berezhiani:2005vv}%
  \BibitemOpen
  \bibfield  {author} {\bibinfo {author} {\bibfnamefont {Z.}~\bibnamefont
  {Berezhiani}}, \bibinfo {author} {\bibfnamefont {S.}~\bibnamefont {Cassisi}},
  \bibinfo {author} {\bibfnamefont {P.}~\bibnamefont {Ciarcelluti}},\ and\
  \bibinfo {author} {\bibfnamefont {A.}~\bibnamefont {Pietrinferni}},\
  }\bibfield  {title} {\bibinfo {title} {{Evolutionary and structural
  properties of mirror star MACHOs}},\ }\href
  {https://doi.org/10.1016/j.astropartphys.2005.10.002} {\bibfield  {journal}
  {\bibinfo  {journal} {Astropart. Phys.}\ }\textbf {\bibinfo {volume} {24}},\
  \bibinfo {pages} {495} (\bibinfo {year} {2006})},\ \Eprint
  {https://arxiv.org/abs/astro-ph/0507153} {arXiv:astro-ph/0507153}
  \BibitemShut {NoStop}%
\bibitem [{\citenamefont {Curtin}\ and\ \citenamefont
  {Setford}(2020{\natexlab{a}})}]{Curtin:2019lhm}%
  \BibitemOpen
  \bibfield  {author} {\bibinfo {author} {\bibfnamefont {D.}~\bibnamefont
  {Curtin}}\ and\ \bibinfo {author} {\bibfnamefont {J.}~\bibnamefont
  {Setford}},\ }\bibfield  {title} {\bibinfo {title} {{How To Discover Mirror
  Stars}},\ }\href {https://doi.org/10.1016/j.physletb.2020.135391} {\bibfield
  {journal} {\bibinfo  {journal} {Phys. Lett. B}\ }\textbf {\bibinfo {volume}
  {804}},\ \bibinfo {pages} {135391} (\bibinfo {year} {2020}{\natexlab{a}})},\
  \Eprint {https://arxiv.org/abs/1909.04071} {arXiv:1909.04071 [hep-ph]}
  \BibitemShut {NoStop}%
\bibitem [{\citenamefont {Curtin}\ and\ \citenamefont
  {Setford}(2020{\natexlab{b}})}]{Curtin:2019ngc}%
  \BibitemOpen
  \bibfield  {author} {\bibinfo {author} {\bibfnamefont {D.}~\bibnamefont
  {Curtin}}\ and\ \bibinfo {author} {\bibfnamefont {J.}~\bibnamefont
  {Setford}},\ }\bibfield  {title} {\bibinfo {title} {{Signatures of Mirror
  Stars}},\ }\href {https://doi.org/10.1007/JHEP03(2020)041} {\bibfield
  {journal} {\bibinfo  {journal} {JHEP}\ }\textbf {\bibinfo {volume} {03}},\
  \bibinfo {pages} {041}},\ \Eprint {https://arxiv.org/abs/1909.04072}
  {arXiv:1909.04072 [hep-ph]} \BibitemShut {NoStop}%
\bibitem [{\citenamefont {Curtin}\ and\ \citenamefont
  {Setford}(2020{\natexlab{c}})}]{Curtin:2020tkm}%
  \BibitemOpen
  \bibfield  {author} {\bibinfo {author} {\bibfnamefont {D.}~\bibnamefont
  {Curtin}}\ and\ \bibinfo {author} {\bibfnamefont {J.}~\bibnamefont
  {Setford}},\ }\bibfield  {title} {\bibinfo {title} {{Direct Detection of
  Atomic Dark Matter in White Dwarfs}},\ }\href@noop {} {\  (\bibinfo {year}
  {2020}{\natexlab{c}})},\ \Eprint {https://arxiv.org/abs/2010.00601}
  {arXiv:2010.00601 [hep-ph]} \BibitemShut {NoStop}%
\bibitem [{\citenamefont {Fan}\ \emph {et~al.}(2013{\natexlab{a}})\citenamefont
  {Fan}, \citenamefont {Katz}, \citenamefont {Randall},\ and\ \citenamefont
  {Reece}}]{Fan:2013yva}%
  \BibitemOpen
  \bibfield  {author} {\bibinfo {author} {\bibfnamefont {J.}~\bibnamefont
  {Fan}}, \bibinfo {author} {\bibfnamefont {A.}~\bibnamefont {Katz}}, \bibinfo
  {author} {\bibfnamefont {L.}~\bibnamefont {Randall}},\ and\ \bibinfo {author}
  {\bibfnamefont {M.}~\bibnamefont {Reece}},\ }\bibfield  {title} {\bibinfo
  {title} {{Double-Disk Dark Matter}},\ }\href
  {https://doi.org/10.1016/j.dark.2013.07.001} {\bibfield  {journal} {\bibinfo
  {journal} {Phys. Dark Univ.}\ }\textbf {\bibinfo {volume} {2}},\ \bibinfo
  {pages} {139} (\bibinfo {year} {2013}{\natexlab{a}})},\ \Eprint
  {https://arxiv.org/abs/1303.1521} {arXiv:1303.1521 [astro-ph.CO]}
  \BibitemShut {NoStop}%
\bibitem [{\citenamefont {Fan}\ \emph {et~al.}(2013{\natexlab{b}})\citenamefont
  {Fan}, \citenamefont {Katz}, \citenamefont {Randall},\ and\ \citenamefont
  {Reece}}]{Fan:2013tia}%
  \BibitemOpen
  \bibfield  {author} {\bibinfo {author} {\bibfnamefont {J.}~\bibnamefont
  {Fan}}, \bibinfo {author} {\bibfnamefont {A.}~\bibnamefont {Katz}}, \bibinfo
  {author} {\bibfnamefont {L.}~\bibnamefont {Randall}},\ and\ \bibinfo {author}
  {\bibfnamefont {M.}~\bibnamefont {Reece}},\ }\bibfield  {title} {\bibinfo
  {title} {{Dark-Disk Universe}},\ }\href
  {https://doi.org/10.1103/PhysRevLett.110.211302} {\bibfield  {journal}
  {\bibinfo  {journal} {Phys. Rev. Lett.}\ }\textbf {\bibinfo {volume} {110}},\
  \bibinfo {pages} {211302} (\bibinfo {year} {2013}{\natexlab{b}})},\ \Eprint
  {https://arxiv.org/abs/1303.3271} {arXiv:1303.3271 [hep-ph]} \BibitemShut
  {NoStop}%
\bibitem [{\citenamefont {Buckley}\ and\ \citenamefont
  {DiFranzo}(2018)}]{Buckley:2017ttd}%
  \BibitemOpen
  \bibfield  {author} {\bibinfo {author} {\bibfnamefont {M.~R.}\ \bibnamefont
  {Buckley}}\ and\ \bibinfo {author} {\bibfnamefont {A.}~\bibnamefont
  {DiFranzo}},\ }\bibfield  {title} {\bibinfo {title} {{Collapsed Dark Matter
  Structures}},\ }\href {https://doi.org/10.1103/PhysRevLett.120.051102}
  {\bibfield  {journal} {\bibinfo  {journal} {Phys. Rev. Lett.}\ }\textbf
  {\bibinfo {volume} {120}},\ \bibinfo {pages} {051102} (\bibinfo {year}
  {2018})},\ \Eprint {https://arxiv.org/abs/1707.03829} {arXiv:1707.03829
  [hep-ph]} \BibitemShut {NoStop}%
\bibitem [{\citenamefont {Tulin}\ and\ \citenamefont
  {Yu}(2018)}]{Tulin:2017ara}%
  \BibitemOpen
  \bibfield  {author} {\bibinfo {author} {\bibfnamefont {S.}~\bibnamefont
  {Tulin}}\ and\ \bibinfo {author} {\bibfnamefont {H.-B.}\ \bibnamefont {Yu}},\
  }\bibfield  {title} {\bibinfo {title} {{Dark Matter Self-interactions and
  Small Scale Structure}},\ }\href
  {https://doi.org/10.1016/j.physrep.2017.11.004} {\bibfield  {journal}
  {\bibinfo  {journal} {Phys. Rept.}\ }\textbf {\bibinfo {volume} {730}},\
  \bibinfo {pages} {1} (\bibinfo {year} {2018})},\ \Eprint
  {https://arxiv.org/abs/1705.02358} {arXiv:1705.02358 [hep-ph]} \BibitemShut
  {NoStop}%
\bibitem [{\citenamefont {Tseliakhovich}\ and\ \citenamefont
  {Hirata}(2010)}]{PhysRevD.82.083520}%
  \BibitemOpen
  \bibfield  {author} {\bibinfo {author} {\bibfnamefont {D.}~\bibnamefont
  {Tseliakhovich}}\ and\ \bibinfo {author} {\bibfnamefont {C.}~\bibnamefont
  {Hirata}},\ }\bibfield  {title} {\bibinfo {title} {Relative velocity of dark
  matter and baryonic fluids and the formation of the first structures},\
  }\href {https://doi.org/10.1103/PhysRevD.82.083520} {\bibfield  {journal}
  {\bibinfo  {journal} {Phys. Rev. D}\ }\textbf {\bibinfo {volume} {82}},\
  \bibinfo {pages} {083520} (\bibinfo {year} {2010})}\BibitemShut {NoStop}%
\bibitem [{\citenamefont {Tseliakhovich}\ \emph {et~al.}(2011)\citenamefont
  {Tseliakhovich}, \citenamefont {Barkana},\ and\ \citenamefont
  {Hirata}}]{10.1111/j.1365-2966.2011.19541.x}%
  \BibitemOpen
  \bibfield  {author} {\bibinfo {author} {\bibfnamefont {D.}~\bibnamefont
  {Tseliakhovich}}, \bibinfo {author} {\bibfnamefont {R.}~\bibnamefont
  {Barkana}},\ and\ \bibinfo {author} {\bibfnamefont {C.~M.}\ \bibnamefont
  {Hirata}},\ }\bibfield  {title} {\bibinfo {title} {{Suppression and spatial
  variation of early galaxies and minihaloes}},\ }\href
  {https://doi.org/10.1111/j.1365-2966.2011.19541.x} {\bibfield  {journal}
  {\bibinfo  {journal} {Monthly Notices of the Royal Astronomical Society}\
  }\textbf {\bibinfo {volume} {418}},\ \bibinfo {pages} {906} (\bibinfo {year}
  {2011})},\ \Eprint
  {https://arxiv.org/abs/https://academic.oup.com/mnras/article-pdf/418/2/906/3692552/mnras0418-0906.pdf}
  {https://academic.oup.com/mnras/article-pdf/418/2/906/3692552/mnras0418-0906.pdf}
  \BibitemShut {NoStop}%
\bibitem [{\citenamefont {Greif}\ \emph {et~al.}(2011)\citenamefont {Greif},
  \citenamefont {White}, \citenamefont {Klessen},\ and\ \citenamefont
  {Springel}}]{Greif_2011}%
  \BibitemOpen
  \bibfield  {author} {\bibinfo {author} {\bibfnamefont {T.~H.}\ \bibnamefont
  {Greif}}, \bibinfo {author} {\bibfnamefont {S.~D.~M.}\ \bibnamefont {White}},
  \bibinfo {author} {\bibfnamefont {R.~S.}\ \bibnamefont {Klessen}},\ and\
  \bibinfo {author} {\bibfnamefont {V.}~\bibnamefont {Springel}},\ }\bibfield
  {title} {\bibinfo {title} {{The} {Delay} {of} {Population} {III} {Star}
  {Formation} {by} {Supersonic} {Streaming} {Velocities}},\ }\href
  {https://doi.org/10.1088/0004-637x/736/2/147} {\bibfield  {journal} {\bibinfo
   {journal} {The Astrophysical Journal}\ }\textbf {\bibinfo {volume} {736}},\
  \bibinfo {pages} {147} (\bibinfo {year} {2011})}\BibitemShut {NoStop}%
\bibitem [{\citenamefont {Naoz}\ \emph
  {et~al.}(2012{\natexlab{a}})\citenamefont {Naoz}, \citenamefont {Yoshida},\
  and\ \citenamefont {Gnedin}}]{Naoz_2012}%
  \BibitemOpen
  \bibfield  {author} {\bibinfo {author} {\bibfnamefont {S.}~\bibnamefont
  {Naoz}}, \bibinfo {author} {\bibfnamefont {N.}~\bibnamefont {Yoshida}},\ and\
  \bibinfo {author} {\bibfnamefont {N.~Y.}\ \bibnamefont {Gnedin}},\ }\bibfield
   {title} {\bibinfo {title} {{Simulations} {of} {Early} {Baryonic} {Structure}
  {Formation} {with} {Stream} {Velocity}. {I}. {Halo} {Abundance}},\ }\href
  {https://doi.org/10.1088/0004-637x/747/2/128} {\bibfield  {journal} {\bibinfo
   {journal} {The Astrophysical Journal}\ }\textbf {\bibinfo {volume} {747}},\
  \bibinfo {pages} {128} (\bibinfo {year} {2012}{\natexlab{a}})}\BibitemShut
  {NoStop}%
\bibitem [{\citenamefont {Naoz}\ \emph
  {et~al.}(2012{\natexlab{b}})\citenamefont {Naoz}, \citenamefont {Yoshida},\
  and\ \citenamefont {Gnedin}}]{Naoz_2012b}%
  \BibitemOpen
  \bibfield  {author} {\bibinfo {author} {\bibfnamefont {S.}~\bibnamefont
  {Naoz}}, \bibinfo {author} {\bibfnamefont {N.}~\bibnamefont {Yoshida}},\ and\
  \bibinfo {author} {\bibfnamefont {N.~Y.}\ \bibnamefont {Gnedin}},\ }\bibfield
   {title} {\bibinfo {title} {{Simulations} {of} {Early} {Baryonic} {Structure}
  {Formation} {with} {Stream} {Velocity}. {II}. {The} {Gas} {Fraction}},\
  }\href {https://doi.org/10.1088/0004-637x/763/1/27} {\bibfield  {journal}
  {\bibinfo  {journal} {The Astrophysical Journal}\ }\textbf {\bibinfo {volume}
  {763}},\ \bibinfo {pages} {27} (\bibinfo {year}
  {2012}{\natexlab{b}})}\BibitemShut {NoStop}%
\bibitem [{\citenamefont {Schauer}\ \emph {et~al.}(2019)\citenamefont
  {Schauer}, \citenamefont {Glover}, \citenamefont {Klessen},\ and\
  \citenamefont {Ceverino}}]{10.1093/mnras/stz013}%
  \BibitemOpen
  \bibfield  {author} {\bibinfo {author} {\bibfnamefont {A.~T.~P.}\
  \bibnamefont {Schauer}}, \bibinfo {author} {\bibfnamefont {S.~C.~O.}\
  \bibnamefont {Glover}}, \bibinfo {author} {\bibfnamefont {R.~S.}\
  \bibnamefont {Klessen}},\ and\ \bibinfo {author} {\bibfnamefont
  {D.}~\bibnamefont {Ceverino}},\ }\bibfield  {title} {\bibinfo {title} {{The
  influence of streaming velocities on the formation of the first stars}},\
  }\href {https://doi.org/10.1093/mnras/stz013} {\bibfield  {journal} {\bibinfo
   {journal} {Monthly Notices of the Royal Astronomical Society}\ }\textbf
  {\bibinfo {volume} {484}},\ \bibinfo {pages} {3510} (\bibinfo {year}
  {2019})},\ \Eprint
  {https://arxiv.org/abs/https://academic.oup.com/mnras/article-pdf/484/3/3510/27714961/stz013.pdf}
  {https://academic.oup.com/mnras/article-pdf/484/3/3510/27714961/stz013.pdf}
  \BibitemShut {NoStop}%
\bibitem [{\citenamefont {Schauer}\ \emph {et~al.}(2020)\citenamefont
  {Schauer}, \citenamefont {Glover}, \citenamefont {Klessen},\ and\
  \citenamefont {Clark}}]{schauer2020influence}%
  \BibitemOpen
  \bibfield  {author} {\bibinfo {author} {\bibfnamefont {A.~T.}\ \bibnamefont
  {Schauer}}, \bibinfo {author} {\bibfnamefont {S.~C.}\ \bibnamefont {Glover}},
  \bibinfo {author} {\bibfnamefont {R.~S.}\ \bibnamefont {Klessen}},\ and\
  \bibinfo {author} {\bibfnamefont {P.}~\bibnamefont {Clark}},\ }\bibfield
  {title} {\bibinfo {title} {The influence of streaming velocities and
  lyman-werner radiation on the formation of the first stars},\ }\href@noop {}
  {\bibfield  {journal} {\bibinfo  {journal} {arXiv:2008.05663}\ } (\bibinfo
  {year} {2020})}\BibitemShut {NoStop}%
\bibitem [{\citenamefont {Greig}\ and\ \citenamefont
  {Mesinger}(2017)}]{Greig:2016wjs}%
  \BibitemOpen
  \bibfield  {author} {\bibinfo {author} {\bibfnamefont {B.}~\bibnamefont
  {Greig}}\ and\ \bibinfo {author} {\bibfnamefont {A.}~\bibnamefont
  {Mesinger}},\ }\bibfield  {title} {\bibinfo {title} {{The global history of
  reionization}},\ }\href {https://doi.org/10.1093/mnras/stw3026} {\bibfield
  {journal} {\bibinfo  {journal} {Mon. Not. Roy. Astron. Soc.}\ }\textbf
  {\bibinfo {volume} {465}},\ \bibinfo {pages} {4838} (\bibinfo {year}
  {2017})},\ \Eprint {https://arxiv.org/abs/1605.05374} {arXiv:1605.05374
  [astro-ph.CO]} \BibitemShut {NoStop}%
\bibitem [{\citenamefont {Bouwens}\ \emph {et~al.}(2015)\citenamefont
  {Bouwens}, \citenamefont {Illingworth}, \citenamefont {Oesch}, \citenamefont
  {Caruana}, \citenamefont {Holwerda}, \citenamefont {Smit},\ and\
  \citenamefont {Wilkins}}]{Bouwens:2015vha}%
  \BibitemOpen
  \bibfield  {author} {\bibinfo {author} {\bibfnamefont {R.}~\bibnamefont
  {Bouwens}}, \bibinfo {author} {\bibfnamefont {G.}~\bibnamefont
  {Illingworth}}, \bibinfo {author} {\bibfnamefont {P.}~\bibnamefont {Oesch}},
  \bibinfo {author} {\bibfnamefont {J.}~\bibnamefont {Caruana}}, \bibinfo
  {author} {\bibfnamefont {B.}~\bibnamefont {Holwerda}}, \bibinfo {author}
  {\bibfnamefont {R.}~\bibnamefont {Smit}},\ and\ \bibinfo {author}
  {\bibfnamefont {S.}~\bibnamefont {Wilkins}},\ }\bibfield  {title} {\bibinfo
  {title} {{Reionization after Planck: The Derived Growth of the Cosmic
  Ionizing Emissivity now matches the Growth of the Galaxy UV Luminosity
  Density}},\ }\href {https://doi.org/10.1088/0004-637X/811/2/140} {\bibfield
  {journal} {\bibinfo  {journal} {Astrophys. J.}\ }\textbf {\bibinfo {volume}
  {811}},\ \bibinfo {pages} {140} (\bibinfo {year} {2015})},\ \Eprint
  {https://arxiv.org/abs/1503.08228} {arXiv:1503.08228 [astro-ph.CO]}
  \BibitemShut {NoStop}%
\bibitem [{\citenamefont {Mitra}\ \emph {et~al.}(2015)\citenamefont {Mitra},
  \citenamefont {Choudhury},\ and\ \citenamefont {Ferrara}}]{Mitra:2015yqa}%
  \BibitemOpen
  \bibfield  {author} {\bibinfo {author} {\bibfnamefont {S.}~\bibnamefont
  {Mitra}}, \bibinfo {author} {\bibfnamefont {T.~R.}\ \bibnamefont
  {Choudhury}},\ and\ \bibinfo {author} {\bibfnamefont {A.}~\bibnamefont
  {Ferrara}},\ }\bibfield  {title} {\bibinfo {title} {{Cosmic reionization
  after Planck}},\ }\href {https://doi.org/10.1093/mnrasl/slv134} {\bibfield
  {journal} {\bibinfo  {journal} {Mon. Not. Roy. Astron. Soc.}\ }\textbf
  {\bibinfo {volume} {454}},\ \bibinfo {pages} {L76} (\bibinfo {year}
  {2015})},\ \Eprint {https://arxiv.org/abs/1505.05507} {arXiv:1505.05507
  [astro-ph.CO]} \BibitemShut {NoStop}%
\bibitem [{\citenamefont {Gorce}\ \emph {et~al.}(2018)\citenamefont {Gorce},
  \citenamefont {Douspis}, \citenamefont {Aghanim},\ and\ \citenamefont
  {Langer}}]{Gorce:2017glg}%
  \BibitemOpen
  \bibfield  {author} {\bibinfo {author} {\bibfnamefont {A.}~\bibnamefont
  {Gorce}}, \bibinfo {author} {\bibfnamefont {M.}~\bibnamefont {Douspis}},
  \bibinfo {author} {\bibfnamefont {N.}~\bibnamefont {Aghanim}},\ and\ \bibinfo
  {author} {\bibfnamefont {M.}~\bibnamefont {Langer}},\ }\bibfield  {title}
  {\bibinfo {title} {{Observational constraints on key-parameters of cosmic
  reionisation history}},\ }\href {https://doi.org/10.1051/0004-6361/201629661}
  {\bibfield  {journal} {\bibinfo  {journal} {Astron. Astrophys.}\ }\textbf
  {\bibinfo {volume} {616}},\ \bibinfo {pages} {A113} (\bibinfo {year}
  {2018})},\ \Eprint {https://arxiv.org/abs/1710.04152} {arXiv:1710.04152
  [astro-ph.CO]} \BibitemShut {NoStop}%
\bibitem [{\citenamefont {Park}\ \emph {et~al.}(2019)\citenamefont {Park},
  \citenamefont {Mesinger}, \citenamefont {Greig},\ and\ \citenamefont
  {Gillet}}]{Park:2018ljd}%
  \BibitemOpen
  \bibfield  {author} {\bibinfo {author} {\bibfnamefont {J.}~\bibnamefont
  {Park}}, \bibinfo {author} {\bibfnamefont {A.}~\bibnamefont {Mesinger}},
  \bibinfo {author} {\bibfnamefont {B.}~\bibnamefont {Greig}},\ and\ \bibinfo
  {author} {\bibfnamefont {N.}~\bibnamefont {Gillet}},\ }\bibfield  {title}
  {\bibinfo {title} {{Inferring the astrophysics of reionization and cosmic
  dawn from galaxy luminosity functions and the 21-cm signal}},\ }\href
  {https://doi.org/10.1093/mnras/stz032} {\bibfield  {journal} {\bibinfo
  {journal} {Mon. Not. Roy. Astron. Soc.}\ }\textbf {\bibinfo {volume} {484}},\
  \bibinfo {pages} {933} (\bibinfo {year} {2019})},\ \Eprint
  {https://arxiv.org/abs/1809.08995} {arXiv:1809.08995 [astro-ph.GA]}
  \BibitemShut {NoStop}%
\bibitem [{\citenamefont {Fukugita}\ \emph {et~al.}(1998)\citenamefont
  {Fukugita}, \citenamefont {Hogan},\ and\ \citenamefont
  {Peebles}}]{Fukugita:1997bi}%
  \BibitemOpen
  \bibfield  {author} {\bibinfo {author} {\bibfnamefont {M.}~\bibnamefont
  {Fukugita}}, \bibinfo {author} {\bibfnamefont {C.}~\bibnamefont {Hogan}},\
  and\ \bibinfo {author} {\bibfnamefont {P.}~\bibnamefont {Peebles}},\
  }\bibfield  {title} {\bibinfo {title} {{The Cosmic baryon budget}},\ }\href
  {https://doi.org/10.1086/306025} {\bibfield  {journal} {\bibinfo  {journal}
  {Astrophys. J.}\ }\textbf {\bibinfo {volume} {503}},\ \bibinfo {pages} {518}
  (\bibinfo {year} {1998})},\ \Eprint {https://arxiv.org/abs/astro-ph/9712020}
  {arXiv:astro-ph/9712020} \BibitemShut {NoStop}%
\bibitem [{\citenamefont {Fukugita}\ and\ \citenamefont
  {Peebles}(2004)}]{Fukugita:2004ee}%
  \BibitemOpen
  \bibfield  {author} {\bibinfo {author} {\bibfnamefont {M.}~\bibnamefont
  {Fukugita}}\ and\ \bibinfo {author} {\bibfnamefont {P.~E.}\ \bibnamefont
  {Peebles}},\ }\bibfield  {title} {\bibinfo {title} {{The Cosmic energy
  inventory}},\ }\href {https://doi.org/10.1086/425155} {\bibfield  {journal}
  {\bibinfo  {journal} {Astrophys. J.}\ }\textbf {\bibinfo {volume} {616}},\
  \bibinfo {pages} {643} (\bibinfo {year} {2004})},\ \Eprint
  {https://arxiv.org/abs/astro-ph/0406095} {arXiv:astro-ph/0406095}
  \BibitemShut {NoStop}%
\bibitem [{\citenamefont {{Shull}}\ \emph {et~al.}(2012)\citenamefont
  {{Shull}}, \citenamefont {{Smith}},\ and\ \citenamefont
  {{Danforth}}}]{2012ApJ...759...23S}%
  \BibitemOpen
  \bibfield  {author} {\bibinfo {author} {\bibfnamefont {J.~M.}\ \bibnamefont
  {{Shull}}}, \bibinfo {author} {\bibfnamefont {B.~D.}\ \bibnamefont
  {{Smith}}},\ and\ \bibinfo {author} {\bibfnamefont {C.~W.}\ \bibnamefont
  {{Danforth}}},\ }\bibfield  {title} {\bibinfo {title} {{The Baryon Census in
  a Multiphase Intergalactic Medium: 30\% of the Baryons May Still be
  Missing}},\ }\href {https://doi.org/10.1088/0004-637X/759/1/23} {\bibfield
  {journal} {\bibinfo  {journal} {Astrophys. J.}\ }\textbf {\bibinfo {volume}
  {759}},\ \bibinfo {eid} {23} (\bibinfo {year} {2012})},\ \Eprint
  {https://arxiv.org/abs/1112.2706} {arXiv:1112.2706 [astro-ph.CO]}
  \BibitemShut {NoStop}%
\bibitem [{\citenamefont {Nicastro}\ \emph {et~al.}(2018)\citenamefont
  {Nicastro} \emph {et~al.}}]{2018Natur.558..406N}%
  \BibitemOpen
  \bibfield  {author} {\bibinfo {author} {\bibfnamefont {F.}~\bibnamefont
  {Nicastro}} \emph {et~al.},\ }\bibfield  {title} {\bibinfo {title}
  {{Observations of the missing baryons in the warm-hot intergalactic
  medium}},\ }\href {https://doi.org/10.1038/s41586-018-0204-1} {\bibfield
  {journal} {\bibinfo  {journal} {Nature}\ }\textbf {\bibinfo {volume} {558}},\
  \bibinfo {pages} {406} (\bibinfo {year} {2018})},\ \Eprint
  {https://arxiv.org/abs/1806.08395} {arXiv:1806.08395 [astro-ph.GA]}
  \BibitemShut {NoStop}%
\bibitem [{\citenamefont {{de Graaff, Anna}}\ \emph {et~al.}(2019)\citenamefont
  {{de Graaff, Anna}}, \citenamefont {{Cai, Yan-Chuan}}, \citenamefont
  {{Heymans, Catherine}},\ and\ \citenamefont {{Peacock, John
  A.}}}]{deGraaff2019}%
  \BibitemOpen
  \bibfield  {author} {\bibinfo {author} {\bibnamefont {{de Graaff, Anna}}},
  \bibinfo {author} {\bibnamefont {{Cai, Yan-Chuan}}}, \bibinfo {author}
  {\bibnamefont {{Heymans, Catherine}}},\ and\ \bibinfo {author} {\bibnamefont
  {{Peacock, John A.}}},\ }\bibfield  {title} {\bibinfo {title} {Probing the
  missing baryons with the sunyaev-zel\'{}dovich effect from filaments},\
  }\href {https://doi.org/10.1051/0004-6361/201935159} {\bibfield  {journal}
  {\bibinfo  {journal} {A\&A}\ }\textbf {\bibinfo {volume} {624}},\ \bibinfo
  {pages} {A48} (\bibinfo {year} {2019})}\BibitemShut {NoStop}%
\bibitem [{\citenamefont {Johnson}\ \emph {et~al.}(2019)\citenamefont {Johnson}
  \emph {et~al.}}]{Johnson_2019}%
  \BibitemOpen
  \bibfield  {author} {\bibinfo {author} {\bibfnamefont {S.~D.}\ \bibnamefont
  {Johnson}} \emph {et~al.},\ }\bibfield  {title} {\bibinfo {title} {The
  physical origins of the identified and still missing components of the
  warm{\textendash}hot intergalactic medium: Insights from deep surveys in the
  field of blazar 1es1553+113},\ }\href
  {https://doi.org/10.3847/2041-8213/ab479a} {\bibfield  {journal} {\bibinfo
  {journal} {The Astrophysical Journal}\ }\textbf {\bibinfo {volume} {884}},\
  \bibinfo {pages} {L31} (\bibinfo {year} {2019})}\BibitemShut {NoStop}%
\bibitem [{\citenamefont {Kov{\'{a}}cs}\ \emph {et~al.}(2019)\citenamefont
  {Kov{\'{a}}cs}, \citenamefont {Bogd{\'{a}}n}, \citenamefont {Smith},
  \citenamefont {Kraft},\ and\ \citenamefont {Forman}}]{Kov_cs_2019}%
  \BibitemOpen
  \bibfield  {author} {\bibinfo {author} {\bibfnamefont {O.~E.}\ \bibnamefont
  {Kov{\'{a}}cs}}, \bibinfo {author} {\bibfnamefont {{\'{A}}.}~\bibnamefont
  {Bogd{\'{a}}n}}, \bibinfo {author} {\bibfnamefont {R.~K.}\ \bibnamefont
  {Smith}}, \bibinfo {author} {\bibfnamefont {R.~P.}\ \bibnamefont {Kraft}},\
  and\ \bibinfo {author} {\bibfnamefont {W.~R.}\ \bibnamefont {Forman}},\
  }\bibfield  {title} {\bibinfo {title} {Detection of the missing baryons
  toward the sightline of h1821+643},\ }\href
  {https://doi.org/10.3847/1538-4357/aaef78} {\bibfield  {journal} {\bibinfo
  {journal} {The Astrophysical Journal}\ }\textbf {\bibinfo {volume} {872}},\
  \bibinfo {pages} {83} (\bibinfo {year} {2019})}\BibitemShut {NoStop}%
\bibitem [{\citenamefont {Macquart}\ \emph {et~al.}(2020)\citenamefont
  {Macquart} \emph {et~al.}}]{Macquart:2020lln}%
  \BibitemOpen
  \bibfield  {author} {\bibinfo {author} {\bibfnamefont {J.-P.}\ \bibnamefont
  {Macquart}} \emph {et~al.},\ }\bibfield  {title} {\bibinfo {title} {{A census
  of baryons in the Universe from localized fast radio bursts}},\ }\href
  {https://doi.org/10.1038/s41586-020-2300-2} {\bibfield  {journal} {\bibinfo
  {journal} {Nature}\ }\textbf {\bibinfo {volume} {581}},\ \bibinfo {pages}
  {391} (\bibinfo {year} {2020})},\ \Eprint {https://arxiv.org/abs/2005.13161}
  {arXiv:2005.13161 [astro-ph.CO]} \BibitemShut {NoStop}%
\end{thebibliography}%

\end{document}